\documentclass[
  reprint,
  twocolumn,
  prx,
  aps,
  superscriptaddress,
  nofootinbib,
  amsmath,
  amssymb,
  showpacs=false
]{revtex4-2}

\usepackage[utf8]{inputenc}
\usepackage[T1]{fontenc}
\usepackage{lmodern}
\usepackage{graphicx}
\usepackage{bm}
\usepackage{dsfont}
\usepackage{xcolor}
\usepackage{booktabs}
\usepackage{placeins}

\usepackage[colorlinks=true,linkcolor=blue,citecolor=blue,urlcolor=blue]{hyperref}
\usepackage{microtype}
\usepackage{enumitem}

\newcommand{\vx}{\mathbf{x}}
\newcommand{\mA}{\mathbf{A}}
\newcommand{\mB}{\mathbf{B}}
\newcommand{\mC}{\mathbf{C}}
\newcommand{\mGamma}{\boldsymbol{\Gamma}}
\newcommand{\vxi}{\boldsymbol{\xi}}

\begin{document}

\title{No persistent circadian oscillator at genome resolution:\\
       pseudo-coherence in gut microbiome dynamics}

\author{V.~Troude}
\affiliation{Institute of Risks Analysis, Prediction and Management,
  Southern University of Science and Technology, Shenzhen, China}
\author{L.~Takayasu}
\affiliation{Meinig School of Biomedical Engineering, Cornell University, Ithaca, NY, USA}
\author{R.~Maskawa}
\affiliation{Graduate School of Informatics, Nagoya University, Nagoya, Japan}
\author{W.~Suda}
\affiliation{RIKEN Center for Integrative Medical Sciences, Yokohama, Japan}
\author{D.~Sornette}
\affiliation{Institute of Risks Analysis, Prediction and Management,
  Southern University of Science and Technology, Shenzhen, China}
\author{H.~Takayasu}
\affiliation{Department of Computer Science, Institute of Science Tokyo, Yokohama, Japan}
\author{M.~Takayasu}
\affiliation{Department of Computer Science, Institute of Science Tokyo, Yokohama, Japan}

\date{\today}

\begin{abstract}
Diurnal rhythms in the gut microbiome are commonly read as evidence of
host-driven entrainment or of microbial oscillators that synchronise to a
common clock. We reanalyse hourly genome-resolved (MAG-level) mouse-gut
time series with diagnostics tailored to test that interpretation. At this
resolution and for both animals in the dataset, the time-frequency
representation carries no persistent ridge; the time-averaged spectrum is
enhanced at low frequencies and depleted at intermediate frequencies; the
lagged covariance is markedly time-asymmetric, with a global imbalance peak
near tens of hours; and an amplitude-adjusted Fourier surrogate test
identifies a weak time-averaged construction in the candidate circadian
band, never as a fixed time-frequency ridge. The two functional guilds
that carry the inferred non-normal amplification are identified independently by the rankings of two inferred dynamical modes (the reaction mode, into which fluctuations are transiently amplified, and the non-normal mode, which injects them), and
recover the primary polysaccharide degraders of Bacteroidota and the
secondary butyrate and propionate fermenters of Bacillota A without
invoking any phase information. The conjunction of these signatures
matches a stable but strongly non-normal stochastic regime, that is,
pseudo-coherence: geometric amplification reshapes stochastic fluctuations
onto a low-dimensional reaction subspace, producing intermittent
synchronisation-like episodes, broken time-reversal symmetry, and
emergent time-averaged characteristic scales without an underlying
oscillator. We propose a falsifiable test via high-resolution clock-gene-knockout cohorts.
\end{abstract}

\maketitle

\section{Introduction}\label{sec:intro}

Diurnal patterns in microbiome abundance and host-microbe coupling are
extensively documented \cite{Thaiss2014Transkingdom,Zarrinpar2014Diet,Liang2015Rhythmicity,Leone2015Diurnal,Asher2015TimeForFood}.
Synchronisation, clustering, and time-frequency structure of microbial
composition are routinely interpreted with the conceptual toolkit of
coupled phase oscillators \cite{kuramoto1975self,acebron2005kuramoto,pikovsky2003synchronization}.
This toolkit makes a strong physical assumption: the system contains
intrinsic or effectively entrained oscillators, possibly close to a Hopf
bifurcation, that phase-lock under common forcing. Once oscillators are
postulated, observed rhythmicity, anti-synchronised clusters, and
finite-time spectral peaks are read as evidence of phase locking.

We ask a different question. Must the time-frequency structure of a
genome-resolved gut microbiome originate from oscillators, or is it
consistent with a stable, oscillator-free stochastic dynamics in which
geometric properties of the interaction matrix, rather than eigenvalue
criticality, organise the collective behaviour?

The relevant geometric mechanism is non-normal transient amplification.
For a stable linear stochastic system $\dot\vx = \mA\vx + \vxi$, when the
operator $\mA$ has non-orthogonal eigenvectors, perturbations can be
transiently amplified by factors far larger than predicted by modal
stability theory \cite{trefethen2005spectra,Farrell1996GST1,trefethen1993hydrodynamic,asllani2018theory}.
In the stochastic case, the amplification funnels fluctuations onto a
low-dimensional reaction subspace, producing intermittent collective
excursions, irreversible probability currents, and finite-window spectral
peaks that drift in time. Ref.~\cite{troude2026pseudoco,troude2025Unifying,troude2025illusion}
identifies a sharp geometric transition (pseudo-criticality) at which
these features turn on while the system remains spectrally stable and
admits no Hopf bifurcation. Broader physical implications of non-normal architectures
have recently been argued from a non-equilibrium-steady-state standpoint
\cite{sornette2025life}, where life itself is interpreted through the non-normal amplification of fluxes, organising asymmetric reaction networks to amplify free-energy throughput and entropy export. The gut microbiome,
embedded in a host that imposes circadian and dietary flux, is a natural
empirical substrate for this picture.

Asymmetric processes also dominate the microbiome at the interaction
level: directed polysaccharide-utilisation cascades among Bacteroides and
related primary degraders \cite{koropatkin2012,rakoffnahoum2014,rakoffnahoum2016,mahowald2009,flint2012gm,sonnenburg2014};
one-directional cross-feeding of metabolic by-products (acetate, lactate,
formate) into secondary fermenters such as \emph{Faecalibacterium} and
Lachnospiraceae \cite{belenguer2006,falony2006,louis2017,denbesten2013};
predation by \emph{Bdellovibrio bacteriovorus} \cite{iebba2013};
bacteriocin warfare \cite{riley2002}; and bile-acid mediated chemical
inhibition \cite{wahlstrom2016}. Time-series inference of generalised
Lotka-Volterra interaction matrices on murine and human gut data returns
non-symmetric off-diagonal coefficients \cite{stein2013,bucci2014,coyte2015};
statistical-physics models of directed cross-feeding networks generate
multistability and bursty collective dynamics
\cite{goyal2018prl,goyal2018isme}. ``Reactivity'', the canonical ecological
footprint of non-normal transient growth, has been formalised for decades
\cite{neubert1997}.

A long tradition in stochastic ecology and stochastic neuroscience has
explored noise-driven collective rhythms in linearly stable systems
without intrinsic oscillators. Demographic-noise quasi-cycles in
predator-prey communities \cite{nisbet1976mechanism,mckane2005predatorprey}
exploit a complex-eigenvalue pair with negative real part: the linear
Jacobian carries a damped oscillation at finite frequency and white
noise resonantly amplifies this intrinsic mode, producing persistent
spectral peaks. Non-normality enters this picture as a multiplier of
the amplification: Refs.~\cite{nicoletti2018nonnormal,muolo2019patterns}
widen and shift the resulting quasi-cycle peaks via the numerical
abscissa of the non-normal Jacobian, but the pre-existing complex
eigenvalues are still required. Ref.~\cite{biancalani2017giant} uses
non-normality to amplify the spatial amplitude of fluctuation-induced
Turing patterns, with a steady-state spatial wavenumber as the
observable. Recent work on directed neural networks
\cite{poggialini2025nonreciprocal,Hennequin2014,Murphy2009,Ganguli2008}
extends the framework to mixed-spectrum non-reciprocal architectures
with reactive transients and effective dimensionality reduction.
These works establish that non-normality is a generic amplifier of
fluctuation-driven structure in stable linear stochastic systems.
The framework we use here \cite{troude2026pseudoco,troude2025Unifying}
differs in two respects from this tradition: (i)~the linearised
operator that produces pseudo-coherent organisation is taken to have
strictly real negative eigenvalues, so there is no pre-existing
complex mode to resonate with, and (ii)~the collective observable is
a time-resolved cluster order parameter, not a spatial wavenumber or
a stationary peak. Pseudo-coherence is the regime in which non-normal
amplification by itself reorganises the imaginary pseudospectrum
sufficiently to produce drifting finite-window spectral peaks,
transient phase alignment, and broken time-reversal symmetry, in a setting with a strictly real spectrum, additive noise, and autonomous linear dynamics. The
present empirical analysis tests whether this regime is realised in
genome-resolved gut-microbiome dynamics, using observables (cluster
order parameter, lead-lag imbalance, wavelet coherence between
support and synchronization, co-membership cluster recovery) that
are diagnostic of pseudo-coherence and that do not require the
detection of an intrinsic frequency.

Using the genome-resolved (MAG-level) mouse-gut dataset of
Ref.~\cite{microbiome2025}, we test whether the system carries the joint
signatures of pseudo-coherence:
\begin{enumerate}[label=(\roman*),leftmargin=*]
\item intermittent cluster-level phase alignment, governed by the spatial
  extent of the inferred reaction mode rather than by eigenvalue
  criticality;
\item broken time-reversal symmetry quantified by a lagged covariance
  imbalance, with a global maximum at intermediate lag;
\item a time-averaged spectrum with low-frequency enhancement and depleted
  intermediate frequencies, no persistent time-frequency ridge, and
  strong temporal intermittency in the bands carrying the largest power;
\item independent identification, from the reaction and non-normal mode
  rankings, of the two functional guilds that drive and absorb the
  amplification.
\end{enumerate}
All four signatures are observed in both animals. They are jointly
consistent with pseudo-coherence and jointly inconsistent with persistent
oscillator-based synchronisation at the MAG scale. Where the surrogate
analysis admits a weak modulation, it is at the time-averaged level only,
with no fixed time-frequency ridge, which is the operational signature
predicted by the theory. We close with a falsifiable prediction for
clock-gene knockout cohorts.

The claim is scale-specific. The host circadian clock is not denied, and
rhythmic modulation at the coarse bacterial-abundance level is not denied.
What is denied is circadian \emph{dominance} at the genome-resolved level,
and an alternative null (pseudo-coherence) is offered that fits the data.

\section{Geometric setup}\label{sec:theory}

\paragraph{Minimal non-normal model.}
Consider a linear overdamped stochastic dynamics
\begin{equation}
  \dot\vx(t) = \mA\vx(t) + \vxi(t),\qquad \mA \in \mathbb{R}^{N\times N},
  \label{eq:VAR}
\end{equation}
with $\vxi$ a zero-mean white noise and all eigenvalues of $\mA$ in the
strict left half-plane. When $\mA$ is normal, relaxation is purely
Ornstein-Uhlenbeck and the spectrum is featureless. For non-normal $\mA$
($[\mA,\mA^\top]\neq 0$), eigenvectors are non-orthogonal and finite-time
perturbations may grow transiently even though every eigenvalue is stable
\cite{trefethen2005spectra,Farrell1996GST1,trefethen1993hydrodynamic}.

When non-normality is strong, all leading transient amplification is
confined to a two-dimensional subspace spanned by a pair of maximally
aligned left and right singular directions of $\mA$. Diagonalisation of
the commutator
\begin{equation}
  \mB \;=\; \mA\mA^\top - \mA^\top \mA,
  \label{eq:commutator}
\end{equation}
which is real symmetric and traceless, returns this subspace as its
rank-two eigenstructure. Projection of $\mA$ onto the subspace yields a
$2\times 2$ matrix
\begin{equation}
  \mGamma \;=\;
  \begin{pmatrix} -\alpha & \beta\kappa \\ \beta/\kappa & -\alpha
  \end{pmatrix},\qquad \alpha>\beta>0,\;\kappa\ge 1,
  \label{eq:Gamma}
\end{equation}
whose eigenvalues
\begin{equation}
  \lambda_\pm \;=\; -\alpha \pm \beta
  \label{eq:lambda}
\end{equation}
remain strictly real and negative for all $\kappa$. The parameter $\kappa$
measures eigenvector alignment ($\kappa=1$ normal, $\kappa\gg 1$ strongly
non-normal). The two eigendirections of $\Gamma$ in \eqref{eq:Gamma} define the modes: the non-normal mode $\hat{\mathbf{n}}$, which absorbs the stochastic forcing, and the reaction mode $\hat{\mathbf{r}}$ into which
perturbations are transiently redirected and amplified. We use the
non-normality index
\begin{equation}
  K \;=\; \tfrac{1}{2}\bigl(\kappa - \kappa^{-1}\bigr),
  \label{eq:K}
\end{equation}
and the geometric threshold
\begin{equation}
  K_c \;=\; \sqrt{\frac{\sqrt{1-\delta^2}}{1-\sqrt{1-\delta^2}}},
  \qquad \delta = \biggl|\frac{\beta}{\alpha}\biggr|,
  \label{eq:Kc}
\end{equation}
above which the system is reactive: a small perturbation transiently grows before it decays, so that the maximal transient gain $G=\max_{t>0}\lVert e^{\Gamma t}\rVert$ exceeds one even though the spectrum stays real and stable \cite{troude2025Unifying}.

\paragraph{Pseudo-coherence.}
Above $K/K_c = 1$, the theoretical work
\cite{troude2026pseudoco,troude2025Unifying,troude2025illusion} establishes
a sharp geometric transition: the reaction mode acquires extensive support
across system components; stochastic fluctuations concentrate onto this
subspace; the imaginary pseudospectrum reshapes the marginal stochastic
spectrum to amplify slow components and suppress intermediate frequencies;
the lagged covariance becomes asymmetric and a finite imbalance appears;
and finite-window spectra develop drifting peaks that disappear under
stationary, long-time averaging. No eigenvalue crosses the imaginary axis,
no Hopf bifurcation occurs, and no oscillator is required. The empirical
fingerprint of the regime is the conjunction of (i)-(iv) of
Sec.~\ref{sec:intro}.

\paragraph{Per-MAG participation and mode support.}
Once the rank-two subspace is identified, the reaction and non-normal
modes are vectors in the original genome-resolved coordinate system. Their
per-MAG magnitudes
\begin{equation}
  |r_i(t)|\;\;\text{and}\;\;|n_i(t)|
  \label{eq:participation}
\end{equation}
give, at every time $t$, the absolute loading of MAG $i$ on the reaction mode and on the non-normal mode. Because each mode is normalised to unit Euclidean norm ($\sum_i r_i^2=1$), these loadings are components of a unit vector, not fractional shares; their time average $\langle|r_i|\rangle_t$ is the mean absolute mode loading reported below. The
commutator $\mB$ has eigenvectors defined up to a common sign, so only
$|r_i|$ and $|n_i|$ are biologically interpretable: the sign of each mode
is a calibration gauge and is not used. The global \emph{support} of each
mode, that is, how broadly the participation spreads across the $N$ MAGs,
is summarised by
\begin{equation}
  s_r(t) \;=\; \frac{1}{\sqrt{N}}\sum_{i=1}^{N} |r_i(t)|,
  \qquad
  s_n(t) \;=\; \frac{1}{\sqrt{N}}\sum_{i=1}^{N} |n_i(t)|,
  \label{eq:support}
\end{equation}
with $s_{r,n} \to 1$ for a uniformly distributed mode and
$s_{r,n} \to 1/\sqrt{N}$ for a fully localised one. The accompanying
theoretical work \cite{troude2026pseudoco} shows that as $s_r$ and $s_n$
grow, the amplified subspace becomes extensive, irreversibility and
entropy production rise jointly, and macroscopic phase coherence emerges
as a secondary consequence: support, not eigenvalue proximity to
instability, is the geometric order parameter of pseudo-coherence.

\paragraph{Cluster-resolved phase coherence.}
We follow the non-parametric phase determination of
Ref.~\cite{microbiome2025} (briefly recalled in the Methods) to obtain a
phase $\theta_i(t)$ for each MAG $i$. The cyc7plus assignment of MAGs into
two phase clusters $\mathcal{C}_1$ and $\mathcal{C}_2$
\cite{microbiome2025}, obtained by average-linkage hierarchical
clustering on pairwise circular-phase distance, is taken as the input to
two cluster-resolved Kuramoto-like order parameters
\begin{equation}
  R_{\mathcal{C}}(t)
  \;=\;
  \left| \frac{1}{|\mathcal{C}|} \sum_{i\in\mathcal{C}}
         e^{\mathrm{i}\,\theta_i(t)} \right|,
  \qquad \mathcal{C}\in\{\mathcal{C}_1,\mathcal{C}_2\},
  \label{eq:Rcluster}
\end{equation}
which take values in $[0,1]$: $R_{\mathcal{C}}=1$ when all MAGs in cluster
$\mathcal{C}$ share the same instantaneous phase and $R_{\mathcal{C}}\to 0$
when phases are uniformly scattered around the unit circle. The two
clusters typically realise opposite values of $\cos\theta$ at a given time
and read as anti-synchronised; in the pseudo-coherent framework this
opposition is the geometric sign structure of the reaction mode rather
than evidence of competing oscillator populations
\cite{troude2026pseudoco,troude2025Unifying}.

\begin{figure}[t]
  \centering
  \includegraphics[width=0.85\columnwidth]{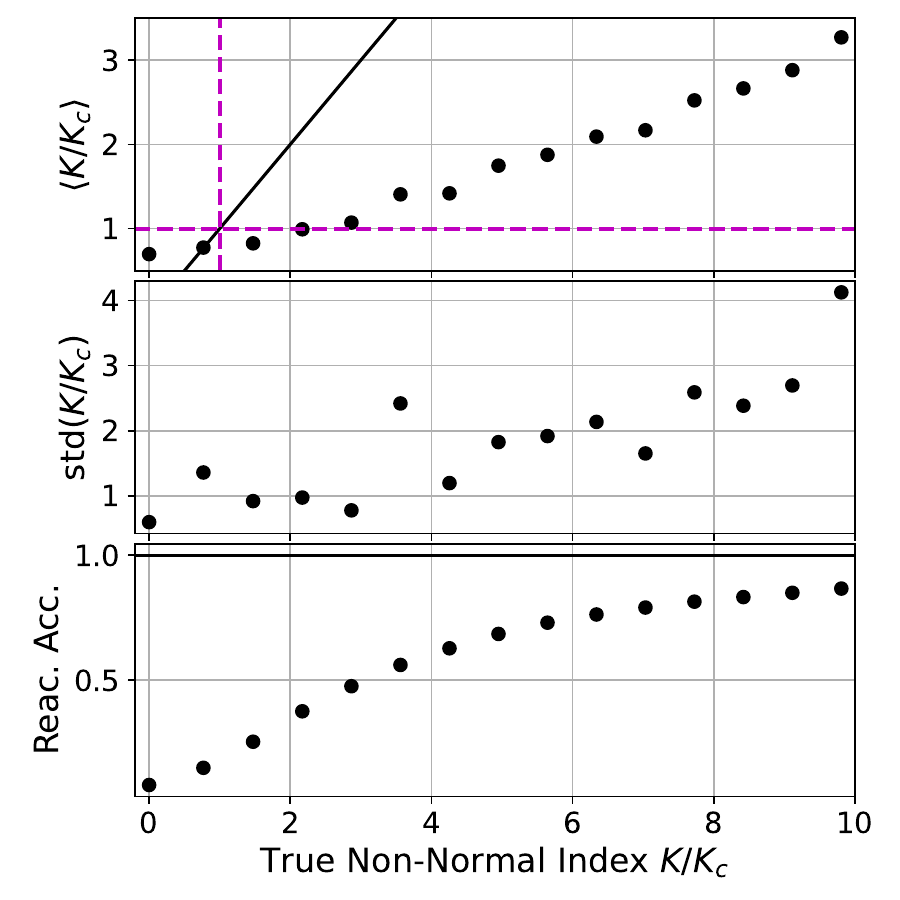}
  \caption{\textbf{Synthetic calibration of the local non-normal
    inference.} Inferred $\langle K/K_c\rangle$ (top), estimator standard
    deviation (middle), and reaction-mode recovery accuracy (bottom)
    versus true $K/K_c$.}
  \label{fig:calibration}
\end{figure}

\paragraph{Time-resolved inference.}
Over two successive sampling steps the local Jacobian is taken constant
and is estimated from four consecutive observations
$(\vx_{k-3},\vx_{k-2},\vx_{k-1},\vx_k)$ by the least-squares update
\begin{equation}
  \widehat{\mA}_k \;=\; \mathbf{Y}_k\,\mathbf{X}_k^{+},
  \label{eq:Ahat}
\end{equation}
with $\mathbf{Y}_k = (\vx_k,\vx_{k-1},\vx_{k-2})$,
$\mathbf{X}_k = (\vx_{k-1},\vx_{k-2},\vx_{k-3})$, and $\mathbf{X}_k^{+}$
the Moore-Penrose pseudo-inverse. The full Jacobian $\widehat{\mA}_k$ is
high-dimensional and ill-conditioned, but the non-normal amplification is
rank two and is robustly extracted by diagonalising the commutator
$\mB_k = [\widehat{\mA}_k,\widehat{\mA}_k^\top]$ from
Eq.~\eqref{eq:commutator}, which is small and well-conditioned even when
$\widehat{\mA}_k$ is itself nearly degenerate. Synthetic calibration
(Methods, Fig.~\ref{fig:calibration}) shows that the inferred $K/K_c$
never produces false positives, that estimator variance grows with the
true $K/K_c$ and itself serves as a diagnostic, and that reaction-mode
recovery becomes essentially perfect for $K/K_c \gtrsim 2$.

\section{Results}\label{sec:results}

We apply the pipeline of Sec.~\ref{sec:theory} to the genome-resolved
mouse-gut time series of Ref.~\cite{microbiome2025}, hourly samples across
two weeks of two animals (Mouse A and Mouse B). The two animals are
treated in parallel throughout, and every diagnostic is reported for both.

\subsection{Cluster order parameters and non-normal diagnostics through time}\label{sec:dynamics}

\begin{figure*}[t]
  \centering
  \includegraphics[width=\textwidth]{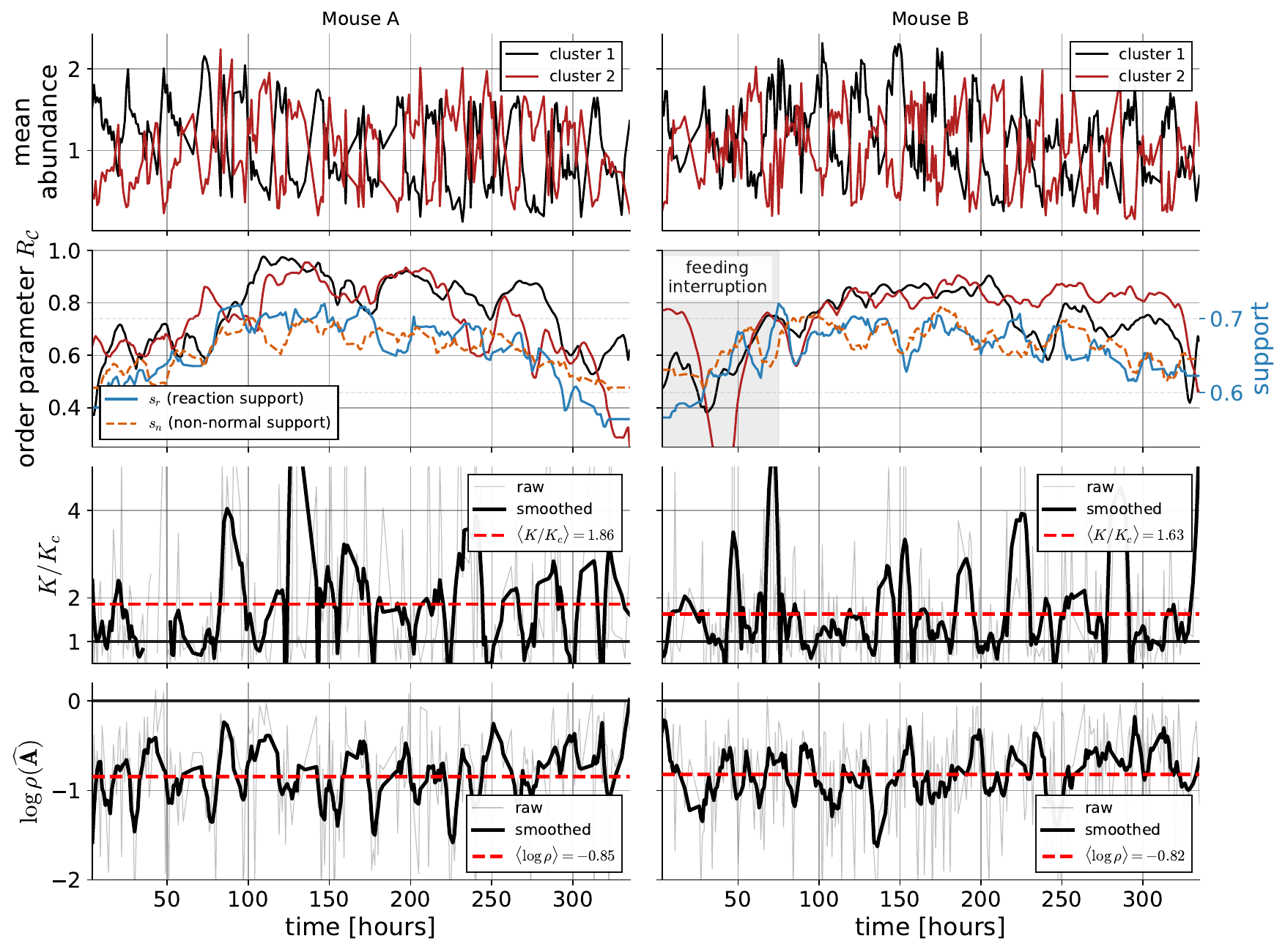}
  \caption{\textbf{Cluster order parameters, mode support, and non-normal
    diagnostics through time, Mouse A (left) and Mouse B (right).}
    Rows from top to bottom: cluster-mean abundances; cluster-resolved
    Kuramoto-like order parameters (black, firebrick) overlaid with the
    reaction- and non-normal-mode support $s_r(t)$, $s_n(t)$ of
    Eq.~\eqref{eq:support} on the twin axis (blue solid, orange dashed);
    time-resolved $K/K_c$ with the geometric threshold $K/K_c = 1$ as a
    thick black line and the record-mean as a red dashed line on top
    (annotated value at the right edge); log spectral radius of the
    locally inferred Jacobian with the stability threshold
    $\log\rho = 0$ as a thick black line and the record-mean as a red
    dashed line. The first $75\,\mathrm{h}$ of the Mouse~B record carry
    a documented nutritional perturbation (grey shaded window).}
  \label{fig:series}
\end{figure*}

We follow the non-parametric phase determination (NPPD) of
Ref.~\cite{microbiome2025} (recalled in the Methods) to obtain a phase
$\theta_i(t)$ for each MAG $i$. The cyc7plus assignment of MAGs to two
phase clusters $\mathcal{C}_1, \mathcal{C}_2$ \cite{microbiome2025},
obtained by average-linkage hierarchical clustering on circular-phase
distance, gives the two cluster-resolved Kuramoto-like order parameters
of Eq.~\eqref{eq:Rcluster}, and our local non-normal calibration
returns the time series $K/K_c(t)$, the log spectral radius
$\log \rho(\widehat{\mA}(t))$, and the mode supports $s_r(t)$, $s_n(t)$
of Eq.~\eqref{eq:support}. All four observables are shown side by side
for both animals in Fig.~\ref{fig:series}.

The system is linearly stable at every time in both animals
($\log \rho < 0$ throughout). The record-mean $K/K_c$ sits clearly above
unity (red dashed lines on the third row of Fig.~\ref{fig:series}): the
inferred local Jacobian is in a strongly non-normal regime on average,
without ever crossing into spectral instability. The cluster
order-parameter excursions in the second row line up with the support
excursions (twin axis, blue and orange) and with the $K/K_c$ excursions
in the third row: macroscopic phase coherence rises when the amplified
subspace becomes more extensive. In Mouse B, the first 75 h of the
record carry a known nutritional perturbation that disrupted phase
synchronisation transiently (grey shaded window); the local non-normal
observables in the lower panels are unaffected and remain informative
through that interval.

\paragraph{Synchronisation tracks reaction-mode support.}

\begin{figure}[t]
  \centering
  \includegraphics[width=\columnwidth]{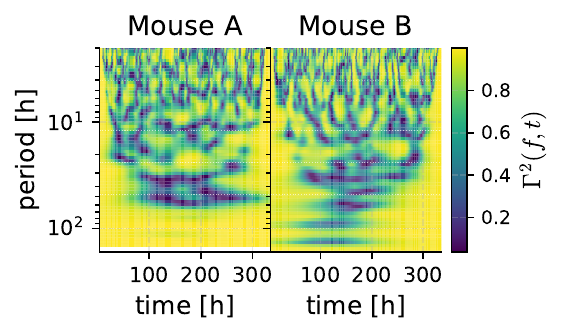}
  \caption{\textbf{Wavelet coherence between the reaction-mode support
  $s_r(t)$ and the cluster-averaged order parameter
  $\langle R_{\mathcal{C}}\rangle(t)$, per mouse.} Mean cross-Morlet
  coherence $\Gamma^2(f, t)$ for Mouse~A (left) and Mouse~B (right)
  on the same period axis with a single shared colorbar.}
  \label{fig:coherence}
\end{figure}

\begin{figure}[t]
  \centering
  \includegraphics[width=\columnwidth]{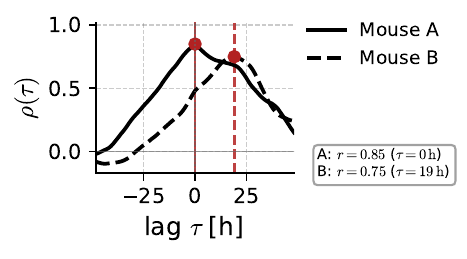}
  \caption{\textbf{Lagged cross-correlation between $s_r(t)$ and
  $\langle R_{\mathcal{C}}\rangle(t)$, per mouse.}
  $\rho(\tau)$ for $\tau \in [-48, +48]\,\mathrm{h}$; Mouse~A in
  solid black, Mouse~B in dashed black; the peak $(\rho, \tau)$ is
  marked with a red vertical guide and annotated. Here $\rho(\tau)$ correlates $s_r(t)$ with $\langle R_{\mathcal{C}}\rangle(t+\tau)$, so a positive peak lag means the support $s_r$ reaches its maximum before (leads) the cluster synchronization.}
  \label{fig:xcorr}
\end{figure}

The supports $s_r(t)$ and $s_n(t)$ on the twin axis of the second row of
Fig.~\ref{fig:series} grow in the same time windows in which the cluster
order parameters peak. To make the link quantitative without imposing
an instantaneous-regression assumption we compute the wavelet coherence
between $s_r(t)$ and the cluster-averaged order parameter
$\langle R_{\mathcal{C}}\rangle(t) = \tfrac{1}{2}\bigl[R_{\mathcal{C}_1}(t) + R_{\mathcal{C}_2}(t)\bigr]$
on the same Morlet scale grid as the marginal spectrum. In both
animals the mean coherence sits in the range $[0.55, 0.70]$ across
every resolved frequency band, with a between-mice gap of at most
$0.06$ (Table~\ref{tab:coherence}). The link is therefore present at
every timescale we resolve and is of comparable strength in the two
animals studied (Fig.~\ref{fig:coherence} and
Table~\ref{tab:coherence}). Linear Granger causality at lag
$4\;\mathrm{h}$ is consistent with the directionality $s_r \to
\langle R_{\mathcal{C}}\rangle$ in both mice (Mouse~A $p = 0.033$;
Mouse~B $p = 0.061$; reverse direction not significant in either,
$p = 0.156$ and $p = 0.131$), and the lagged cross-correlation
(Fig.~\ref{fig:xcorr}) peaks at zero lag in
Mouse~A and at $+19\;\mathrm{h}$ in Mouse~B,  the reaction-mode support reaches its maximum roughly one day before the cluster synchronization in Mouse~B. 
Macroscopic coherence is controlled by how broadly the amplified
excursion spreads across genomes, not by how strongly one could in
principle amplify. This is the geometric signature of pseudo-coherence
\cite{troude2026pseudoco},  order is driven by
support, not by proximity to a spectral instability that never occurs.

\begin{table}[h]
  \centering\small
  \caption{\textbf{Wavelet coherence between the reaction-mode support
  $s_r(t)$ and the cluster-averaged order parameter
  $\langle R_{\mathcal{C}}\rangle(t)$, in four frequency bands.}
  Mean cross-Morlet coherence on the same scale grid as the marginal
  spectrum. The coherence sits in $[0.55, 0.70]$ in every band of
  both animals, with a between-mice gap of at most $0.06$,
  demonstrating that the support--coherence link is present at every
  resolved timescale and is of comparable strength in the two mice
  studied.}
  \label{tab:coherence}
  \begin{tabular}{lcccc}
  \hline
  Mouse & $4$--$12$\,h & $12$--$24$\,h & $24$--$48$\,h & $48$--$100$\,h \\
  \hline
  A & $0.61$ & $0.67$ & $0.58$ & $0.70$ \\
  B & $0.65$ & $0.66$ & $0.55$ & $0.64$ \\
  \hline
  \end{tabular}
\end{table}

\paragraph{Summary.}
 Order is driven by the spatial extent of the inferred reaction subspace,
in both animals. Support and entropy production are linked in the theory
\cite{troude2026pseudoco,sornette2025life}, and the empirical $s_r$, $s_n$
traces here are the dynamical observable behind that link.

\subsection{The marginal spectrum carries a weak time-averaged construction without a persistent ridge}\label{sec:spectrum}

\begin{figure*}[t]
  \centering
  \includegraphics[width=\textwidth]{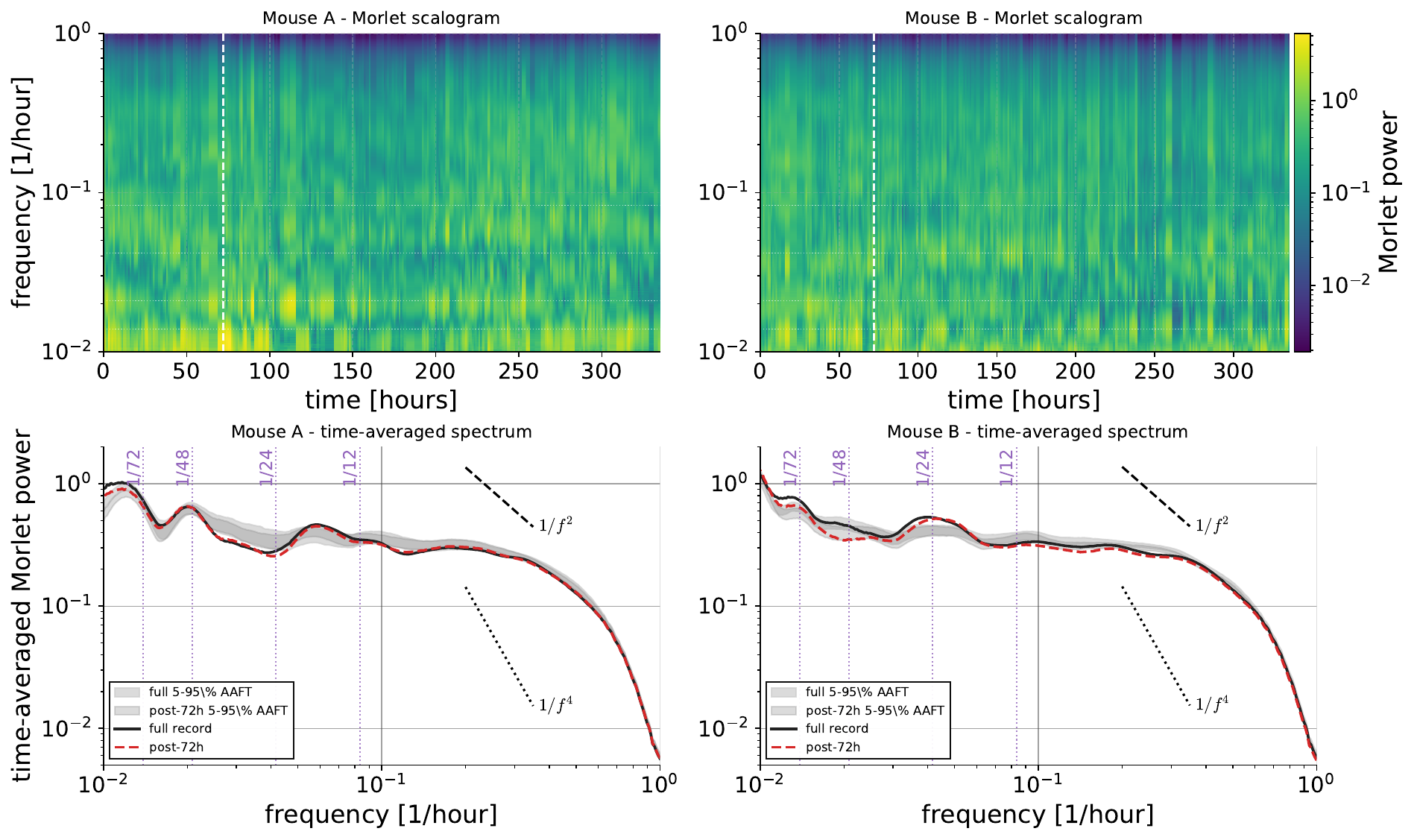}
  \caption{\textbf{Time-frequency structure for Mouse A (left) and Mouse B
    (right).} Top row: Morlet wavelet scalogram of the MAG-averaged power
    on the full record; the vertical dashed line marks the
    $72\,\mathrm{h}$ cage-transfer cutoff. Bottom row: time-averaged
    Morlet spectrum, solid black for the full record and dashed red for
    the post-$72\,\mathrm{h}$ window, superposed on the $5\%$--$95\%$
    envelope of $250$ amplitude-adjusted Fourier surrogates per MAG
    \cite{theiler1992,schreiber1996,schreiber2000,lancaster2018}.
    Vertical dotted lines mark the candidate harmonic frequencies
    $1/12, 1/24, 1/48, 1/72\,\mathrm{h}^{-1}$; a dotted black guide
    shows the $1/f^{4}$ reference slope. The candidate low-frequency band below $1/10\,\mathrm{h}^{-1}$ sits in the left portion of the bottom-row panels.}
  \label{fig:wavelet}
\end{figure*}

The time-frequency representation (Fig.~\ref{fig:wavelet}, top row)
carries no persistent ridge at any frequency in either animal. The
time-averaged spectrum (bottom row) shows a weak low-frequency
construction in both mice. To test the significance of this construction,
we generate $N_\mathrm{surr}=250$ amplitude-adjusted Fourier (AAFT) surrogates per
MAG \cite{theiler1992,schreiber1996,schreiber2000,lancaster2018}: each
surrogate preserves the per-MAG one-point distribution and approximately
the marginal power spectrum but destroys phase coherence across MAGs and
non-stationary temporal structure. The time-averaged spectrum of the
observed dataset is compared, at each frequency, with the 5\%-95\%
empirical surrogate envelope, with the per-frequency exceedance $p$-values controlled across frequency bins by the Benjamini--Hochberg false-discovery-rate procedure at $q=0.05$, both on the full record and on the
post-72 h window after the cage-transfer transient
\cite{carmody2015,friswell2010,david2014}. The candidate harmonic
locations $1/12$, $1/24$, $1/48$, $1/72\;\mathrm{h}^{-1}$ are marked.

The observed spectrum carries a construction in the low-frequency band
in both mice. The peak location falls in the range between roughly
$1/24$ and $1/16\;\mathrm{h}^{-1}$ in either animal (resolved in the low-frequency part of the bottom-row panels of Fig.~\ref{fig:wavelet}), and is not pinned to exactly
$1/24\;\mathrm{h}^{-1}$. The MAG-level data are obtained after a long
compression chain from raw reads to relative abundance to phase, and the
band averaging implicit in that pipeline can shift the constructive peak
by a few hours, so we read the peak location as a soft band rather than
a sharp frequency. What the surrogate test does establish in both
animals is that the low-frequency band carries a marginal excess over
the AAFT null. The scalograms in the top row of Fig.~\ref{fig:wavelet}
simultaneously show that this excess does \emph{not} take the form of a
persistent ridge in the time-frequency plane. The high-frequency tail
in both animals follows the $1/f^{4}$ guide line shown in the bottom
panels, which is the slope predicted by non-normal spectral reshaping
for a system with a strong reaction-mode amplification
\cite{troude2026pseudoco}: the tail therefore validates the
pseudo-coherent picture independently of the low-frequency band.

\paragraph{Summary.}
 A time-averaged characteristic time scale exists, in both animals, but no
persistent ridge does. This dissociation, a global construction without
a stationary ridge, is the operational signature of pseudo-coherence:
finite-window spectral concentrations drift in time, and only the average
across many windows shows the construction.

\subsection{Broken time-reversal symmetry with a global imbalance peak at intermediate lag}\label{sec:irreversibility}

\begin{figure}[t]
  \centering
  \includegraphics[width=\columnwidth]{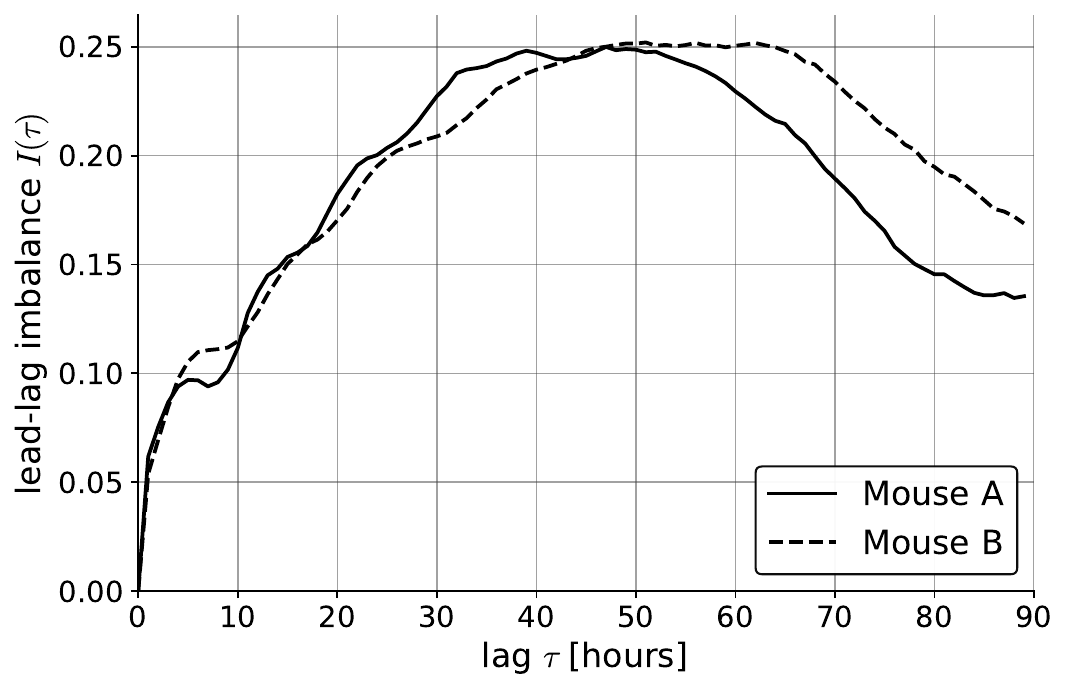}
  \caption{\textbf{Lead-lag covariance imbalance.}
    $I(\tau) = \|\mC(\tau) - \mC(\tau)^\top\|_F / \sqrt{2}$ as a function
    of lag, computed on the full state including the residual subspace;
    Mouse~A in solid black, Mouse~B in dashed black.}
  \label{fig:imbalance}
\end{figure}

The data $\vx(t)$ are centered to zero time-mean before the
covariance is computed, so the empirical estimator is unbiased under
time-reversal. The lagged covariance
\begin{equation}
  \mC(\tau) \;=\; \langle \vx(t)\,\vx(t+\tau)^\top \rangle,
  \label{eq:Ctau}
\end{equation}
with $\vx(t) = \widetilde{\vx}(t) - \langle\widetilde{\vx}\rangle_t$,
would be symmetric under time reversal; its Frobenius antisymmetric
component
\begin{equation}
  I(\tau) \;=\; \frac{1}{\sqrt{2}}\,
  \bigl\|\mC(\tau) - \mC(\tau)^\top\bigr\|_F
  \label{eq:Itau}
\end{equation}
measures the deviation. In both Mouse A and Mouse B
(Fig.~\ref{fig:imbalance}), $I(\tau)$ is strictly positive, with a clear
maximum at intermediate lag and a gradual decay at larger lag. A finite
imbalance peak is direct evidence of irreversibility and circulating
probability currents at the community level
\cite{seifert2012stochastic,gnesotto2018broken,fyodorov2025nonorthogonal,troude2026pseudoco}.

\paragraph{Why the peak sits at tens of hours.}
The peak of $I(\tau)$ does not localise a microscopic interaction time;
it localises a macroscopic memory. The biological inputs that the
microbiome processes are themselves multi-scale: intestinal transit (of
order a day in the mouse colon), the enterohepatic bile-acid axis
\cite{wahlstrom2016}, mucus turnover
\cite{johansson2008mucus,johansson2013}, and bacterial replication times
of $1$-$7\;\mathrm{h}$ depending on phylum
\cite{korem2015,brown2016irep}. The directed cross-feeding chain of
Sec.~\ref{sec:biology} convolves these scales: a perturbation entering
the primary-degrader guild traverses several biochemical layers before
appearing in the downstream fermenters, so the global imbalance peaks at
the length of the pipeline rather than at any single step. Diet response
in the murine gut sits at $1$-$3$ days \cite{carmody2015,david2014},
which matches the order of magnitude of the observed peak. The local
imaginary-pseudospectrum geometry of the inferred Jacobian
\cite{troude2026pseudoco} provides the corresponding microscopic
contribution.

\paragraph{Empirical entropy production rate.}

The small-$\tau$ expansion of Eq.~\eqref{eq:Itau_closed} defines a
directly measurable entropy-production proxy
$\Sigma_{\rm local}(t) = \left.\partial_\tau I(\tau, t)\right|_{\tau = 0}$
(Eq.~\eqref{eq:Sigma_local}). We estimate $\Sigma_{\rm local}(t)$ by a
linear fit to $I(\tau, t)$ on $\tau \in [0, 3]\;\mathrm{h}$ within a
$72$~h centred rolling window. The fit recovers the predicted
small-$\tau$ linearity with mean coefficient of determination
$R^2 = 0.957$ (Mouse~A) and $R^2 = 0.968$ (Mouse~B), so the
linear-coefficient identification of Eq.~\eqref{eq:Sigma_local} is the
correct empirical reduction of Eq.~\eqref{eq:Itau_closed} at hourly
sampling. The recovered $\Sigma_{\rm local}(t)$ is strictly positive at
every measurement window of both animals studied ($n = 2$; $100\%$ of
$172$ windows in Mouse~A and $209$ windows in Mouse~B) and
quasi-stationary in time, with coefficient of variation $0.18$ in
Mouse~A and $0.12$ in Mouse~B. The time-mean
$\langle\Sigma_{\rm local}\rangle$ agrees across the two animals to
$0.5\%$ (Table~\ref{tab:sigma}) despite their different feeding
histories and different time-mean
$\langle K/K_c\rangle$ ($1.86$ in Mouse~A, $1.63$ in Mouse~B). The
corresponding $\langle\Sigma_{\rm local}^2\rangle$ is proportional to
the non-equilibrium entropy production rate $\Phi$ of
Eq.~\eqref{eq:Phi_closed} through the isotropic-noise identity
$\Sigma_{\rm local}^2 = (\sigma^4 / 4\alpha)\,\Phi$. Both animals
therefore carry a strictly positive, stable, and cross-mouse-consistent
entropy production rate of the rank-two reaction subspace, and the
system is in a non-equilibrium steady state rather than near
equilibrium throughout the record~\cite{troude2026pseudoco}.

\begin{table}[h]
  \centering\small
  \caption{\textbf{Empirical entropy-production proxy
  $\Sigma_{\rm local}(t)$ summary statistics per mouse}, computed
  from the linear-fit slope of $I(\tau, t)$ on $\tau \in [0, 3]\;\mathrm{h}$
  in a $72$\,h centred rolling window. The time-mean
  $\langle\Sigma_{\rm local}\rangle$ agrees across the two animals
  studied to $0.5\%$. The proxy is strictly positive at $100\%$ of
  measurement windows in both mice (fifth column), establishing a
  strict non-equilibrium steady state throughout the record.}
  \label{tab:sigma}
  \begin{tabular}{lccccc}
  \hline
  Mouse & $\langle\Sigma_{\rm local}\rangle$ &
  $\langle\Sigma_{\rm local}^2\rangle$ & CV &
  $n$ windows & fraction $> 0$ \\
  \hline
  A & $0.0635$ & $0.00417$ & $0.18$ & $172$ & $100\%$ \\
  B & $0.0632$ & $0.00405$ & $0.12$ & $209$ & $100\%$ \\
  \hline
  \end{tabular}
\end{table}

\begin{figure}[h]
  \centering
  \includegraphics[width=\columnwidth]{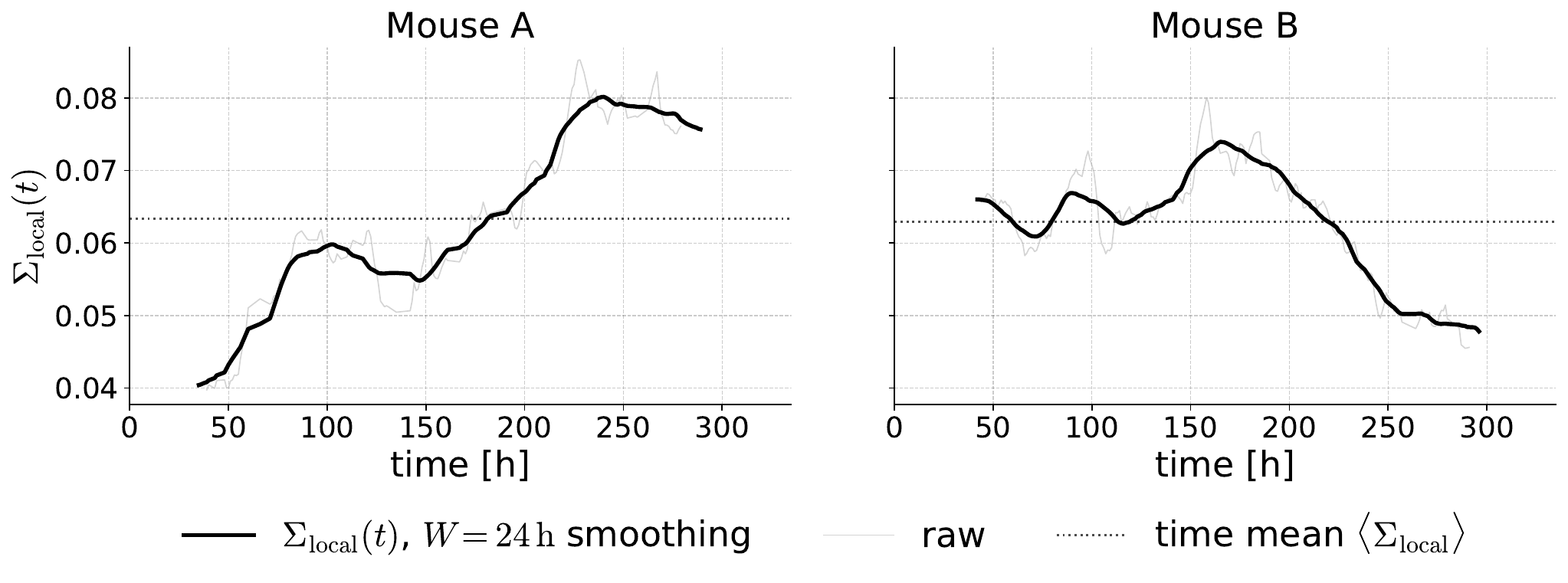}
  \caption{\textbf{Empirical entropy-production proxy
  $\Sigma_{\rm local}(t)$ time series, per mouse.} Light grey: raw
  per-step $\Sigma_{\rm local}(t) = \partial_\tau I(\tau, t)|_{\tau=0}$.
  Solid black: centred rolling-mean smoothing at $W = 24\,\mathrm{h}$.
  Black dotted: time mean.}
  \label{fig:sigma_timeseries}
\end{figure}

\paragraph{Summary.}
 A finite irreversibility scale appears in both animals at intermediate
lag, as expected from multi-step propagation along a directed
cross-feeding pipeline; the slope of the lead-lag imbalance at the
origin returns a strictly positive, quasi-stationary, and
cross-mouse-consistent measurement of the non-equilibrium entropy
production rate that sustains the pseudo-coherent steady state.

\subsection{Phase-agnostic cluster recovery and biological composition}\label{sec:biology}\label{sec:signcluster}

The commutator diagonalisation of Eq.~\eqref{eq:commutator} fixes the
sign of each axis of the rank-two subspace only up to a common sign per
calibration step. Ref.~\cite{troude2026pseudoco,troude2025Unifying}
shows that the sign of the per-MAG reaction-mode component $r_i(t)$
encodes the cluster identity of
the pseudo-coherent regime: components with $r_i > 0$ co-evolve as one
cluster, components with $r_i < 0$ as the anti-synchronised partner. To
use this within a sign-degenerate calibration, we  define
$s_i(t) = \mathrm{sign}(r_i(t))$ and form the per-step co-membership
matrix
\begin{equation}
  C_{ij} \;=\;
  \bigl\langle \mathds{1}[s_i(t) = s_j(t)]\bigr\rangle_t
  \;=\; \tfrac{1}{2}\bigl(1 + \langle s_i(t)\,s_j(t)\rangle_t\bigr).
  \label{eq:comem}
\end{equation}
 The quantity $C_{ij}$ is invariant
under a global sign flip of $r$ at any time $t$ and is therefore unaffected
by the calibration sign gauge. Average-linkage hierarchical clustering
on $D_{ij} = 1 - C_{ij}$ returns a two-cluster partition of the MAGs that
uses no instantaneous phase $\theta_i$, no NPPD pipeline, and no
Kuramoto-like order parameter. The per-MAG confidence
$\varphi_i = \bar C_{\rm own}(i) - \bar C_{\rm opp}(i)$ (mean
co-membership with own cluster minus mean co-membership with the
opposite cluster) ranks MAGs from the most to the least confidently
classified.

\begin{figure}[t]
  \centering
  \includegraphics[width=\columnwidth]{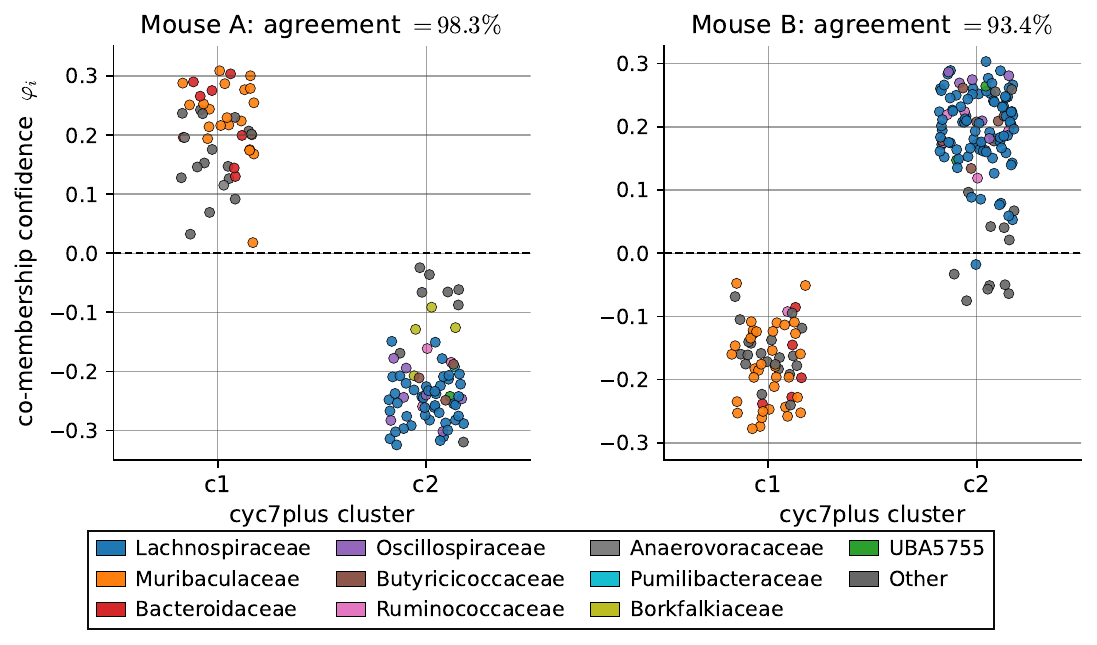}
  \caption{\textbf{Co-membership cluster recovery.}
    Per-MAG signed co-membership confidence
    $\pm \varphi_i = \pm(\bar C_{\rm own}(i) - \bar C_{\rm opp}(i))$
     (positive for MAGs assigned to recovered cluster~1, negative
    for cluster~2), plotted against the cyc7plus cluster label of
    Ref.~\cite{microbiome2025} for Mouse~A (left) and Mouse~B (right);
    each marker is colour-coded by taxonomic family (shared legend
    below). The horizontal dashed line is the decision boundary at
    zero.}
  \label{fig:signcluster}
\end{figure}

Empirically (Fig.~\ref{fig:signcluster}), the co-membership partition
defined by Eq.~\eqref{eq:comem} agrees with the cyc7plus NPPD-based
partition of Ref.~\cite{microbiome2025} for $98.3\%$ of all $118$
labelled MAGs in Mouse~A and $93.4\%$ of all $181$ labelled MAGs in
Mouse~B (chance baseline: $50\%$). The two partitions are constructed
by two operationally distinct pipelines whose inputs do not share the
same dynamical observable: cyc7plus uses circular-phase distance on the
NPPD-derived $\theta_i(t)$, while the co-membership statistic uses only
the per-step sign of $r_i(t)$ inferred from the commutator of the local
Jacobian, with no phase information and no Kuramoto-like order
parameter. The recovered cluster sizes are $(44, 74)$ in Mouse~A and
$(108, 73)$ in Mouse~B; the cyc7plus sizes are $(46, 72)$ and
$(120, 61)$, respectively. Restricting to the top~$90$ MAGs by
per-MAG confidence $\varphi_i$, the agreement with cyc7plus reaches
$100\%$ in both animals, so all disagreement is concentrated in the
low-confidence tail.

An amplitude-adjusted Fourier surrogate null on the per-MAG
abundance series (Methods) returns a chance
agreement of $60.2\%$ in Mouse~A and $63.5\%$ in Mouse~B, with the
empirical $98.3\%/93.4\%$ values sitting $38$ and $24$ percentage
points above the null maximum, respectively, and empirical
$p < 5\times 10^{-3}$ in both animals. The two pipelines are
therefore complementary: they share the same low-rank organisation of
the data, and the size of the agreement above the surrogate null is
itself empirical evidence that the system carries the rank-two
reaction-mode structure assumed by the pseudo-coherent reading
\cite{troude2026pseudoco}; were the dynamics not concentrated on a
low-dimensional non-normal subspace, the commutator-based reaction
mode would not stably extract a sign series that recovers a partition
built independently from the phase signal.

The agreement is conditional on the per-MAG calibration confidence
$\varphi_i$ in a way that is sharp and quantitatively interpretable.
The NPPD pipeline retains only the MAGs that pass a phase-significance
filter, so its cluster labels are most reliable for the MAGs with the
strongest cyclic signal; symmetrically, the co-membership confidence
$\varphi_i$ is large only for MAGs whose participation in the reaction
mode is well above the noise floor. Restricting the comparison to the
top $K$ MAGs by $\varphi_i$ is therefore the natural conditioning. In
both animals the top $90$ MAGs reach $100\%$ agreement with the
cyc7plus partition. The few MAGs on which the two partitions disagree
all sit in the low-confidence tail and contribute marginally to the
order parameter and to the spectral signature. The reaction-mode
geometry therefore recovers, through a pipeline that does not use the
phase $\theta_i(t)$, the same two-cluster organisation that the
original phase-based analysis extracts; at the per-MAG-confidence
level the recovery is exact for the high-confidence top decile.

\begin{figure*}[t]
  \centering
  \includegraphics[width=\textwidth]{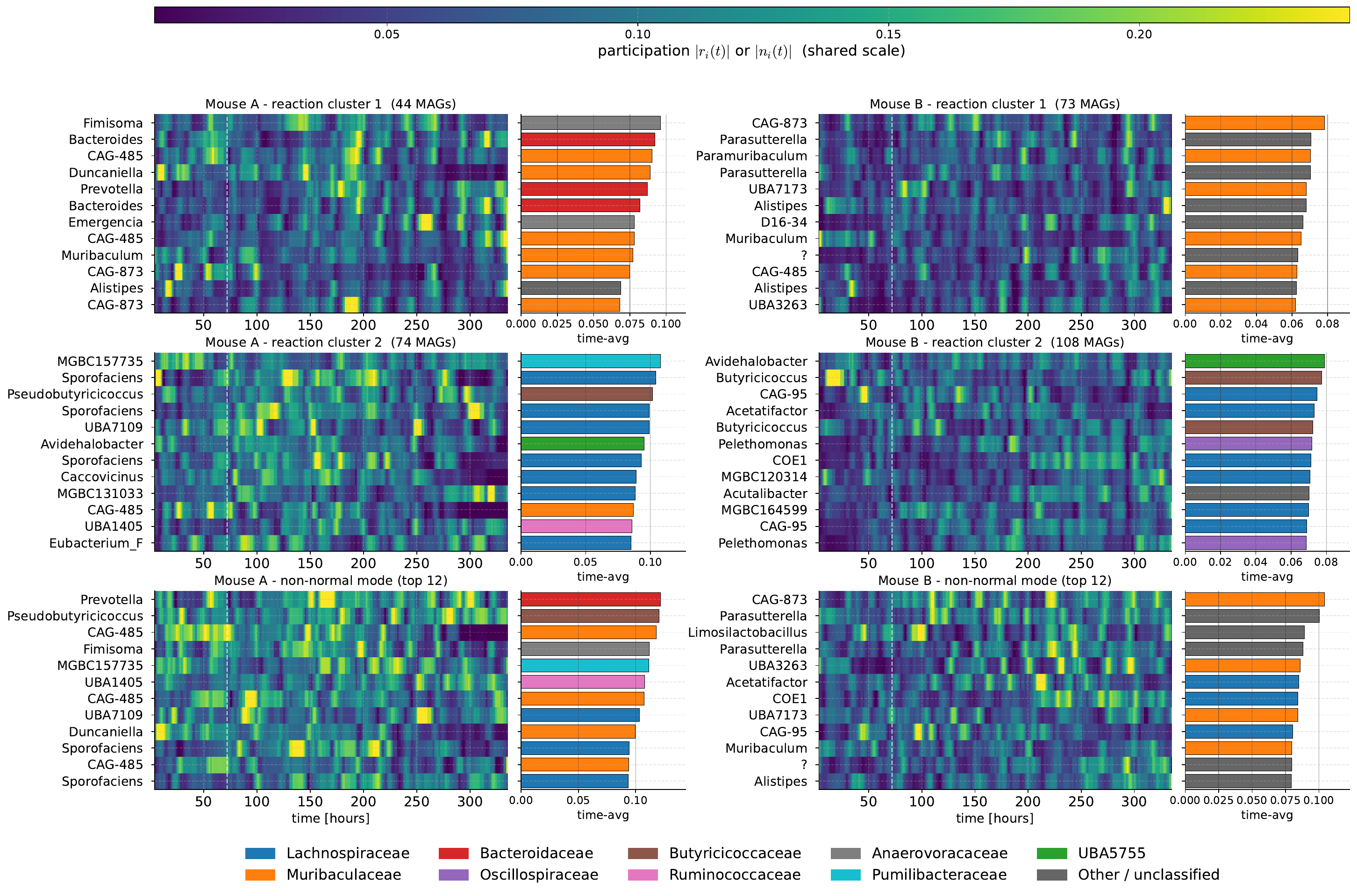}
  \caption{\textbf{Time-resolved per-MAG absolute mode loading, organised by the co-membership cluster recovery.} Each row pairs the smoothed loading heatmap of the top-$12$ MAGs with the time-averaged absolute loading as a horizontal bar coloured by taxonomic family. Top
    row: reaction-mode cluster~1 in each mouse; middle row:
    reaction-mode cluster~2 in each mouse; bottom row: the top-$12$
    contributors to the non-normal mode in each mouse. Cluster labels
    are those returned by hierarchical clustering on the co-membership
    matrix $C_{ij}$ of Eq.~\eqref{eq:comem}, aligned to cyc7plus by the
    global sign flip that maximises agreement; partition sizes are
    $(44, 74)$ in Mouse~A and $(108, 73)$ in Mouse~B. A single colour
    scale (top bar) applies to all six heatmaps.}
  \label{fig:composition}
\end{figure*}

\paragraph{Biological composition of the two recovered clusters and of the non-normal mode.}
Fig.~\ref{fig:composition} shows the per-MAG reaction-mode time series
and the time-averaged participation, separately for the two clusters in
each animal, and adds the non-normal-mode top contributors as a third row. Before naming the loaded taxa we quantify how concentrated the modes are. With $p_i(t)=r_i(t)^2$ the per-MAG energy fraction of the unit-norm mode, the inverse participation ratio $N_{\mathrm{eff}}(t)=1/\sum_i p_i(t)^2$ gives the effective number of MAGs carrying the mode; a random unit vector in $N$ dimensions has $N_{\mathrm{eff}}\approx N/3$. Averaged over time, the reaction and non-normal modes give $N_{\mathrm{eff}}\approx 22$ (Mouse~A, $N=118$) and $\approx 29$ (Mouse~B, $N=181$), about half the random baselines ($41$ and $62$), and the $20$ largest-loading MAGs carry $77\%$ (Mouse~A) and $67\%$ (Mouse~B) of the mode energy against $59\%$ and $46\%$ for a random mode. The modes are therefore moderately concentrated, about twice beyond chance, on a few tens of MAGs spanning the two guilds, rather than dominated by a few species or spread uniformly across the community. The two reaction-mode clusters are well populated in both
animals: $(44, 74)$ in Mouse~A and $(108, 73)$ in Mouse~B; the
cyc7plus labelling of the same MAGs reads as $(46, 72)$ and
$(120, 61)$, respectively. In each animal, one cluster is dominated by
Bacteroidota lineages: Muribaculaceae and Bacteroidaceae in Mouse~A,
Muribaculaceae together with Rikenellaceae-type (\emph{Alistipes})
and Sutterellaceae-type (\emph{Parasutterella}) genera in Mouse~B,
all of which belong to the upstream guild of primary polysaccharide
degraders and dietary-fibre fermenters. The other cluster is dominated
by Bacillota~A families (Lachnospiraceae, Oscillospiraceae,
Butyricicoccaceae, Ruminococcaceae), the canonical secondary
fermenters of the short-chain fatty-acid cascade
\cite{koropatkin2012,flint2012gm,rakoffnahoum2014,belenguer2006,falony2006,louis2017,denbesten2013}.
The directional cross-feeding architecture of the colon is therefore
reproduced, in both animals and at the genus level visible in
Fig.~\ref{fig:composition}, by a partition constructed from a strictly
phase-agnostic calibration. The non-normal-mode top contributors in
the bottom row of Fig.~\ref{fig:composition} are more mixed than the
two reaction-mode clusters: primary-degrader Muribaculaceae lineages
are over-represented in both mice, but Lachnospiraceae secondary
fermenters also contribute substantially. This is consistent with the
reading that the non-normal mode reads the spatial gradient that drives
the amplification, set primarily by the upstream donors but also
loaded on the strongest downstream recipients; the reaction
mode encodes its coherent two-cluster output side.

\paragraph{Why species memberships are not identical across animals.}
The two animals share the families most strongly loaded on the amplification (Bacteroidota primary degraders and Bacillota~A secondary fermenters),
but the exact genus-level membership of the top-12 ranking differs
between mice; for example, \emph{Prevotella} and \emph{Pseudobutyricicoccus}
appear in Mouse~A but not Mouse~B, while \emph{Parasutterella} and
\emph{Alistipes} are prominent in Mouse~B and absent from Mouse~A.
Three contributions to this variability are worth naming: each animal
hosts an individually unique community at the strain and genus level
(sample-to-sample biological variation in the mouse microbiome is well
documented \cite{microbiome2025}), the metagenomic assembly and binning
pipeline used to construct the MAGs is itself stochastic and recovers
slightly different binnings in the two mice from independent sequencing
runs, and the rank-two calibration places MAGs into the top of the
ranking on the basis of their participation in the inferred reaction
direction (which itself depends on the local Jacobian of the actual
community, not on which genera are most abundant). The fact that the
family-level identity of the upstream and downstream guilds is
preserved across the two animals while the genus-level identity is not
is therefore the expected pattern: the directional trophic
architecture is a community-wide property; the particular genera that
occupy each position in that architecture are animal-specific.

\paragraph{Summary.}
 A calibration that uses only the local Jacobian geometry recovers the
two-cluster organisation of Ref.~\cite{microbiome2025}: $98.3\%$ of
all $118$ labelled MAGs in Mouse~A and $93.4\%$ of all $181$ labelled
MAGs in Mouse~B, with the top $90$ MAGs by per-MAG co-membership
confidence reaching $100\%$ agreement in both animals. The two
recovered clusters carry biologically distinct family compositions
(Fig.~\ref{fig:composition}): Bacteroidota primary degraders vs
Bacillota~A secondary fermenters, the same upstream/downstream trophic
split identified earlier in Sec.~\ref{sec:biology}.

\section{Discussion}\label{sec:discussion}

The four expected fingerprints of pseudo-coherence are present jointly, and consistently across both animals, in the genome-resolved mouse-gut data. These four are: intermittent cluster-level phase alignment that tracks the spatial support of the amplified mode rather than proximity to an instability; a markedly time-asymmetric lagged covariance with a global imbalance peak at intermediate lag; a time-averaged spectrum enhanced at low frequency with drifting, non-stationary peaks rather than a fixed ridge; and reaction- and non-normal-mode rankings that recover the two trophic guilds without phase information. Cluster-level phase alignment is intermittent and tracks the
spatial extent of the inferred non-normal amplification rather than
proximity to a spectral instability that never occurs. The lagged
covariance is markedly time-asymmetric, with a global peak at
intermediate lag whose magnitude reflects the directed multi-step
structure of the gut cross-feeding network. The time-averaged spectrum
shows the predicted shape (low-frequency enhancement, depleted
intermediate frequencies, strong temporal intermittency on the same
bands that carry the largest mean power) and no persistent ridge in the
scalograms. Quantitative AAFT surrogate testing
\cite{theiler1992,schreiber1996,schreiber2000,lancaster2018} localises a
weak time-averaged construction in the low-frequency band of both mice
without a fixed ridge in time. The inferred reaction and non-normal
modes independently identify the two canonical functional guilds of the
mouse colon and assign them biologically informative donor-recipient
roles.

These findings strain the dominant oscillator-based reading of microbiome
rhythmicity at the MAG scale. A persistent host-entrained ridge would
have to manifest in the scalogram of every individual; it does not. A
near-Hopf interpretation would predict qualitative reorganisation after
moderate perturbations; the antibiotic dataset reported in
Ref.~\cite{microbiome2025} shows shock-and-recovery rather than
bifurcation crossing. A genuine persistent multi-mouse circadian
component, at exactly $1/24\;\mathrm{h}^{-1}$ and stationary in time, is
not what the data show. By contrast, every signature of pseudo-coherence
is present in both mice and at every diagnostic. We emphasise that the surrogate test on the marginal spectrum is a lower-bound consistency check on excess power, not in itself a discriminator between an oscillator and pseudo-coherence: an amplitude-adjusted surrogate inherits each MAG's marginal spectrum, so a genuine narrow-band oscillator would copy into the surrogate ensemble and erode its own exceedance. The discrimination instead rests on the absence of a persistent scalogram ridge, on the lead--lag imbalance peaking at tens of hours rather than at an oscillator half-period, and on the recovery of two genuine reaction-mode clusters.

This does not deny the host circadian clock, nor does it deny the
existence of rhythmic modulation at the abundance level of bacterial
families. It denies that, at hourly genome resolution, a stable
oscillator is the appropriate null. The appropriate null is a stable,
strongly non-normal stochastic system.

\paragraph{From host-clock labels to a trophic re-reading of the two clusters.}
The two phase clusters $\mathcal{C}_1, \mathcal{C}_2$ in
Ref.~\cite{microbiome2025} are extracted from circular-phase distance
on the NPPD-derived $\theta_i(t)$ and identified there with a
host-clock-aligned night vs day partition. The cluster recovery of
Sec.~\ref{sec:biology} uses an independent route: it builds the
co-membership statistic from the sign of the per-step reaction-mode
component, which by construction is phase-agnostic and
spectrally-agnostic. The fact that the two routes agree on $98.3\%$
(Mouse~A) and $93.4\%$ (Mouse~B) of all labelled MAGs, and on the top
$90$ MAGs of each animal exactly, shows that the same two-cluster
organisation is the natural output of two very different lenses on the
data. The composition of the two clusters
(Fig.~\ref{fig:composition}) makes the interpretation concrete: the
cyc7plus cluster $\mathcal{C}_1$ collects the Bacteroidota primary
polysaccharide degraders (Muribaculaceae and Bacteroidaceae in
Mouse~A, together with Rikenellaceae-like \emph{Alistipes} and
Sutterellaceae-like \emph{Parasutterella} in Mouse~B), while cluster
$\mathcal{C}_2$ collects the Bacillota~A secondary SCFA fermenters
(Lachnospiraceae, Butyricicoccaceae, Oscillospiraceae,
Ruminococcaceae). The two clusters are therefore not naturally a
night/day partition but the upstream and downstream ends of the
cross-feeding cascade.

This re-reading is in fact directly consistent with the host-clock
labels in Ref.~\cite{microbiome2025}: in nocturnal mice the active
feeding phase is the dark phase, dietary fibre arrives at the colon
predominantly during the night, the primary-degrader Bacteroidota
respond first, and the secondary fermenters of the SCFA cascade
respond a multi-hour lag later through cross-feeding. The
cyc7plus night cluster of
Ref.~\cite{microbiome2025} is then exactly the upstream guild that
tracks fibre arrival, and the day cluster is the downstream guild
that tracks the delayed SCFA wave. The apparent
night/day split is a faithful readout of the cross-feeding delay
between primary and secondary fermenters, not evidence of an
autonomous oscillator at the MAG level. The
intermediate-lag peak of $I(\tau)$ in Fig.~\ref{fig:imbalance}
(Sec.~\ref{sec:irreversibility}) gives an independent estimate of this
delay from the genome-resolved data alone. Under this reading, the
24~h modulation of the host feeding cycle remains the natural
entrainment input, but the two-cluster organisation of the MAG-level
dynamics is a property of the trophic architecture of the community,
not of a community-level circadian oscillator. This is a
non-circadian-driven mechanistic account that is fully compatible
with the cyc7plus phenomenology and with the absence of a persistent
24~h ridge in the scalograms of Fig.~\ref{fig:wavelet}.

\paragraph{Support, entropy production, and non-equilibrium steady state.}
A subtler feature of the data is that the supports $s_r$ and $s_n$ of
Eq.~\eqref{eq:support} grow synchronously with the cluster order
parameters, and the band-resolved wavelet coherence between $s_r$ and
$\langle R_{\mathcal{C}}\rangle$ (Table~\ref{tab:coherence}) plus the
slow-component Pearson correlation identify support, not $K/K_c$, as
the dominant geometric driver of macroscopic phase coherence. In the pseudo-coherent framework, support has a
thermodynamic meaning: as the reaction subspace becomes more extensive,
circulating probability currents in phase space grow, the lead-lag
imbalance of Fig.~\ref{fig:imbalance} rises, and the entropy production
rate increases. The community is held away from equilibrium by a
continuous influx of free energy from the host (dietary input, bile-acid
recycling) and exports the corresponding entropy through metabolic
products (SCFAs, gases, microbial cell death). The empirical observation
that support and coherence rise together is therefore the dynamical
signature of a microbiome operating in a non-equilibrium steady state
\cite{seifert2012stochastic,gnesotto2018broken,fyodorov2025nonorthogonal},
sustained by directional cross-feeding rather than by oscillator-mediated
phase locking.

\paragraph{Life as non-normal amplification of fluxes.}
This reading aligns the microbiome with a recent general framework in which life is interpreted through the non-normal amplification of fluxes \cite{sornette2025life}: biological systems evolve asymmetric,
hierarchical reaction networks that amplify the throughput of free energy
without crossing a bifurcation, and the resulting directional
architectures generate transient amplification cycles that maintain the
non-equilibrium steady state. The gut microbiome here provides a direct
empirical realisation of that principle. In the gut, metabolic by-products that leak from one population become resources for others, so resource competition and unidirectional cross-feeding commensalism act together; this combination raises free-energy use, entropy production, and the amplification of metabolic fluxes. The directional cross-feeding chain from Bacteroidota primary degraders to Bacillota A secondary fermenters identified in Sec.~\ref{sec:biology} is exactly the kind of hierarchical asymmetric architecture the framework predicts will generate large flux amplification, and the inferred reaction and non-normal modes localise the upstream and downstream endpoints of the chain from a dynamical observable alone.

\paragraph{Active matter and broader physics.}
At the same level of generality, active matter itself is a proper subfield of non-normal stochastic dynamics~\cite{marchetti2013hydrodynamics,ramaswamy2010mechanics}: its non-reciprocal, hierarchical interactions render the linearised operator non-normal, so the transient amplification, broken time-reversal symmetry, and entropy production documented here also organise self-propelled active systems, of which the gut microbiome is one chemical realisation. Active matter literature has long observed that directional
inter-particle interactions generate band-pass spectral signatures and
broken time-reversal symmetry without underlying
oscillators~\cite{toner1995longrange,cates2015motility,fodor2016how};
the present analysis brings the same machinery to genome-resolved
microbial data and demonstrates that the operational signatures are
present and quantifiable.

Non-normal stochastic systems are common in ecological interaction
networks \cite{stein2013,bucci2014,coyte2015,goyal2018prl,goyal2018isme,neubert1997},
in atmospheric and oceanic dynamics
\cite{Farrell1996GST1,FarrellIoannou2003,trefethen1993hydrodynamic}, and
in balanced neural networks \cite{Hennequin2014,Murphy2009,Ganguli2008}.
Whenever interactions are intrinsically directional, the operator that
governs the dynamics is generically non-normal, and rhythmic structure in
observables of such systems is not in general evidence of oscillators.
In neural systems, the canonical frequency bands of the rhythm of the
brain \cite{buzsaki2004,buzsaki2013} have been interpreted within the
pseudo-coherent framework as finite-time spectral concentrations driven
by non-normal amplification in balanced networks
\cite{troude2026pseudoco,Hennequin2014}. The same diagnostic battery used
here, namely reaction- and non-normal-mode participation, support-based
coherence regression, lead-lag imbalance, and marginal-spectrum surrogate
testing, is portable to such other high-dimensional rhythmic biological
data.

\paragraph{Falsifiable prediction for clock-gene-knockout cohorts.}
The mechanism makes two contrasting empirical predictions for a
genome-resolved hourly cohort of clock-gene-disrupted mice (intestinal
Bmal1 knockout, Per1/Per2 double knockout, Cry1/Cry2 double knockout, or
environmental jet lag \cite{Thaiss2014Transkingdom,mukherji2013,Liang2015Rhythmicity,heddes2022,kuang2019,reitmeier2020,tuganbaev2020,voigt2014,deaver2018,altaha2022,thaiss2016transcr,paulose2016,zheng2020review,teichman2020,voigt2016review,Leone2015Diurnal}).
The non-normal interpretation predicts a selective dissociation between two classes of observables, rather than a binary opposite to an oscillator picture. Quantities tied to the microbial interaction geometry, namely transient amplification, the reaction- and non-normal-mode structure, the lead-lag asymmetry, and the broad guild identities of the highest-loading MAGs, should remain present, although their numerical values may shift if the knockout changes diet, transit, bile acids, or community composition. By contrast, any narrow host-clock-locked $24\;\mathrm{h}$ component should be reduced or lose phase consistency. The decisive signature is therefore the preservation of finite-time asymmetric response structure together with the weakening of host-clock-locked circadian coherence. This separates three pictures: a host-clock-entrainment picture, in which clock disruption primarily degrades phase locking, cluster synchrony, and any $24\;\mathrm{h}$ ridge, with the interaction-geometry diagnostics changing only as consequences of that loss; a microbial-autonomous-oscillator picture, in which a coherent near-$24\;\mathrm{h}$ ridge persists even without the host clock; and the non-normal pseudo-coherence picture, in which the narrow $24\;\mathrm{h}$ component weakens but the transient amplification, lead-lag asymmetry, and mode-guild structure remain. Existing datasets can test coarse circadian abundance rhythms, but not the hourly, MAG-resolved non-normal diagnostics used here; the discriminating experiment therefore requires hourly, genome-resolved sampling under clock disruption.

\section{Methods}\label{sec:methods}

\paragraph{Data.}
We use the genome-resolved mouse-gut metagenomic time series of
Ref.~\cite{microbiome2025}, sampled hourly across two weeks for two
animals (Mouse A and Mouse B). For each animal we use the relative-
abundance and phase tables together with the cyc7plus cluster assignments
of Ref.~\cite{microbiome2025}, joined with the GTDB-Tk taxonomy of each
MAG \cite{chaumeil2019gtdbtk}. The taxonomic family and genus labels
appearing in the figures and text are those assigned by the GTDB-Tk
toolkit on the MAGs provided with Ref.~\cite{microbiome2025}, not
hand-curated; uncultured lineages carry placeholder names of the form
``UBA''/``CAG''/``MGBC''.

\paragraph{Phase extraction (NPPD).}
The non-parametric phase determination is taken verbatim from
Ref.~\cite{microbiome2025}. For a discrete time series $x(t)$ sampled
hourly, a moving-median trend over a $24\;\mathrm{h}$ window is removed,
$x'(t) = x(t) - x_\mathrm{med}(t)$. A local reference $m(t)$ is taken as
the median of $x'$ in the same $24\;\mathrm{h}$ window. Within that
window the signs of $x'$ relative to $m(t)$ are tabulated into a
$2\times 2$ contingency table comparing past and future halves, and a
one-tailed Fisher exact test produces a $p$-value
$P_\mathrm{Fisher}(t)$. The signed local significance score
\begin{equation}
  w(t) \;=\; \mathrm{sign}\bigl[a(t)d(t) - b(t)c(t)\bigr]\,
            \log P_\mathrm{Fisher}(t)
  \label{eq:wt}
\end{equation}
captures the direction and statistical strength of local transitions.
Zero-crossings of $w(t)$ and local extrema between them serve as anchor
points; each cycle is divided into four quadrants
$\phi_T \in \{0, \pi/2, \pi, 3\pi/2\}$, and the instantaneous phase
$\theta(t) = \phi_T(t)\bmod 2\pi$ is obtained by linear interpolation
between anchors followed by wrapping. The procedure is non-parametric,
makes no waveform assumption, and is robust to non-stationarity and
non-sinusoidal shapes.

\paragraph{Cluster construction.}
The cyc7plus partition of MAGs into two clusters (used throughout this
paper and in the original Ref.~\cite{microbiome2025}) is built from the
pairwise circular phase distance between MAGs. After phase unwrapping,
the pairwise mean phase difference $\Delta\phi_{ij}$ is computed and
mapped to a distance $d_{ij} = \tfrac{1}{2}(1-\cos\Delta\phi_{ij})$.
Average-linkage hierarchical clustering on $d_{ij}$ followed by
two-cluster extraction yields the partition reported in
Ref.~\cite{microbiome2025} that we adopt as the input to the Kuramoto-
like order parameters
\begin{equation}
  R_\mathcal{C}(t) \;=\; \biggl|
    \frac{1}{|\mathcal{C}|}\sum_{i\in\mathcal{C}} e^{\mathrm{i}\theta_i(t)}
  \biggr|.
\end{equation}

\paragraph{Local non-normal calibration via the commutator.}
Over four consecutive observations we estimate $\widehat{\mA}_k$ from
Eq.~\eqref{eq:Ahat}. Rather than diagonalising $\widehat{\mA}_k$, we
diagonalise the commutator $\mB_k$ of Eq.~\eqref{eq:commutator}, which
is real symmetric and traceless. When non-normality is significant,
$\mB_k$ has rank two with eigenvalues $\pm\lambda_\mathrm{max}$ and
eigenvectors spanning the non-normal subspace. Projecting
$\widehat{\mA}_k$ onto this subspace yields the reduced $2\times 2$
matrix $\mGamma_k$ of Eq.~\eqref{eq:Gamma}, from which the non-normality
index $K$ from \eqref{eq:K}, the threshold $K_c$ from \eqref{eq:Kc}, the
spectral radius $\rho$, and the per-MAG reaction- and non-normal-mode
loadings of \eqref{eq:participation} are computed. The signs of the
loadings are not used because the eigendecomposition of $\mB_k$ fixes
each axis only up to a common sign flip.

\paragraph{Co-membership cluster recovery.}
The commutator diagonalisation of $\mB_k$ fixes the axis
$\hat{\mathbf{r}}(t)$ of the rank-two reaction subspace only up to a
common sign flip: $(\hat{\mathbf{r}},\hat{\mathbf{n}})$ and
$(-\hat{\mathbf{r}},-\hat{\mathbf{n}})$ are
indistinguishable eigenpairs of the real symmetric $\mB_k$. Any quantity
built from the time series $r_i(t)$ that is not invariant under
this per-step sign gauge is therefore not a well-defined statistic.
In particular, the naive cluster recovery
$c_i = \mathrm{sign}\,\langle r_i(t)\rangle_t$ requires an a priori
sign-alignment of $\hat{\mathbf{r}}(t)$ across consecutive calibration
windows, and is sensitive to spurious sign flips that the alignment
procedure inevitably introduces in low-confidence stretches.

To bypass the gauge entirely we use the per-step co-membership
matrix of Eq.~\eqref{eq:comem}. With $s_i(t)=\mathrm{sign}\,r_i(t)$,
the statistic
\begin{equation}
  C_{ij} \;=\;
  \bigl\langle \mathds{1}\bigl[s_i(t) = s_j(t)\bigr]\bigr\rangle_t
  \;=\; \tfrac{1}{2}\bigl(1 + \langle s_i(t)\,s_j(t)\rangle_t\bigr)
  \label{eq:comem-methods}
\end{equation}
is manifestly invariant under a global sign flip of $\hat{\mathbf{r}}$
at \emph{any} time $t$, because flipping all signs at the same $t$ does
not change the indicator $\mathds{1}[s_i(t)=s_j(t)]$ for any pair
$(i,j)$. $C_{ij}\in[0,1]$ measures the fraction of calibration windows
in which MAGs $i$ and $j$ sit on the same side of zero in the reaction
mode, and is well-defined without any sign-alignment pass.

The partition is obtained by average-linkage hierarchical clustering on
the distance matrix $D_{ij} = 1 - C_{ij}$, symmetrised and with zero
diagonal, with the dendrogram cut at two clusters. Per-MAG confidence
is the gap between mean own-cluster and mean opposite-cluster
co-membership,
\begin{equation}
  \varphi_i \;=\;
  \frac{1}{|\mathcal{C}(i)|-1}\!\!\!\!\sum_{\substack{j \in \mathcal{C}(i)\\j \neq i}}\!\!\! C_{ij}
  \;-\;
  \frac{1}{|\mathcal{C}(i)^c|}\!\!\!\sum_{\,j \in \mathcal{C}(i)^c}\!\!\! C_{ij},
  \label{eq:phi}
\end{equation}
where $\mathcal{C}(i)$ is the cluster containing MAG $i$ and
$\mathcal{C}(i)^c$ its complement. $\varphi_i \to 1$ for a MAG that is
in the same group as all of its own cluster at every time and never in
the same group as the opposite cluster; $\varphi_i \to 0$ for a MAG
whose sign in the reaction mode is uninformative. The top-$K$
conditional agreement reported in Sec.~\ref{sec:biology} is the fraction
of cyc7plus labels reproduced when the $K$ MAGs of largest $\varphi_i$
are retained.

\paragraph{Surrogate test for the marginal Morlet spectrum.}
For each MAG we generate $N_\mathrm{surr}=250$ amplitude-adjusted Fourier
surrogates following Refs.~\cite{theiler1992,schreiber1996,schreiber2000}.
Each surrogate preserves the per-MAG one-point distribution and
approximately the marginal power spectrum, but destroys phase coherence
across MAGs and non-stationary temporal structure. For each surrogate we
compute the time-averaged Morlet spectrum on the same scale grid as the
data (100 scales logarithmically spaced over periods of $1$-$100\;\mathrm{h}$,
cmor1.5-1.0 wavelet) and average across MAGs. Per-frequency exceedance probabilities are corrected for multiple comparisons across frequency bins by the Benjamini--Hochberg procedure at a false-discovery rate of $0.05$; at $N_\mathrm{surr}=250$ the resolvable $p$-value floor ($\approx 1/251$) is small enough to pass this correction, whereas $N_\mathrm{surr}=50$ is not. The full record and the
post-72 h window (after the cage-transfer transient) are analysed
separately.

\paragraph{Lead-lag imbalance: closed form and small-$\tau$ expansion.}
The lagged covariance $\mC(\tau)$ of Eq.~\eqref{eq:Ctau} is estimated
by direct sample average and $I(\tau)$ from Eq.~\eqref{eq:Itau} by the
Frobenius norm of its antisymmetric part. For the reduced two-dimensional
pseudo-coherent dynamics with noise covariance
$\mathbf{B} = \begin{pmatrix}\sigma_1^2 & \rho\sigma_1\sigma_2 \\
\rho\sigma_1\sigma_2 & \sigma_2^2 \end{pmatrix}$,
Ref.~\cite{troude2026pseudoco,troude2025Unifying} derives the closed form
\begin{equation}
I(\tau) =
\frac{\sigma_1\sigma_2}{2\sqrt{2}\,\alpha}\,|K_\sigma|\,
\left|e^{-(\alpha-\beta)\tau} - e^{-(\alpha+\beta)\tau}\right|,
\label{eq:Itau_closed}
\end{equation}
where $\alpha$ and $\beta$ are the diagonal coefficients of the
rotated reduced operator $\mGamma_{\rm rot}$, and $K_\sigma =
\tfrac{1}{2}(\kappa_\sigma - \kappa_\sigma^{-1})$ with
$\kappa_\sigma = \kappa\,\sigma_2/\sigma_1$ is the effective
non-normality index modulated by the noise covariance. In the
isotropic-noise case ($\sigma_1 = \sigma_2$, $\rho = 0$),
$K_\sigma = K$ and the lead-lag imbalance is controlled purely by
the geometric non-normality of the operator.

Expanding Eq.~\eqref{eq:Itau_closed} around $\tau = 0$ gives, to
leading order in $\tau$, 
\begin{equation}
\begin{split}
I(\tau)\;&=\;\Sigma_{\rm local}\,\tau \;+\; \mathcal{O}(\tau^2),\\
\Sigma_{\rm local}\;&\equiv\;\left.\partial_\tau I(\tau)\right|_{\tau = 0}
\;=\;\frac{\sigma_1\sigma_2\,\beta\,|K_\sigma|}{\sqrt{2}\,\alpha}.
\end{split}
\label{eq:Sigma_local}
\end{equation}
The slope of $I(\tau)$ at the origin is therefore a directed-flow
amplitude that grows linearly with the effective non-normality
$|K_\sigma|$ and serves as a local, hour-scale empirical observable.
We estimate $\Sigma_{\rm local}(t)$ from a parabolic fit to $I(\tau,
t)$ on $\tau \in [0, 3]\;\mathrm{h}$ within a $72\;\mathrm{h}$
centred rolling window.

\paragraph{Entropy production rate.}
For the same reduced dynamics, Ref.~\cite{troude2026pseudoco} derives
the closed form for the stationary entropy production rate of the
non-equilibrium steady state,
\begin{equation}
\Phi \;=\;
\frac{2\beta^2}{\alpha}\,\frac{K_\sigma^2}{1 - \rho^2},
\label{eq:Phi_closed}
\end{equation}
which grows quadratically with the effective non-normality and
vanishes in the normal limit. Comparing
Eqs.~\eqref{eq:Sigma_local} and \eqref{eq:Phi_closed} in the
isotropic-noise case yields the identity
$\Sigma_{\rm local}^2 = \tfrac{1}{4}\sigma^4\,\Phi/\alpha$, so that
the empirically accessible $\Sigma_{\rm local}(t)$ is the
$\tau\to 0$ projection of the same non-equilibrium current that is
captured by $\Phi$. Increases of the reaction-mode and non-normal-mode
supports therefore couple directly to increases of $\Phi$ through the
effective non-normality index $K_\sigma$: an extended reaction
subspace amplifies the noise anisotropy seen by the rank-two
projection, raising $K_\sigma$, $\Sigma_{\rm local}$, and $\Phi$
together. This is the thermodynamic content of pseudo-coherence
\cite{troude2026pseudoco,sornette2025life}.

\paragraph{Support--coherence link: three
diagnostics with disjoint sensitivities.}
The link between the reaction-mode support $s_r(t)$ and the cluster-
averaged order parameter $\langle R_{\mathcal{C}}\rangle(t)$ is
quantified by three standard signal-processing tools,
each probing a different aspect of the dependence.

\emph{Wavelet coherence.} Cross-Morlet decomposition with the same
\texttt{cmor1.5-1.0} wavelet and scale grid as the marginal spectrum
yields a time-frequency coherence map $\Gamma^2(f, t) \in [0, 1]$ that
measures the squared magnitude of the local cross-spectrum normalised
by the geometric mean of the two local autospectra; the local
autospectra and cross-spectrum are smoothed in time by a Hann window
of width $3$ periods at each scale before normalisation. We report
the band-averaged mean coherence in four logarithmically spaced bands
of periods $4$--$12$, $12$--$24$, $24$--$48$, and
$48$--$100\;\mathrm{h}$. Edge effects within one period of either
record boundary are excluded by a cone-of-influence mask. This measure
is invariant to constant phase offsets between the two series and is
therefore unaffected by the lead-lag structure that biases the
instantaneous Pearson correlation.

\emph{Lagged cross-correlation.} We standardise $s_r$ and
$\langle R_{\mathcal{C}}\rangle$ to zero mean and unit variance and
compute the cross-correlation
$\rho(\tau) = \mathrm{corr}\bigl[s_r(t), \langle R_{\mathcal{C}}
\rangle(t + \tau)\bigr]$ for $\tau \in [-48, +48]\;\mathrm{h}$ at the
hourly resolution of the data, and report the peak amplitude and the
peak lag. A positive peak lag indicates that $s_r$ leads
$\langle R_{\mathcal{C}}\rangle$.

\emph{Linear Granger causality.} We fit two vector-autoregressive
models of lag $L = 4\;\mathrm{h}$ on the centred series: an
unrestricted model where each series is regressed on its own four
past values and on the four past values of the other, and a
restricted model where the cross-series terms are zero. The Granger
$F$-test of the restricted-vs-unrestricted residual sum of squares
gives a $p$-value for the null ``the second series does not
Granger-cause the first''; we report the $p$-values in both
directions at lag $4\;\mathrm{h}$ on the un-smoothed series; the
test is reported as a directional consistency check, not as a
hypothesis test against a $0.05$ threshold, because the residual
autocorrelation of the data inflates the variance of the
$F$-statistic. Implementation uses
\texttt{statsmodels.tsa.stattools.grangercausalitytests}.

\paragraph{Calibration pipeline at a glance.}
The empirical pipeline that produces every per-step observable used
in this work is the same composition of four operations at each
time-step $t$: (1)~assemble the $4$-sample window $X_t = [\vx(t-3),
\vx(t-2), \vx(t-1), \vx(t)]^\top$; (2)~fit the local Jacobian
$\widehat{A}_t$ from $X_t$ by Eq.~\eqref{eq:Ahat}; (3)~extract the
rank-two reaction direction $\hat{\mathbf{r}}_t$ and non-normal direction
$\hat{\mathbf{n}}_t$ via the eigenvectors of the commutator
$\widehat{B}_t = \widehat{A}_t \widehat{A}_t^\top - \widehat{A}_t^\top
\widehat{A}_t$ (Eq.~\eqref{eq:commutator}); (4)~project per-MAG
components, derive $K/K_c(t)$, $s_r(t)$, $s_n(t)$, and the per-step
reaction-mode sign $s_i(t) = \mathrm{sign}(r_i(t))$ that feeds the
co-membership matrix of Eq.~\eqref{eq:comem}. The same pipeline
feeds both the spectral and lead-lag analyses (through the time
series of $K/K_c$ and the support) and the cluster recovery (through
the per-step signs).

\paragraph{Surrogate-null distribution for the co-membership cluster
agreement.}
To distinguish the agreement between the cyc7plus and the
co-membership partitions from the rank-two coincidence that any two
sign-based pipelines might extract from a near-rank-two abundance
matrix, we generate $B = 200$ amplitude-adjusted Fourier surrogates
of each MAG abundance series. The surrogate procedure preserves the
per-MAG marginal power spectrum and one-point amplitude distribution
by construction while destroying the inter-MAG phase coherence
through an independent phase randomisation per MAG. On each surrogate
ensemble we run the full co-membership pipeline (rank-two commutator
decomposition, per-step sign extraction, $C_{ij}$ matrix
construction, average-linkage hierarchical clustering) and compare
the resulting two-cluster partition to the fixed cyc7plus reference.
We report the empirical distribution of the agreement statistic across
the $B$ surrogates, and the $p$-value of the data against the quantile
of that distribution.

\bibliographystyle{apsrev4-2}
\bibliography{bibliography}

\begin{thebibliography}{81}%
\makeatletter
\providecommand \@ifxundefined [1]{%
 \@ifx{#1\undefined}
}%
\providecommand \@ifnum [1]{%
 \ifnum #1\expandafter \@firstoftwo
 \else \expandafter \@secondoftwo
 \fi
}%
\providecommand \@ifx [1]{%
 \ifx #1\expandafter \@firstoftwo
 \else \expandafter \@secondoftwo
 \fi
}%
\providecommand \natexlab [1]{#1}%
\providecommand \enquote  [1]{``#1''}%
\providecommand \bibnamefont  [1]{#1}%
\providecommand \bibfnamefont [1]{#1}%
\providecommand \citenamefont [1]{#1}%
\providecommand \href@noop [0]{\@secondoftwo}%
\providecommand \href [0]{\begingroup \@sanitize@url \@href}%
\providecommand \@href[1]{\@@startlink{#1}\@@href}%
\providecommand \@@href[1]{\endgroup#1\@@endlink}%
\providecommand \@sanitize@url [0]{\catcode `\\12\catcode `\$12\catcode
  `\&12\catcode `\#12\catcode `\^12\catcode `\_12\catcode `\%12\relax}%
\providecommand \@@startlink[1]{}%
\providecommand \@@endlink[0]{}%
\providecommand \url  [0]{\begingroup\@sanitize@url \@url }%
\providecommand \@url [1]{\endgroup\@href {#1}{\urlprefix }}%
\providecommand \urlprefix  [0]{URL }%
\providecommand \Eprint [0]{\href }%
\providecommand \doibase [0]{https://doi.org/}%
\providecommand \selectlanguage [0]{\@gobble}%
\providecommand \bibinfo  [0]{\@secondoftwo}%
\providecommand \bibfield  [0]{\@secondoftwo}%
\providecommand \translation [1]{[#1]}%
\providecommand \BibitemOpen [0]{}%
\providecommand \bibitemStop [0]{}%
\providecommand \bibitemNoStop [0]{.\EOS\space}%
\providecommand \EOS [0]{\spacefactor3000\relax}%
\providecommand \BibitemShut  [1]{\csname bibitem#1\endcsname}%
\let\auto@bib@innerbib\@empty
\bibitem [{\citenamefont {Thaiss}\ \emph {et~al.}(2014)\citenamefont {Thaiss},
  \citenamefont {Zeevi}, \citenamefont {Levy}, \citenamefont
  {Zilberman-Schapira}, \citenamefont {Suez}, \citenamefont {Tengeler},
  \citenamefont {Abramson}, \citenamefont {Katz}, \citenamefont {Korem},
  \citenamefont {Zmora}, \citenamefont {Kuperman}, \citenamefont {Biton},
  \citenamefont {Gilad}, \citenamefont {Harmelin}, \citenamefont {Shapiro},
  \citenamefont {Halpern}, \citenamefont {Segal},\ and\ \citenamefont
  {Elinav}}]{Thaiss2014Transkingdom}%
  \BibitemOpen
  \bibfield  {author} {\bibinfo {author} {\bibfnamefont {C.}~\bibnamefont
  {Thaiss}}, \bibinfo {author} {\bibfnamefont {D.}~\bibnamefont {Zeevi}},
  \bibinfo {author} {\bibfnamefont {M.}~\bibnamefont {Levy}}, \bibinfo {author}
  {\bibfnamefont {G.}~\bibnamefont {Zilberman-Schapira}}, \bibinfo {author}
  {\bibfnamefont {J.}~\bibnamefont {Suez}}, \bibinfo {author} {\bibfnamefont
  {A.}~\bibnamefont {Tengeler}}, \bibinfo {author} {\bibfnamefont
  {L.}~\bibnamefont {Abramson}}, \bibinfo {author} {\bibfnamefont
  {M.}~\bibnamefont {Katz}}, \bibinfo {author} {\bibfnamefont {T.}~\bibnamefont
  {Korem}}, \bibinfo {author} {\bibfnamefont {N.}~\bibnamefont {Zmora}},
  \bibinfo {author} {\bibfnamefont {Y.}~\bibnamefont {Kuperman}}, \bibinfo
  {author} {\bibfnamefont {I.}~\bibnamefont {Biton}}, \bibinfo {author}
  {\bibfnamefont {S.}~\bibnamefont {Gilad}}, \bibinfo {author} {\bibfnamefont
  {A.}~\bibnamefont {Harmelin}}, \bibinfo {author} {\bibfnamefont
  {H.}~\bibnamefont {Shapiro}}, \bibinfo {author} {\bibfnamefont
  {Z.}~\bibnamefont {Halpern}}, \bibinfo {author} {\bibfnamefont
  {E.}~\bibnamefont {Segal}},\ and\ \bibinfo {author} {\bibfnamefont
  {E.}~\bibnamefont {Elinav}},\ }\href
  {https://doi.org/10.1016/j.cell.2014.09.048} {\bibfield  {journal} {\bibinfo
  {journal} {Cell}\ }\textbf {\bibinfo {volume} {159}},\ \bibinfo {pages} {514}
  (\bibinfo {year} {2014})}\BibitemShut {NoStop}%
\bibitem [{\citenamefont {Zarrinpar}\ \emph {et~al.}(2014)\citenamefont
  {Zarrinpar}, \citenamefont {Chaix}, \citenamefont {Yooseph},\ and\
  \citenamefont {Panda}}]{Zarrinpar2014Diet}%
  \BibitemOpen
  \bibfield  {author} {\bibinfo {author} {\bibfnamefont {A.}~\bibnamefont
  {Zarrinpar}}, \bibinfo {author} {\bibfnamefont {A.}~\bibnamefont {Chaix}},
  \bibinfo {author} {\bibfnamefont {S.}~\bibnamefont {Yooseph}},\ and\ \bibinfo
  {author} {\bibfnamefont {S.}~\bibnamefont {Panda}},\ }\href
  {https://doi.org/10.1016/j.cmet.2014.11.008} {\bibfield  {journal} {\bibinfo
  {journal} {Cell Metabolism}\ }\textbf {\bibinfo {volume} {20}},\ \bibinfo
  {pages} {1006} (\bibinfo {year} {2014})}\BibitemShut {NoStop}%
\bibitem [{\citenamefont {Liang}\ \emph {et~al.}(2015)\citenamefont {Liang},
  \citenamefont {Bushman},\ and\ \citenamefont
  {FitzGerald}}]{Liang2015Rhythmicity}%
  \BibitemOpen
  \bibfield  {author} {\bibinfo {author} {\bibfnamefont {X.}~\bibnamefont
  {Liang}}, \bibinfo {author} {\bibfnamefont {F.~D.}\ \bibnamefont {Bushman}},\
  and\ \bibinfo {author} {\bibfnamefont {G.~A.}\ \bibnamefont {FitzGerald}},\
  }\href {https://doi.org/10.1073/pnas.1501305112} {\bibfield  {journal}
  {\bibinfo  {journal} {Proc. Natl. Acad. Sci. USA}\ }\textbf {\bibinfo
  {volume} {112}},\ \bibinfo {pages} {10479} (\bibinfo {year}
  {2015})}\BibitemShut {NoStop}%
\bibitem [{\citenamefont {Leone}\ \emph {et~al.}(2015)\citenamefont {Leone},
  \citenamefont {Gibbons}, \citenamefont {Martinez}, \citenamefont {Hutchison},
  \citenamefont {Huang}, \citenamefont {Cham}, \citenamefont {Pierre},
  \citenamefont {Heneghan}, \citenamefont {Nadimpalli}, \citenamefont {Hubert},
  \citenamefont {Zale}, \citenamefont {Wang}, \citenamefont {Huang},
  \citenamefont {Theriault}, \citenamefont {Dinner}, \citenamefont {Musch},
  \citenamefont {Kudsk}, \citenamefont {Prendergast}, \citenamefont {Gilbert},\
  and\ \citenamefont {Chang}}]{Leone2015Diurnal}%
  \BibitemOpen
  \bibfield  {author} {\bibinfo {author} {\bibfnamefont {V.}~\bibnamefont
  {Leone}}, \bibinfo {author} {\bibfnamefont {S.}~\bibnamefont {Gibbons}},
  \bibinfo {author} {\bibfnamefont {K.}~\bibnamefont {Martinez}}, \bibinfo
  {author} {\bibfnamefont {A.}~\bibnamefont {Hutchison}}, \bibinfo {author}
  {\bibfnamefont {E.}~\bibnamefont {Huang}}, \bibinfo {author} {\bibfnamefont
  {C.}~\bibnamefont {Cham}}, \bibinfo {author} {\bibfnamefont {J.}~\bibnamefont
  {Pierre}}, \bibinfo {author} {\bibfnamefont {A.}~\bibnamefont {Heneghan}},
  \bibinfo {author} {\bibfnamefont {A.}~\bibnamefont {Nadimpalli}}, \bibinfo
  {author} {\bibfnamefont {N.}~\bibnamefont {Hubert}}, \bibinfo {author}
  {\bibfnamefont {E.}~\bibnamefont {Zale}}, \bibinfo {author} {\bibfnamefont
  {Y.}~\bibnamefont {Wang}}, \bibinfo {author} {\bibfnamefont {Y.}~\bibnamefont
  {Huang}}, \bibinfo {author} {\bibfnamefont {B.}~\bibnamefont {Theriault}},
  \bibinfo {author} {\bibfnamefont {A.}~\bibnamefont {Dinner}}, \bibinfo
  {author} {\bibfnamefont {M.}~\bibnamefont {Musch}}, \bibinfo {author}
  {\bibfnamefont {K.}~\bibnamefont {Kudsk}}, \bibinfo {author} {\bibfnamefont
  {B.}~\bibnamefont {Prendergast}}, \bibinfo {author} {\bibfnamefont
  {J.}~\bibnamefont {Gilbert}},\ and\ \bibinfo {author} {\bibfnamefont
  {E.}~\bibnamefont {Chang}},\ }\href
  {https://doi.org/10.1016/j.chom.2015.03.006} {\bibfield  {journal} {\bibinfo
  {journal} {Cell Host Microbe}\ }\textbf {\bibinfo {volume} {17}},\ \bibinfo
  {pages} {681} (\bibinfo {year} {2015})}\BibitemShut {NoStop}%
\bibitem [{\citenamefont {Asher}\ and\ \citenamefont
  {Sassone-Corsi}(2015)}]{Asher2015TimeForFood}%
  \BibitemOpen
  \bibfield  {author} {\bibinfo {author} {\bibfnamefont {G.}~\bibnamefont
  {Asher}}\ and\ \bibinfo {author} {\bibfnamefont {P.}~\bibnamefont
  {Sassone-Corsi}},\ }\href {https://doi.org/10.1016/j.cell.2015.03.015}
  {\bibfield  {journal} {\bibinfo  {journal} {Cell}\ }\textbf {\bibinfo
  {volume} {161}},\ \bibinfo {pages} {84} (\bibinfo {year} {2015})}\BibitemShut
  {NoStop}%
\bibitem [{\citenamefont {Kuramoto}(1975)}]{kuramoto1975self}%
  \BibitemOpen
  \bibfield  {author} {\bibinfo {author} {\bibfnamefont {Y.}~\bibnamefont
  {Kuramoto}},\ }\href@noop {} {\bibfield  {journal} {\bibinfo  {journal}
  {International Symposium on Mathematical Problems in Theoretical Physics}\ ,\
  \bibinfo {pages} {420}} (\bibinfo {year} {1975})}\BibitemShut {NoStop}%
\bibitem [{\citenamefont {Acebr{\'o}n}\ \emph {et~al.}(2005)\citenamefont
  {Acebr{\'o}n}, \citenamefont {Bonilla}, \citenamefont {Vicente},
  \citenamefont {Ritort},\ and\ \citenamefont {Spigler}}]{acebron2005kuramoto}%
  \BibitemOpen
  \bibfield  {author} {\bibinfo {author} {\bibfnamefont {J.~A.}\ \bibnamefont
  {Acebr{\'o}n}}, \bibinfo {author} {\bibfnamefont {L.~L.}\ \bibnamefont
  {Bonilla}}, \bibinfo {author} {\bibfnamefont {C.~J.~P.}\ \bibnamefont
  {Vicente}}, \bibinfo {author} {\bibfnamefont {F.}~\bibnamefont {Ritort}},\
  and\ \bibinfo {author} {\bibfnamefont {R.}~\bibnamefont {Spigler}},\
  }\href@noop {} {\bibfield  {journal} {\bibinfo  {journal} {Reviews of Modern
  Physics}\ }\textbf {\bibinfo {volume} {77}},\ \bibinfo {pages} {137}
  (\bibinfo {year} {2005})}\BibitemShut {NoStop}%
\bibitem [{\citenamefont {Pikovsky}\ \emph {et~al.}(2001)\citenamefont
  {Pikovsky}, \citenamefont {Rosenblum},\ and\ \citenamefont
  {Kurths}}]{pikovsky2003synchronization}%
  \BibitemOpen
  \bibfield  {author} {\bibinfo {author} {\bibfnamefont {A.}~\bibnamefont
  {Pikovsky}}, \bibinfo {author} {\bibfnamefont {M.}~\bibnamefont
  {Rosenblum}},\ and\ \bibinfo {author} {\bibfnamefont {J.}~\bibnamefont
  {Kurths}},\ }\href@noop {} {\emph {\bibinfo {title} {Synchronization: A
  Universal Concept in Nonlinear Sciences}}}\ (\bibinfo  {publisher} {Cambridge
  University Press},\ \bibinfo {year} {2001})\BibitemShut {NoStop}%
\bibitem [{\citenamefont {Trefethen}\ and\ \citenamefont
  {Embree}(2005)}]{trefethen2005spectra}%
  \BibitemOpen
  \bibfield  {author} {\bibinfo {author} {\bibfnamefont {L.~N.}\ \bibnamefont
  {Trefethen}}\ and\ \bibinfo {author} {\bibfnamefont {M.}~\bibnamefont
  {Embree}},\ }\href@noop {} {\emph {\bibinfo {title} {Spectra and
  Pseudospectra}}}\ (\bibinfo  {publisher} {Princeton University Press},\
  \bibinfo {year} {2005})\BibitemShut {NoStop}%
\bibitem [{\citenamefont {Farrell}\ and\ \citenamefont
  {Ioannou}(1996)}]{Farrell1996GST1}%
  \BibitemOpen
  \bibfield  {author} {\bibinfo {author} {\bibfnamefont {B.~F.}\ \bibnamefont
  {Farrell}}\ and\ \bibinfo {author} {\bibfnamefont {P.~J.}\ \bibnamefont
  {Ioannou}},\ }\href
  {https://doi.org/10.1175/1520-0469(1996)053<2025:GSTPIA>2.0.CO;2} {\bibfield
  {journal} {\bibinfo  {journal} {Journal of the Atmospheric Sciences}\
  }\textbf {\bibinfo {volume} {53}},\ \bibinfo {pages} {2025} (\bibinfo {year}
  {1996})}\BibitemShut {NoStop}%
\bibitem [{\citenamefont {Trefethen}\ \emph {et~al.}(1993)\citenamefont
  {Trefethen}, \citenamefont {Trefethen}, \citenamefont {Reddy},\ and\
  \citenamefont {Driscoll}}]{trefethen1993hydrodynamic}%
  \BibitemOpen
  \bibfield  {author} {\bibinfo {author} {\bibfnamefont {L.~N.}\ \bibnamefont
  {Trefethen}}, \bibinfo {author} {\bibfnamefont {A.~E.}\ \bibnamefont
  {Trefethen}}, \bibinfo {author} {\bibfnamefont {S.~C.}\ \bibnamefont
  {Reddy}},\ and\ \bibinfo {author} {\bibfnamefont {T.~A.}\ \bibnamefont
  {Driscoll}},\ }\href@noop {} {\bibfield  {journal} {\bibinfo  {journal}
  {Science}\ }\textbf {\bibinfo {volume} {261}},\ \bibinfo {pages} {578}
  (\bibinfo {year} {1993})}\BibitemShut {NoStop}%
\bibitem [{\citenamefont {Muolo}\ \emph
  {et~al.}(2019{\natexlab{a}})\citenamefont {Muolo}, \citenamefont {Asllani},
  \citenamefont {Fanelli}, \citenamefont {Maini},\ and\ \citenamefont
  {Carletti}}]{asllani2018theory}%
  \BibitemOpen
  \bibfield  {author} {\bibinfo {author} {\bibfnamefont {R.}~\bibnamefont
  {Muolo}}, \bibinfo {author} {\bibfnamefont {M.}~\bibnamefont {Asllani}},
  \bibinfo {author} {\bibfnamefont {D.}~\bibnamefont {Fanelli}}, \bibinfo
  {author} {\bibfnamefont {P.~K.}\ \bibnamefont {Maini}},\ and\ \bibinfo
  {author} {\bibfnamefont {T.}~\bibnamefont {Carletti}},\ }\href
  {https://doi.org/https://doi.org/10.1016/j.jtbi.2019.07.004} {\bibfield
  {journal} {\bibinfo  {journal} {Journal of Theoretical Biology}\ }\textbf
  {\bibinfo {volume} {480}},\ \bibinfo {pages} {81} (\bibinfo {year}
  {2019}{\natexlab{a}})}\BibitemShut {NoStop}%
\bibitem [{\citenamefont {Troude}\ and\ \citenamefont
  {Sornette}(2026)}]{troude2026pseudoco}%
  \BibitemOpen
  \bibfield  {author} {\bibinfo {author} {\bibfnamefont {V.}~\bibnamefont
  {Troude}}\ and\ \bibinfo {author} {\bibfnamefont {D.}~\bibnamefont
  {Sornette}},\ }\href {https://arxiv.org/abs/2603.07206} {\bibfield  {journal}
  {\bibinfo  {journal} {arXiv preprint arXiv:2603.07206}\ } (\bibinfo {year}
  {2026})},\ \Eprint {https://arxiv.org/abs/2603.07206} {arXiv:2603.07206}
  \BibitemShut {NoStop}%
\bibitem [{\citenamefont {Troude}\ and\ \citenamefont
  {Sornette}(2025)}]{troude2025Unifying}%
  \BibitemOpen
  \bibfield  {author} {\bibinfo {author} {\bibfnamefont {V.}~\bibnamefont
  {Troude}}\ and\ \bibinfo {author} {\bibfnamefont {D.}~\bibnamefont
  {Sornette}},\ }\href {https://doi.org/10.1103/kdgw-shxf} {\bibfield
  {journal} {\bibinfo  {journal} {Phys. Rev. Res.}\ }\textbf {\bibinfo {volume}
  {7}},\ \bibinfo {pages} {L042048} (\bibinfo {year} {2025})}\BibitemShut
  {NoStop}%
\bibitem [{\citenamefont {Troude}\ \emph {et~al.}(2025)\citenamefont {Troude},
  \citenamefont {Lera}, \citenamefont {Wu},\ and\ \citenamefont
  {Sornette}}]{troude2025illusion}%
  \BibitemOpen
  \bibfield  {author} {\bibinfo {author} {\bibfnamefont {V.}~\bibnamefont
  {Troude}}, \bibinfo {author} {\bibfnamefont {S.~C.}\ \bibnamefont {Lera}},
  \bibinfo {author} {\bibfnamefont {K.}~\bibnamefont {Wu}},\ and\ \bibinfo
  {author} {\bibfnamefont {D.}~\bibnamefont {Sornette}},\ }\href
  {https://arxiv.org/abs/2412.01833} {\bibinfo {title} {Illusions of
  criticality: Crises without tipping points}} (\bibinfo {year} {2025}),\
  \Eprint {https://arxiv.org/abs/2412.01833} {arXiv:2412.01833 [nlin.CD]}
  \BibitemShut {NoStop}%
\bibitem [{\citenamefont {Sornette}\ and\ \citenamefont
  {Troude}(2025)}]{sornette2025life}%
  \BibitemOpen
  \bibfield  {author} {\bibinfo {author} {\bibfnamefont {D.}~\bibnamefont
  {Sornette}}\ and\ \bibinfo {author} {\bibfnamefont {V.}~\bibnamefont
  {Troude}},\ }\href {https://arxiv.org/abs/2512.18438} {\bibinfo {title} {Life
  as a non-normal chemical accelerator}} (\bibinfo {year} {2025}),\ \Eprint
  {https://arxiv.org/abs/2512.18438} {arXiv:2512.18438 [cond-mat.stat-mech]}
  \BibitemShut {NoStop}%
\bibitem [{\citenamefont {Koropatkin}\ \emph {et~al.}(2012)\citenamefont
  {Koropatkin}, \citenamefont {Cameron},\ and\ \citenamefont
  {Martens}}]{koropatkin2012}%
  \BibitemOpen
  \bibfield  {author} {\bibinfo {author} {\bibfnamefont {N.~M.}\ \bibnamefont
  {Koropatkin}}, \bibinfo {author} {\bibfnamefont {E.~A.}\ \bibnamefont
  {Cameron}},\ and\ \bibinfo {author} {\bibfnamefont {E.~C.}\ \bibnamefont
  {Martens}},\ }\href {https://doi.org/10.1038/nrmicro2746} {\bibfield
  {journal} {\bibinfo  {journal} {Nat. Rev. Microbiol.}\ }\textbf {\bibinfo
  {volume} {10}},\ \bibinfo {pages} {323} (\bibinfo {year} {2012})}\BibitemShut
  {NoStop}%
\bibitem [{\citenamefont {Rakoff-Nahoum}\ \emph {et~al.}(2014)\citenamefont
  {Rakoff-Nahoum}, \citenamefont {Coyne},\ and\ \citenamefont
  {Comstock}}]{rakoffnahoum2014}%
  \BibitemOpen
  \bibfield  {author} {\bibinfo {author} {\bibfnamefont {S.}~\bibnamefont
  {Rakoff-Nahoum}}, \bibinfo {author} {\bibfnamefont {M.}~\bibnamefont
  {Coyne}},\ and\ \bibinfo {author} {\bibfnamefont {L.}~\bibnamefont
  {Comstock}},\ }\href {https://doi.org/10.1016/j.cub.2013.10.077} {\bibfield
  {journal} {\bibinfo  {journal} {Curr. Biol.}\ }\textbf {\bibinfo {volume}
  {24}},\ \bibinfo {pages} {40} (\bibinfo {year} {2014})}\BibitemShut {NoStop}%
\bibitem [{\citenamefont {Rakoff-Nahoum}\ \emph {et~al.}(2016)\citenamefont
  {Rakoff-Nahoum}, \citenamefont {Foster},\ and\ \citenamefont
  {Comstock}}]{rakoffnahoum2016}%
  \BibitemOpen
  \bibfield  {author} {\bibinfo {author} {\bibfnamefont {S.}~\bibnamefont
  {Rakoff-Nahoum}}, \bibinfo {author} {\bibfnamefont {K.~R.}\ \bibnamefont
  {Foster}},\ and\ \bibinfo {author} {\bibfnamefont {L.~E.}\ \bibnamefont
  {Comstock}},\ }\href {https://doi.org/10.1038/nature17626} {\bibfield
  {journal} {\bibinfo  {journal} {Nature}\ }\textbf {\bibinfo {volume} {533}},\
  \bibinfo {pages} {255} (\bibinfo {year} {2016})}\BibitemShut {NoStop}%
\bibitem [{\citenamefont {Mahowald}\ \emph {et~al.}(2009)\citenamefont
  {Mahowald}, \citenamefont {Rey}, \citenamefont {Seedorf}, \citenamefont
  {Turnbaugh}, \citenamefont {Fulton}, \citenamefont {Wollam}, \citenamefont
  {Shah}, \citenamefont {Wang}, \citenamefont {Magrini}, \citenamefont
  {Wilson}, \citenamefont {Cantarel}, \citenamefont {Coutinho}, \citenamefont
  {Henrissat}, \citenamefont {Crock}, \citenamefont {Russell}, \citenamefont
  {Verberkmoes}, \citenamefont {Hettich},\ and\ \citenamefont
  {Gordon}}]{mahowald2009}%
  \BibitemOpen
  \bibfield  {author} {\bibinfo {author} {\bibfnamefont {M.~A.}\ \bibnamefont
  {Mahowald}}, \bibinfo {author} {\bibfnamefont {F.~E.}\ \bibnamefont {Rey}},
  \bibinfo {author} {\bibfnamefont {H.}~\bibnamefont {Seedorf}}, \bibinfo
  {author} {\bibfnamefont {P.~J.}\ \bibnamefont {Turnbaugh}}, \bibinfo {author}
  {\bibfnamefont {R.~S.}\ \bibnamefont {Fulton}}, \bibinfo {author}
  {\bibfnamefont {A.}~\bibnamefont {Wollam}}, \bibinfo {author} {\bibfnamefont
  {N.}~\bibnamefont {Shah}}, \bibinfo {author} {\bibfnamefont {C.}~\bibnamefont
  {Wang}}, \bibinfo {author} {\bibfnamefont {V.}~\bibnamefont {Magrini}},
  \bibinfo {author} {\bibfnamefont {R.~K.}\ \bibnamefont {Wilson}}, \bibinfo
  {author} {\bibfnamefont {B.~L.}\ \bibnamefont {Cantarel}}, \bibinfo {author}
  {\bibfnamefont {P.~M.}\ \bibnamefont {Coutinho}}, \bibinfo {author}
  {\bibfnamefont {B.}~\bibnamefont {Henrissat}}, \bibinfo {author}
  {\bibfnamefont {L.~W.}\ \bibnamefont {Crock}}, \bibinfo {author}
  {\bibfnamefont {A.}~\bibnamefont {Russell}}, \bibinfo {author} {\bibfnamefont
  {N.~C.}\ \bibnamefont {Verberkmoes}}, \bibinfo {author} {\bibfnamefont
  {R.~L.}\ \bibnamefont {Hettich}},\ and\ \bibinfo {author} {\bibfnamefont
  {J.~I.}\ \bibnamefont {Gordon}},\ }\href
  {https://doi.org/10.1073/pnas.0901529106} {\bibfield  {journal} {\bibinfo
  {journal} {Proc. Natl. Acad. Sci. USA}\ }\textbf {\bibinfo {volume} {106}},\
  \bibinfo {pages} {5859} (\bibinfo {year} {2009})}\BibitemShut {NoStop}%
\bibitem [{\citenamefont {Flint}\ \emph {et~al.}(2012)\citenamefont {Flint},
  \citenamefont {Scott}, \citenamefont {Duncan}, \citenamefont {Louis},\ and\
  \citenamefont {Forano}}]{flint2012gm}%
  \BibitemOpen
  \bibfield  {author} {\bibinfo {author} {\bibfnamefont {H.~J.}\ \bibnamefont
  {Flint}}, \bibinfo {author} {\bibfnamefont {K.~P.}\ \bibnamefont {Scott}},
  \bibinfo {author} {\bibfnamefont {S.~H.}\ \bibnamefont {Duncan}}, \bibinfo
  {author} {\bibfnamefont {P.}~\bibnamefont {Louis}},\ and\ \bibinfo {author}
  {\bibfnamefont {E.}~\bibnamefont {Forano}},\ }\href
  {https://doi.org/10.4161/gmic.19897} {\bibfield  {journal} {\bibinfo
  {journal} {Gut Microbes}\ }\textbf {\bibinfo {volume} {3}},\ \bibinfo {pages}
  {289} (\bibinfo {year} {2012})}\BibitemShut {NoStop}%
\bibitem [{\citenamefont {Sonnenburg}\ and\ \citenamefont
  {Sonnenburg}(2014)}]{sonnenburg2014}%
  \BibitemOpen
  \bibfield  {author} {\bibinfo {author} {\bibfnamefont {E.}~\bibnamefont
  {Sonnenburg}}\ and\ \bibinfo {author} {\bibfnamefont {J.}~\bibnamefont
  {Sonnenburg}},\ }\href {https://doi.org/10.1016/j.cmet.2014.07.003}
  {\bibfield  {journal} {\bibinfo  {journal} {Cell Metab.}\ }\textbf {\bibinfo
  {volume} {20}},\ \bibinfo {pages} {779} (\bibinfo {year} {2014})}\BibitemShut
  {NoStop}%
\bibitem [{\citenamefont {Belenguer}\ \emph {et~al.}(2006)\citenamefont
  {Belenguer}, \citenamefont {Duncan}, \citenamefont {Calder}, \citenamefont
  {Holtrop}, \citenamefont {Louis}, \citenamefont {Lobley},\ and\ \citenamefont
  {Flint}}]{belenguer2006}%
  \BibitemOpen
  \bibfield  {author} {\bibinfo {author} {\bibfnamefont {A.}~\bibnamefont
  {Belenguer}}, \bibinfo {author} {\bibfnamefont {S.~H.}\ \bibnamefont
  {Duncan}}, \bibinfo {author} {\bibfnamefont {A.~G.}\ \bibnamefont {Calder}},
  \bibinfo {author} {\bibfnamefont {G.}~\bibnamefont {Holtrop}}, \bibinfo
  {author} {\bibfnamefont {P.}~\bibnamefont {Louis}}, \bibinfo {author}
  {\bibfnamefont {G.~E.}\ \bibnamefont {Lobley}},\ and\ \bibinfo {author}
  {\bibfnamefont {H.~J.}\ \bibnamefont {Flint}},\ }\href
  {https://doi.org/10.1128/aem.72.5.3593-3599.2006} {\bibfield  {journal}
  {\bibinfo  {journal} {Appl. Environ. Microbiol.}\ }\textbf {\bibinfo {volume}
  {72}},\ \bibinfo {pages} {3593} (\bibinfo {year} {2006})}\BibitemShut
  {NoStop}%
\bibitem [{\citenamefont {Falony}\ \emph {et~al.}(2006)\citenamefont {Falony},
  \citenamefont {Vlachou}, \citenamefont {Verbrugghe},\ and\ \citenamefont
  {Vuyst}}]{falony2006}%
  \BibitemOpen
  \bibfield  {author} {\bibinfo {author} {\bibfnamefont {G.}~\bibnamefont
  {Falony}}, \bibinfo {author} {\bibfnamefont {A.}~\bibnamefont {Vlachou}},
  \bibinfo {author} {\bibfnamefont {K.}~\bibnamefont {Verbrugghe}},\ and\
  \bibinfo {author} {\bibfnamefont {L.~D.}\ \bibnamefont {Vuyst}},\ }\href
  {https://doi.org/10.1128/aem.01296-06} {\bibfield  {journal} {\bibinfo
  {journal} {Appl. Environ. Microbiol.}\ }\textbf {\bibinfo {volume} {72}},\
  \bibinfo {pages} {7835} (\bibinfo {year} {2006})}\BibitemShut {NoStop}%
\bibitem [{\citenamefont {Louis}\ and\ \citenamefont
  {Flint}(2016)}]{louis2017}%
  \BibitemOpen
  \bibfield  {author} {\bibinfo {author} {\bibfnamefont {P.}~\bibnamefont
  {Louis}}\ and\ \bibinfo {author} {\bibfnamefont {H.~J.}\ \bibnamefont
  {Flint}},\ }\href {https://doi.org/10.1111/1462-2920.13589} {\bibfield
  {journal} {\bibinfo  {journal} {Environ. Microbiol.}\ }\textbf {\bibinfo
  {volume} {19}},\ \bibinfo {pages} {29} (\bibinfo {year} {2016})}\BibitemShut
  {NoStop}%
\bibitem [{\citenamefont {den Besten}\ \emph {et~al.}(2013)\citenamefont {den
  Besten}, \citenamefont {van Eunen}, \citenamefont {Groen}, \citenamefont
  {Venema}, \citenamefont {Reijngoud},\ and\ \citenamefont
  {Bakker}}]{denbesten2013}%
  \BibitemOpen
  \bibfield  {author} {\bibinfo {author} {\bibfnamefont {G.}~\bibnamefont {den
  Besten}}, \bibinfo {author} {\bibfnamefont {K.}~\bibnamefont {van Eunen}},
  \bibinfo {author} {\bibfnamefont {A.~K.}\ \bibnamefont {Groen}}, \bibinfo
  {author} {\bibfnamefont {K.}~\bibnamefont {Venema}}, \bibinfo {author}
  {\bibfnamefont {D.-J.}\ \bibnamefont {Reijngoud}},\ and\ \bibinfo {author}
  {\bibfnamefont {B.~M.}\ \bibnamefont {Bakker}},\ }\href
  {https://doi.org/10.1194/jlr.r036012} {\bibfield  {journal} {\bibinfo
  {journal} {J. Lipid Res.}\ }\textbf {\bibinfo {volume} {54}},\ \bibinfo
  {pages} {2325} (\bibinfo {year} {2013})}\BibitemShut {NoStop}%
\bibitem [{\citenamefont {Iebba}\ \emph {et~al.}(2013)\citenamefont {Iebba},
  \citenamefont {Santangelo}, \citenamefont {Totino}, \citenamefont
  {Nicoletti}, \citenamefont {Gagliardi}, \citenamefont {Biase}, \citenamefont
  {Cucchiara}, \citenamefont {Nencioni}, \citenamefont {Conte},\ and\
  \citenamefont {Schippa}}]{iebba2013}%
  \BibitemOpen
  \bibfield  {author} {\bibinfo {author} {\bibfnamefont {V.}~\bibnamefont
  {Iebba}}, \bibinfo {author} {\bibfnamefont {F.}~\bibnamefont {Santangelo}},
  \bibinfo {author} {\bibfnamefont {V.}~\bibnamefont {Totino}}, \bibinfo
  {author} {\bibfnamefont {M.}~\bibnamefont {Nicoletti}}, \bibinfo {author}
  {\bibfnamefont {A.}~\bibnamefont {Gagliardi}}, \bibinfo {author}
  {\bibfnamefont {R.~V.~D.}\ \bibnamefont {Biase}}, \bibinfo {author}
  {\bibfnamefont {S.}~\bibnamefont {Cucchiara}}, \bibinfo {author}
  {\bibfnamefont {L.}~\bibnamefont {Nencioni}}, \bibinfo {author}
  {\bibfnamefont {M.~P.}\ \bibnamefont {Conte}},\ and\ \bibinfo {author}
  {\bibfnamefont {S.}~\bibnamefont {Schippa}},\ }\href
  {https://doi.org/10.1371/journal.pone.0061608} {\bibfield  {journal}
  {\bibinfo  {journal} {PLoS ONE}\ }\textbf {\bibinfo {volume} {8}},\ \bibinfo
  {pages} {e61608} (\bibinfo {year} {2013})}\BibitemShut {NoStop}%
\bibitem [{\citenamefont {Riley}\ and\ \citenamefont
  {Wertz}(2002)}]{riley2002}%
  \BibitemOpen
  \bibfield  {author} {\bibinfo {author} {\bibfnamefont {M.~A.}\ \bibnamefont
  {Riley}}\ and\ \bibinfo {author} {\bibfnamefont {J.~E.}\ \bibnamefont
  {Wertz}},\ }\href {https://doi.org/10.1146/annurev.micro.56.012302.161024}
  {\bibfield  {journal} {\bibinfo  {journal} {Annu. Rev. Microbiol.}\ }\textbf
  {\bibinfo {volume} {56}},\ \bibinfo {pages} {117} (\bibinfo {year}
  {2002})}\BibitemShut {NoStop}%
\bibitem [{\citenamefont {Wahlström}\ \emph {et~al.}(2016)\citenamefont
  {Wahlström}, \citenamefont {Sayin}, \citenamefont {Marschall},\ and\
  \citenamefont {Bäckhed}}]{wahlstrom2016}%
  \BibitemOpen
  \bibfield  {author} {\bibinfo {author} {\bibfnamefont {A.}~\bibnamefont
  {Wahlström}}, \bibinfo {author} {\bibfnamefont {S.}~\bibnamefont {Sayin}},
  \bibinfo {author} {\bibfnamefont {H.-U.}\ \bibnamefont {Marschall}},\ and\
  \bibinfo {author} {\bibfnamefont {F.}~\bibnamefont {Bäckhed}},\ }\href
  {https://doi.org/10.1016/j.cmet.2016.05.005} {\bibfield  {journal} {\bibinfo
  {journal} {Cell Metab.}\ }\textbf {\bibinfo {volume} {24}},\ \bibinfo {pages}
  {41} (\bibinfo {year} {2016})}\BibitemShut {NoStop}%
\bibitem [{\citenamefont {Stein}\ \emph {et~al.}(2013)\citenamefont {Stein},
  \citenamefont {Bucci}, \citenamefont {Toussaint}, \citenamefont {Buffie},
  \citenamefont {Rätsch}, \citenamefont {Pamer}, \citenamefont {Sander},\ and\
  \citenamefont {Xavier}}]{stein2013}%
  \BibitemOpen
  \bibfield  {author} {\bibinfo {author} {\bibfnamefont {R.~R.}\ \bibnamefont
  {Stein}}, \bibinfo {author} {\bibfnamefont {V.}~\bibnamefont {Bucci}},
  \bibinfo {author} {\bibfnamefont {N.~C.}\ \bibnamefont {Toussaint}}, \bibinfo
  {author} {\bibfnamefont {C.~G.}\ \bibnamefont {Buffie}}, \bibinfo {author}
  {\bibfnamefont {G.}~\bibnamefont {Rätsch}}, \bibinfo {author} {\bibfnamefont
  {E.~G.}\ \bibnamefont {Pamer}}, \bibinfo {author} {\bibfnamefont
  {C.}~\bibnamefont {Sander}},\ and\ \bibinfo {author} {\bibfnamefont {J.~B.}\
  \bibnamefont {Xavier}},\ }\href
  {https://doi.org/10.1371/journal.pcbi.1003388} {\bibfield  {journal}
  {\bibinfo  {journal} {PLoS Comput. Biol.}\ }\textbf {\bibinfo {volume} {9}},\
  \bibinfo {pages} {e1003388} (\bibinfo {year} {2013})}\BibitemShut {NoStop}%
\bibitem [{\citenamefont {Bucci}\ and\ \citenamefont
  {Xavier}(2014)}]{bucci2014}%
  \BibitemOpen
  \bibfield  {author} {\bibinfo {author} {\bibfnamefont {V.}~\bibnamefont
  {Bucci}}\ and\ \bibinfo {author} {\bibfnamefont {J.~B.}\ \bibnamefont
  {Xavier}},\ }\href {https://doi.org/10.1016/j.jmb.2014.03.017} {\bibfield
  {journal} {\bibinfo  {journal} {J. Mol. Biol.}\ }\textbf {\bibinfo {volume}
  {426}},\ \bibinfo {pages} {3907} (\bibinfo {year} {2014})}\BibitemShut
  {NoStop}%
\bibitem [{\citenamefont {Coyte}\ \emph {et~al.}(2015)\citenamefont {Coyte},
  \citenamefont {Schluter},\ and\ \citenamefont {Foster}}]{coyte2015}%
  \BibitemOpen
  \bibfield  {author} {\bibinfo {author} {\bibfnamefont {K.~Z.}\ \bibnamefont
  {Coyte}}, \bibinfo {author} {\bibfnamefont {J.}~\bibnamefont {Schluter}},\
  and\ \bibinfo {author} {\bibfnamefont {K.~R.}\ \bibnamefont {Foster}},\
  }\href {https://doi.org/10.1126/science.aad2602} {\bibfield  {journal}
  {\bibinfo  {journal} {Science}\ }\textbf {\bibinfo {volume} {350}},\ \bibinfo
  {pages} {663} (\bibinfo {year} {2015})}\BibitemShut {NoStop}%
\bibitem [{\citenamefont {Goyal}\ and\ \citenamefont
  {Maslov}(2018)}]{goyal2018prl}%
  \BibitemOpen
  \bibfield  {author} {\bibinfo {author} {\bibfnamefont {A.}~\bibnamefont
  {Goyal}}\ and\ \bibinfo {author} {\bibfnamefont {S.}~\bibnamefont {Maslov}},\
  }\bibfield  {journal} {\bibinfo  {journal} {Phys. Rev. Lett.}\ }\textbf
  {\bibinfo {volume} {120}},\ \href
  {https://doi.org/10.1103/physrevlett.120.158102}
  {10.1103/physrevlett.120.158102} (\bibinfo {year} {2018})\BibitemShut
  {NoStop}%
\bibitem [{\citenamefont {Goyal}\ \emph {et~al.}(2018)\citenamefont {Goyal},
  \citenamefont {Dubinkina},\ and\ \citenamefont {Maslov}}]{goyal2018isme}%
  \BibitemOpen
  \bibfield  {author} {\bibinfo {author} {\bibfnamefont {A.}~\bibnamefont
  {Goyal}}, \bibinfo {author} {\bibfnamefont {V.}~\bibnamefont {Dubinkina}},\
  and\ \bibinfo {author} {\bibfnamefont {S.}~\bibnamefont {Maslov}},\ }\href
  {https://doi.org/10.1038/s41396-018-0222-x} {\bibfield  {journal} {\bibinfo
  {journal} {ISME J.}\ }\textbf {\bibinfo {volume} {12}},\ \bibinfo {pages}
  {2823} (\bibinfo {year} {2018})}\BibitemShut {NoStop}%
\bibitem [{\citenamefont {Neubert}\ and\ \citenamefont
  {Caswell}(1997)}]{neubert1997}%
  \BibitemOpen
  \bibfield  {author} {\bibinfo {author} {\bibfnamefont {M.~G.}\ \bibnamefont
  {Neubert}}\ and\ \bibinfo {author} {\bibfnamefont {H.}~\bibnamefont
  {Caswell}},\ }\href {https://doi.org/10.2307/2266047} {\bibfield  {journal}
  {\bibinfo  {journal} {Ecology}\ }\textbf {\bibinfo {volume} {78}},\ \bibinfo
  {pages} {653} (\bibinfo {year} {1997})}\BibitemShut {NoStop}%
\bibitem [{\citenamefont {Nisbet}\ and\ \citenamefont
  {Gurney}(1976)}]{nisbet1976mechanism}%
  \BibitemOpen
  \bibfield  {author} {\bibinfo {author} {\bibfnamefont {R.~M.}\ \bibnamefont
  {Nisbet}}\ and\ \bibinfo {author} {\bibfnamefont {W.~S.~C.}\ \bibnamefont
  {Gurney}},\ }\href {https://doi.org/10.1038/263319a0} {\bibfield  {journal}
  {\bibinfo  {journal} {Nature}\ }\textbf {\bibinfo {volume} {263}},\ \bibinfo
  {pages} {319} (\bibinfo {year} {1976})}\BibitemShut {NoStop}%
\bibitem [{\citenamefont {McKane}\ and\ \citenamefont
  {Newman}(2005)}]{mckane2005predatorprey}%
  \BibitemOpen
  \bibfield  {author} {\bibinfo {author} {\bibfnamefont {A.~J.}\ \bibnamefont
  {McKane}}\ and\ \bibinfo {author} {\bibfnamefont {T.~J.}\ \bibnamefont
  {Newman}},\ }\href {https://doi.org/10.1103/PhysRevLett.94.218102} {\bibfield
   {journal} {\bibinfo  {journal} {Physical Review Letters}\ }\textbf {\bibinfo
  {volume} {94}},\ \bibinfo {pages} {218102} (\bibinfo {year}
  {2005})}\BibitemShut {NoStop}%
\bibitem [{\citenamefont {Nicoletti}\ \emph {et~al.}(2018)\citenamefont
  {Nicoletti}, \citenamefont {Zagli}, \citenamefont {Fanelli}, \citenamefont
  {Livi}, \citenamefont {Carletti},\ and\ \citenamefont
  {Innocenti}}]{nicoletti2018nonnormal}%
  \BibitemOpen
  \bibfield  {author} {\bibinfo {author} {\bibfnamefont {S.}~\bibnamefont
  {Nicoletti}}, \bibinfo {author} {\bibfnamefont {N.}~\bibnamefont {Zagli}},
  \bibinfo {author} {\bibfnamefont {D.}~\bibnamefont {Fanelli}}, \bibinfo
  {author} {\bibfnamefont {R.}~\bibnamefont {Livi}}, \bibinfo {author}
  {\bibfnamefont {T.}~\bibnamefont {Carletti}},\ and\ \bibinfo {author}
  {\bibfnamefont {G.}~\bibnamefont {Innocenti}},\ }\href
  {https://doi.org/10.1103/PhysRevE.98.032214} {\bibfield  {journal} {\bibinfo
  {journal} {Physical Review E}\ }\textbf {\bibinfo {volume} {98}},\ \bibinfo
  {pages} {032214} (\bibinfo {year} {2018})}\BibitemShut {NoStop}%
\bibitem [{\citenamefont {Muolo}\ \emph
  {et~al.}(2019{\natexlab{b}})\citenamefont {Muolo}, \citenamefont {Asllani},
  \citenamefont {Fanelli}, \citenamefont {Maini},\ and\ \citenamefont
  {Carletti}}]{muolo2019patterns}%
  \BibitemOpen
  \bibfield  {author} {\bibinfo {author} {\bibfnamefont {R.}~\bibnamefont
  {Muolo}}, \bibinfo {author} {\bibfnamefont {M.}~\bibnamefont {Asllani}},
  \bibinfo {author} {\bibfnamefont {D.}~\bibnamefont {Fanelli}}, \bibinfo
  {author} {\bibfnamefont {P.~K.}\ \bibnamefont {Maini}},\ and\ \bibinfo
  {author} {\bibfnamefont {T.}~\bibnamefont {Carletti}},\ }\href
  {https://doi.org/10.1016/j.jtbi.2019.07.004} {\bibfield  {journal} {\bibinfo
  {journal} {Journal of Theoretical Biology}\ }\textbf {\bibinfo {volume}
  {480}},\ \bibinfo {pages} {81} (\bibinfo {year}
  {2019}{\natexlab{b}})}\BibitemShut {NoStop}%
\bibitem [{\citenamefont {Biancalani}\ \emph {et~al.}(2017)\citenamefont
  {Biancalani}, \citenamefont {Jafarpour},\ and\ \citenamefont
  {Goldenfeld}}]{biancalani2017giant}%
  \BibitemOpen
  \bibfield  {author} {\bibinfo {author} {\bibfnamefont {T.}~\bibnamefont
  {Biancalani}}, \bibinfo {author} {\bibfnamefont {F.}~\bibnamefont
  {Jafarpour}},\ and\ \bibinfo {author} {\bibfnamefont {N.}~\bibnamefont
  {Goldenfeld}},\ }\href {https://doi.org/10.1103/PhysRevLett.118.018101}
  {\bibfield  {journal} {\bibinfo  {journal} {Physical Review Letters}\
  }\textbf {\bibinfo {volume} {118}},\ \bibinfo {pages} {018101} (\bibinfo
  {year} {2017})}\BibitemShut {NoStop}%
\bibitem [{\citenamefont {Poggialini}\ \emph {et~al.}(2025)\citenamefont
  {Poggialini}, \citenamefont {Santo}, \citenamefont {Villegas}, \citenamefont
  {Gabrielli},\ and\ \citenamefont {{n}oz}}]{poggialini2025nonreciprocal}%
  \BibitemOpen
  \bibfield  {author} {\bibinfo {author} {\bibfnamefont {A.}~\bibnamefont
  {Poggialini}}, \bibinfo {author} {\bibfnamefont {S.~D.}\ \bibnamefont
  {Santo}}, \bibinfo {author} {\bibfnamefont {P.}~\bibnamefont {Villegas}},
  \bibinfo {author} {\bibfnamefont {A.}~\bibnamefont {Gabrielli}},\ and\
  \bibinfo {author} {\bibfnamefont {M.~A.~M.}\ \bibnamefont {{n}oz}},\
  }\bibfield  {journal} {\bibinfo  {journal} {arXiv preprint arXiv:2507.19127}\
  }\href {https://doi.org/10.48550/arXiv.2507.19127}
  {10.48550/arXiv.2507.19127} (\bibinfo {year} {2025})\BibitemShut {NoStop}%
\bibitem [{\citenamefont {Hennequin}\ \emph {et~al.}(2014)\citenamefont
  {Hennequin}, \citenamefont {Vogels},\ and\ \citenamefont
  {Gerstner}}]{Hennequin2014}%
  \BibitemOpen
  \bibfield  {author} {\bibinfo {author} {\bibfnamefont {G.}~\bibnamefont
  {Hennequin}}, \bibinfo {author} {\bibfnamefont {T.~P.}\ \bibnamefont
  {Vogels}},\ and\ \bibinfo {author} {\bibfnamefont {W.}~\bibnamefont
  {Gerstner}},\ }\href@noop {} {\bibfield  {journal} {\bibinfo  {journal}
  {Neuron}\ }\textbf {\bibinfo {volume} {82}},\ \bibinfo {pages} {1394}
  (\bibinfo {year} {2014})}\BibitemShut {NoStop}%
\bibitem [{\citenamefont {Murphy}\ and\ \citenamefont
  {Miller}(2009)}]{Murphy2009}%
  \BibitemOpen
  \bibfield  {author} {\bibinfo {author} {\bibfnamefont {B.~K.}\ \bibnamefont
  {Murphy}}\ and\ \bibinfo {author} {\bibfnamefont {K.~D.}\ \bibnamefont
  {Miller}},\ }\href@noop {} {\bibfield  {journal} {\bibinfo  {journal}
  {Neuron}\ }\textbf {\bibinfo {volume} {61}},\ \bibinfo {pages} {635}
  (\bibinfo {year} {2009})}\BibitemShut {NoStop}%
\bibitem [{\citenamefont {Ganguli}\ \emph {et~al.}(2008)\citenamefont
  {Ganguli}, \citenamefont {Huh},\ and\ \citenamefont
  {Sompolinsky}}]{Ganguli2008}%
  \BibitemOpen
  \bibfield  {author} {\bibinfo {author} {\bibfnamefont {S.}~\bibnamefont
  {Ganguli}}, \bibinfo {author} {\bibfnamefont {D.}~\bibnamefont {Huh}},\ and\
  \bibinfo {author} {\bibfnamefont {H.}~\bibnamefont {Sompolinsky}},\
  }\href@noop {} {\bibfield  {journal} {\bibinfo  {journal} {Proceedings of the
  National Academy of Sciences}\ }\textbf {\bibinfo {volume} {105}},\ \bibinfo
  {pages} {18970} (\bibinfo {year} {2008})}\BibitemShut {NoStop}%
\bibitem [{\citenamefont {Kurokawa}\ \emph {et~al.}(2026)\citenamefont
  {Kurokawa}, \citenamefont {Maskawa}, \citenamefont {Arakawa}, \citenamefont
  {Masuoka}, \citenamefont {Takayasu}, \citenamefont {Yoshikawa}, \citenamefont
  {Raihan}, \citenamefont {Shindo}, \citenamefont {Kaida}, \citenamefont
  {Takagi}, \citenamefont {Tanokura}, \citenamefont {Takayasu}, \citenamefont
  {Takayasu},\ and\ \citenamefont {Suda}}]{microbiome2025}%
  \BibitemOpen
  \bibfield  {author} {\bibinfo {author} {\bibfnamefont {R.}~\bibnamefont
  {Kurokawa}}, \bibinfo {author} {\bibfnamefont {R.}~\bibnamefont {Maskawa}},
  \bibinfo {author} {\bibfnamefont {M.}~\bibnamefont {Arakawa}}, \bibinfo
  {author} {\bibfnamefont {H.}~\bibnamefont {Masuoka}}, \bibinfo {author}
  {\bibfnamefont {H.}~\bibnamefont {Takayasu}}, \bibinfo {author}
  {\bibfnamefont {Y.}~\bibnamefont {Yoshikawa}}, \bibinfo {author}
  {\bibfnamefont {T.}~\bibnamefont {Raihan}}, \bibinfo {author} {\bibfnamefont
  {C.}~\bibnamefont {Shindo}}, \bibinfo {author} {\bibfnamefont
  {K.}~\bibnamefont {Kaida}}, \bibinfo {author} {\bibfnamefont
  {M.}~\bibnamefont {Takagi}}, \bibinfo {author} {\bibfnamefont
  {M.}~\bibnamefont {Tanokura}}, \bibinfo {author} {\bibfnamefont
  {L.}~\bibnamefont {Takayasu}}, \bibinfo {author} {\bibfnamefont
  {M.}~\bibnamefont {Takayasu}},\ and\ \bibinfo {author} {\bibfnamefont
  {W.}~\bibnamefont {Suda}},\ }\bibfield  {journal} {\bibinfo  {journal}
  {bioRxiv}\ }\href {https://doi.org/10.64898/2026.03.26.714232}
  {10.64898/2026.03.26.714232} (\bibinfo {year} {2026}),\ \bibinfo {note}
  {preprint, under review at Microbiome (2026).}\BibitemShut {Stop}%
\bibitem [{\citenamefont {Theiler}\ \emph {et~al.}(1992)\citenamefont
  {Theiler}, \citenamefont {Eubank}, \citenamefont {Longtin}, \citenamefont
  {Galdrikian},\ and\ \citenamefont {Farmer}}]{theiler1992}%
  \BibitemOpen
  \bibfield  {author} {\bibinfo {author} {\bibfnamefont {J.}~\bibnamefont
  {Theiler}}, \bibinfo {author} {\bibfnamefont {S.}~\bibnamefont {Eubank}},
  \bibinfo {author} {\bibfnamefont {A.}~\bibnamefont {Longtin}}, \bibinfo
  {author} {\bibfnamefont {B.}~\bibnamefont {Galdrikian}},\ and\ \bibinfo
  {author} {\bibfnamefont {J.~D.}\ \bibnamefont {Farmer}},\ }\href
  {https://doi.org/10.1016/0167-2789(92)90102-s} {\bibfield  {journal}
  {\bibinfo  {journal} {Physica D}\ }\textbf {\bibinfo {volume} {58}},\
  \bibinfo {pages} {77} (\bibinfo {year} {1992})}\BibitemShut {NoStop}%
\bibitem [{\citenamefont {Schreiber}\ and\ \citenamefont
  {Schmitz}(1996)}]{schreiber1996}%
  \BibitemOpen
  \bibfield  {author} {\bibinfo {author} {\bibfnamefont {T.}~\bibnamefont
  {Schreiber}}\ and\ \bibinfo {author} {\bibfnamefont {A.}~\bibnamefont
  {Schmitz}},\ }\href {https://doi.org/10.1103/physrevlett.77.635} {\bibfield
  {journal} {\bibinfo  {journal} {Phys. Rev. Lett.}\ }\textbf {\bibinfo
  {volume} {77}},\ \bibinfo {pages} {635} (\bibinfo {year} {1996})}\BibitemShut
  {NoStop}%
\bibitem [{\citenamefont {Schreiber}\ and\ \citenamefont
  {Schmitz}(2000)}]{schreiber2000}%
  \BibitemOpen
  \bibfield  {author} {\bibinfo {author} {\bibfnamefont {T.}~\bibnamefont
  {Schreiber}}\ and\ \bibinfo {author} {\bibfnamefont {A.}~\bibnamefont
  {Schmitz}},\ }\href {https://doi.org/10.1016/s0167-2789(00)00043-9}
  {\bibfield  {journal} {\bibinfo  {journal} {Physica D}\ }\textbf {\bibinfo
  {volume} {142}},\ \bibinfo {pages} {346} (\bibinfo {year}
  {2000})}\BibitemShut {NoStop}%
\bibitem [{\citenamefont {Lancaster}\ \emph {et~al.}(2018)\citenamefont
  {Lancaster}, \citenamefont {Iatsenko}, \citenamefont {Pidde}, \citenamefont
  {Ticcinelli},\ and\ \citenamefont {Stefanovska}}]{lancaster2018}%
  \BibitemOpen
  \bibfield  {author} {\bibinfo {author} {\bibfnamefont {G.}~\bibnamefont
  {Lancaster}}, \bibinfo {author} {\bibfnamefont {D.}~\bibnamefont {Iatsenko}},
  \bibinfo {author} {\bibfnamefont {A.}~\bibnamefont {Pidde}}, \bibinfo
  {author} {\bibfnamefont {V.}~\bibnamefont {Ticcinelli}},\ and\ \bibinfo
  {author} {\bibfnamefont {A.}~\bibnamefont {Stefanovska}},\ }\href
  {https://doi.org/10.1016/j.physrep.2018.06.001} {\bibfield  {journal}
  {\bibinfo  {journal} {Phys. Rep.}\ }\textbf {\bibinfo {volume} {748}},\
  \bibinfo {pages} {1} (\bibinfo {year} {2018})}\BibitemShut {NoStop}%
\bibitem [{\citenamefont {Carmody}\ \emph {et~al.}(2015)\citenamefont
  {Carmody}, \citenamefont {Gerber}, \citenamefont {Luevano}, \citenamefont
  {Gatti}, \citenamefont {Somes}, \citenamefont {Svenson},\ and\ \citenamefont
  {Turnbaugh}}]{carmody2015}%
  \BibitemOpen
  \bibfield  {author} {\bibinfo {author} {\bibfnamefont {R.}~\bibnamefont
  {Carmody}}, \bibinfo {author} {\bibfnamefont {G.}~\bibnamefont {Gerber}},
  \bibinfo {author} {\bibfnamefont {J.}~\bibnamefont {Luevano}}, \bibinfo
  {author} {\bibfnamefont {D.}~\bibnamefont {Gatti}}, \bibinfo {author}
  {\bibfnamefont {L.}~\bibnamefont {Somes}}, \bibinfo {author} {\bibfnamefont
  {K.}~\bibnamefont {Svenson}},\ and\ \bibinfo {author} {\bibfnamefont
  {P.}~\bibnamefont {Turnbaugh}},\ }\href
  {https://doi.org/10.1016/j.chom.2014.11.010} {\bibfield  {journal} {\bibinfo
  {journal} {Cell Host Microbe}\ }\textbf {\bibinfo {volume} {17}},\ \bibinfo
  {pages} {72} (\bibinfo {year} {2015})}\BibitemShut {NoStop}%
\bibitem [{\citenamefont {Friswell}\ \emph {et~al.}(2010)\citenamefont
  {Friswell}, \citenamefont {Gika}, \citenamefont {Stratford}, \citenamefont
  {Theodoridis}, \citenamefont {Telfer}, \citenamefont {Wilson},\ and\
  \citenamefont {McBain}}]{friswell2010}%
  \BibitemOpen
  \bibfield  {author} {\bibinfo {author} {\bibfnamefont {M.~K.}\ \bibnamefont
  {Friswell}}, \bibinfo {author} {\bibfnamefont {H.}~\bibnamefont {Gika}},
  \bibinfo {author} {\bibfnamefont {I.~J.}\ \bibnamefont {Stratford}}, \bibinfo
  {author} {\bibfnamefont {G.}~\bibnamefont {Theodoridis}}, \bibinfo {author}
  {\bibfnamefont {B.}~\bibnamefont {Telfer}}, \bibinfo {author} {\bibfnamefont
  {I.~D.}\ \bibnamefont {Wilson}},\ and\ \bibinfo {author} {\bibfnamefont
  {A.~J.}\ \bibnamefont {McBain}},\ }\href
  {https://doi.org/10.1371/journal.pone.0008584} {\bibfield  {journal}
  {\bibinfo  {journal} {PLoS ONE}\ }\textbf {\bibinfo {volume} {5}},\ \bibinfo
  {pages} {e8584} (\bibinfo {year} {2010})}\BibitemShut {NoStop}%
\bibitem [{\citenamefont {David}\ \emph {et~al.}(2013)\citenamefont {David},
  \citenamefont {Maurice}, \citenamefont {Carmody}, \citenamefont {Gootenberg},
  \citenamefont {Button}, \citenamefont {Wolfe}, \citenamefont {Ling},
  \citenamefont {Devlin}, \citenamefont {Varma}, \citenamefont {Fischbach},
  \citenamefont {Biddinger}, \citenamefont {Dutton},\ and\ \citenamefont
  {Turnbaugh}}]{david2014}%
  \BibitemOpen
  \bibfield  {author} {\bibinfo {author} {\bibfnamefont {L.~A.}\ \bibnamefont
  {David}}, \bibinfo {author} {\bibfnamefont {C.~F.}\ \bibnamefont {Maurice}},
  \bibinfo {author} {\bibfnamefont {R.~N.}\ \bibnamefont {Carmody}}, \bibinfo
  {author} {\bibfnamefont {D.~B.}\ \bibnamefont {Gootenberg}}, \bibinfo
  {author} {\bibfnamefont {J.~E.}\ \bibnamefont {Button}}, \bibinfo {author}
  {\bibfnamefont {B.~E.}\ \bibnamefont {Wolfe}}, \bibinfo {author}
  {\bibfnamefont {A.~V.}\ \bibnamefont {Ling}}, \bibinfo {author}
  {\bibfnamefont {A.~S.}\ \bibnamefont {Devlin}}, \bibinfo {author}
  {\bibfnamefont {Y.}~\bibnamefont {Varma}}, \bibinfo {author} {\bibfnamefont
  {M.~A.}\ \bibnamefont {Fischbach}}, \bibinfo {author} {\bibfnamefont {S.~B.}\
  \bibnamefont {Biddinger}}, \bibinfo {author} {\bibfnamefont {R.~J.}\
  \bibnamefont {Dutton}},\ and\ \bibinfo {author} {\bibfnamefont {P.~J.}\
  \bibnamefont {Turnbaugh}},\ }\href {https://doi.org/10.1038/nature12820}
  {\bibfield  {journal} {\bibinfo  {journal} {Nature}\ }\textbf {\bibinfo
  {volume} {505}},\ \bibinfo {pages} {559} (\bibinfo {year}
  {2013})}\BibitemShut {NoStop}%
\bibitem [{\citenamefont {Seifert}(2012)}]{seifert2012stochastic}%
  \BibitemOpen
  \bibfield  {author} {\bibinfo {author} {\bibfnamefont {U.}~\bibnamefont
  {Seifert}},\ }\href@noop {} {\bibfield  {journal} {\bibinfo  {journal}
  {Reports on Progress in Physics}\ }\textbf {\bibinfo {volume} {75}},\
  \bibinfo {pages} {126001} (\bibinfo {year} {2012})}\BibitemShut {NoStop}%
\bibitem [{\citenamefont {Gnesotto}\ \emph {et~al.}(2018)\citenamefont
  {Gnesotto}, \citenamefont {Mura}, \citenamefont {Gladrow},\ and\
  \citenamefont {Broedersz}}]{gnesotto2018broken}%
  \BibitemOpen
  \bibfield  {author} {\bibinfo {author} {\bibfnamefont {F.~S.}\ \bibnamefont
  {Gnesotto}}, \bibinfo {author} {\bibfnamefont {F.}~\bibnamefont {Mura}},
  \bibinfo {author} {\bibfnamefont {J.}~\bibnamefont {Gladrow}},\ and\ \bibinfo
  {author} {\bibfnamefont {C.~P.}\ \bibnamefont {Broedersz}},\ }\href@noop {}
  {\bibfield  {journal} {\bibinfo  {journal} {Reports on Progress in Physics}\
  }\textbf {\bibinfo {volume} {81}},\ \bibinfo {pages} {066601} (\bibinfo
  {year} {2018})}\BibitemShut {NoStop}%
\bibitem [{\citenamefont {Fyodorov}\ \emph {et~al.}(2025)\citenamefont
  {Fyodorov}, \citenamefont {Gudowska-Nowak}, \citenamefont {Nowak},\ and\
  \citenamefont {Tarnowski}}]{fyodorov2025nonorthogonal}%
  \BibitemOpen
  \bibfield  {author} {\bibinfo {author} {\bibfnamefont {Y.~V.}\ \bibnamefont
  {Fyodorov}}, \bibinfo {author} {\bibfnamefont {E.}~\bibnamefont
  {Gudowska-Nowak}}, \bibinfo {author} {\bibfnamefont {M.~A.}\ \bibnamefont
  {Nowak}},\ and\ \bibinfo {author} {\bibfnamefont {W.}~\bibnamefont
  {Tarnowski}},\ }\href {https://doi.org/10.1103/PhysRevLett.134.087102}
  {\bibfield  {journal} {\bibinfo  {journal} {Phys. Rev. Lett.}\ }\textbf
  {\bibinfo {volume} {134}},\ \bibinfo {pages} {087102} (\bibinfo {year}
  {2025})}\BibitemShut {NoStop}%
\bibitem [{\citenamefont {Johansson}\ \emph {et~al.}(2008)\citenamefont
  {Johansson}, \citenamefont {Phillipson}, \citenamefont {Petersson},
  \citenamefont {Velcich}, \citenamefont {Holm},\ and\ \citenamefont
  {Hansson}}]{johansson2008mucus}%
  \BibitemOpen
  \bibfield  {author} {\bibinfo {author} {\bibfnamefont {M.~E.~V.}\
  \bibnamefont {Johansson}}, \bibinfo {author} {\bibfnamefont {M.}~\bibnamefont
  {Phillipson}}, \bibinfo {author} {\bibfnamefont {J.}~\bibnamefont
  {Petersson}}, \bibinfo {author} {\bibfnamefont {A.}~\bibnamefont {Velcich}},
  \bibinfo {author} {\bibfnamefont {L.}~\bibnamefont {Holm}},\ and\ \bibinfo
  {author} {\bibfnamefont {G.~C.}\ \bibnamefont {Hansson}},\ }\href
  {https://doi.org/10.1073/pnas.0803124105} {\bibfield  {journal} {\bibinfo
  {journal} {Proc. Natl. Acad. Sci. USA}\ }\textbf {\bibinfo {volume} {105}},\
  \bibinfo {pages} {15064} (\bibinfo {year} {2008})}\BibitemShut {NoStop}%
\bibitem [{\citenamefont {Johansson}\ and\ \citenamefont
  {Hansson}(2013)}]{johansson2013}%
  \BibitemOpen
  \bibfield  {author} {\bibinfo {author} {\bibfnamefont {M.~E.}\ \bibnamefont
  {Johansson}}\ and\ \bibinfo {author} {\bibfnamefont {G.~C.}\ \bibnamefont
  {Hansson}},\ }\href {https://doi.org/10.1159/000354683} {\bibfield  {journal}
  {\bibinfo  {journal} {Dig. Dis.}\ }\textbf {\bibinfo {volume} {31}},\
  \bibinfo {pages} {305} (\bibinfo {year} {2013})}\BibitemShut {NoStop}%
\bibitem [{\citenamefont {Korem}\ \emph {et~al.}(2015)\citenamefont {Korem},
  \citenamefont {Zeevi}, \citenamefont {Suez}, \citenamefont {Weinberger},
  \citenamefont {Avnit-Sagi}, \citenamefont {Pompan-Lotan}, \citenamefont
  {Matot}, \citenamefont {Jona}, \citenamefont {Harmelin}, \citenamefont
  {Cohen}, \citenamefont {Sirota-Madi}, \citenamefont {Thaiss}, \citenamefont
  {Pevsner-Fischer}, \citenamefont {Sorek}, \citenamefont {Xavier},
  \citenamefont {Elinav},\ and\ \citenamefont {Segal}}]{korem2015}%
  \BibitemOpen
  \bibfield  {author} {\bibinfo {author} {\bibfnamefont {T.}~\bibnamefont
  {Korem}}, \bibinfo {author} {\bibfnamefont {D.}~\bibnamefont {Zeevi}},
  \bibinfo {author} {\bibfnamefont {J.}~\bibnamefont {Suez}}, \bibinfo {author}
  {\bibfnamefont {A.}~\bibnamefont {Weinberger}}, \bibinfo {author}
  {\bibfnamefont {T.}~\bibnamefont {Avnit-Sagi}}, \bibinfo {author}
  {\bibfnamefont {M.}~\bibnamefont {Pompan-Lotan}}, \bibinfo {author}
  {\bibfnamefont {E.}~\bibnamefont {Matot}}, \bibinfo {author} {\bibfnamefont
  {G.}~\bibnamefont {Jona}}, \bibinfo {author} {\bibfnamefont {A.}~\bibnamefont
  {Harmelin}}, \bibinfo {author} {\bibfnamefont {N.}~\bibnamefont {Cohen}},
  \bibinfo {author} {\bibfnamefont {A.}~\bibnamefont {Sirota-Madi}}, \bibinfo
  {author} {\bibfnamefont {C.~A.}\ \bibnamefont {Thaiss}}, \bibinfo {author}
  {\bibfnamefont {M.}~\bibnamefont {Pevsner-Fischer}}, \bibinfo {author}
  {\bibfnamefont {R.}~\bibnamefont {Sorek}}, \bibinfo {author} {\bibfnamefont
  {R.~J.}\ \bibnamefont {Xavier}}, \bibinfo {author} {\bibfnamefont
  {E.}~\bibnamefont {Elinav}},\ and\ \bibinfo {author} {\bibfnamefont
  {E.}~\bibnamefont {Segal}},\ }\href {https://doi.org/10.1126/science.aac4812}
  {\bibfield  {journal} {\bibinfo  {journal} {Science}\ }\textbf {\bibinfo
  {volume} {349}},\ \bibinfo {pages} {1101} (\bibinfo {year}
  {2015})}\BibitemShut {NoStop}%
\bibitem [{\citenamefont {Brown}\ \emph {et~al.}(2016)\citenamefont {Brown},
  \citenamefont {Olm}, \citenamefont {Thomas},\ and\ \citenamefont
  {Banfield}}]{brown2016irep}%
  \BibitemOpen
  \bibfield  {author} {\bibinfo {author} {\bibfnamefont {C.~T.}\ \bibnamefont
  {Brown}}, \bibinfo {author} {\bibfnamefont {M.~R.}\ \bibnamefont {Olm}},
  \bibinfo {author} {\bibfnamefont {B.~C.}\ \bibnamefont {Thomas}},\ and\
  \bibinfo {author} {\bibfnamefont {J.~F.}\ \bibnamefont {Banfield}},\ }\href
  {https://doi.org/10.1038/nbt.3704} {\bibfield  {journal} {\bibinfo  {journal}
  {Nat. Biotechnol.}\ }\textbf {\bibinfo {volume} {34}},\ \bibinfo {pages}
  {1256} (\bibinfo {year} {2016})}\BibitemShut {NoStop}%
\bibitem [{\citenamefont {Marchetti}\ \emph {et~al.}(2013)\citenamefont
  {Marchetti}, \citenamefont {Joanny}, \citenamefont {Ramaswamy}, \citenamefont
  {Liverpool}, \citenamefont {Prost}, \citenamefont {Rao},\ and\ \citenamefont
  {Simha}}]{marchetti2013hydrodynamics}%
  \BibitemOpen
  \bibfield  {author} {\bibinfo {author} {\bibfnamefont {M.~C.}\ \bibnamefont
  {Marchetti}}, \bibinfo {author} {\bibfnamefont {J.~F.}\ \bibnamefont
  {Joanny}}, \bibinfo {author} {\bibfnamefont {S.}~\bibnamefont {Ramaswamy}},
  \bibinfo {author} {\bibfnamefont {T.~B.}\ \bibnamefont {Liverpool}}, \bibinfo
  {author} {\bibfnamefont {J.}~\bibnamefont {Prost}}, \bibinfo {author}
  {\bibfnamefont {M.}~\bibnamefont {Rao}},\ and\ \bibinfo {author}
  {\bibfnamefont {R.~A.}\ \bibnamefont {Simha}},\ }\href
  {https://doi.org/10.1103/RevModPhys.85.1143} {\bibfield  {journal} {\bibinfo
  {journal} {Reviews of Modern Physics}\ }\textbf {\bibinfo {volume} {85}},\
  \bibinfo {pages} {1143} (\bibinfo {year} {2013})}\BibitemShut {NoStop}%
\bibitem [{\citenamefont {Ramaswamy}(2010)}]{ramaswamy2010mechanics}%
  \BibitemOpen
  \bibfield  {author} {\bibinfo {author} {\bibfnamefont {S.}~\bibnamefont
  {Ramaswamy}},\ }\href
  {https://doi.org/10.1146/annurev-conmatphys-070909-104101} {\bibfield
  {journal} {\bibinfo  {journal} {Annual Review of Condensed Matter Physics}\
  }\textbf {\bibinfo {volume} {1}},\ \bibinfo {pages} {323} (\bibinfo {year}
  {2010})}\BibitemShut {NoStop}%
\bibitem [{\citenamefont {Toner}\ and\ \citenamefont
  {Tu}(1995)}]{toner1995longrange}%
  \BibitemOpen
  \bibfield  {author} {\bibinfo {author} {\bibfnamefont {J.}~\bibnamefont
  {Toner}}\ and\ \bibinfo {author} {\bibfnamefont {Y.}~\bibnamefont {Tu}},\
  }\href {https://doi.org/10.1103/PhysRevLett.75.4326} {\bibfield  {journal}
  {\bibinfo  {journal} {Physical Review Letters}\ }\textbf {\bibinfo {volume}
  {75}},\ \bibinfo {pages} {4326} (\bibinfo {year} {1995})}\BibitemShut
  {NoStop}%
\bibitem [{\citenamefont {Cates}\ and\ \citenamefont
  {Tailleur}(2015)}]{cates2015motility}%
  \BibitemOpen
  \bibfield  {author} {\bibinfo {author} {\bibfnamefont {M.~E.}\ \bibnamefont
  {Cates}}\ and\ \bibinfo {author} {\bibfnamefont {J.}~\bibnamefont
  {Tailleur}},\ }\href
  {https://doi.org/10.1146/annurev-conmatphys-031214-014710} {\bibfield
  {journal} {\bibinfo  {journal} {Annual Review of Condensed Matter Physics}\
  }\textbf {\bibinfo {volume} {6}},\ \bibinfo {pages} {219} (\bibinfo {year}
  {2015})}\BibitemShut {NoStop}%
\bibitem [{\citenamefont {Fodor}\ \emph {et~al.}(2016)\citenamefont {Fodor},
  \citenamefont {Nardini}, \citenamefont {Cates}, \citenamefont {Tailleur},
  \citenamefont {Visco},\ and\ \citenamefont {van Wijland}}]{fodor2016how}%
  \BibitemOpen
  \bibfield  {author} {\bibinfo {author} {\bibfnamefont {{\'E}.}~\bibnamefont
  {Fodor}}, \bibinfo {author} {\bibfnamefont {C.}~\bibnamefont {Nardini}},
  \bibinfo {author} {\bibfnamefont {M.~E.}\ \bibnamefont {Cates}}, \bibinfo
  {author} {\bibfnamefont {J.}~\bibnamefont {Tailleur}}, \bibinfo {author}
  {\bibfnamefont {P.}~\bibnamefont {Visco}},\ and\ \bibinfo {author}
  {\bibfnamefont {F.}~\bibnamefont {van Wijland}},\ }\href
  {https://doi.org/10.1103/PhysRevLett.117.038103} {\bibfield  {journal}
  {\bibinfo  {journal} {Physical Review Letters}\ }\textbf {\bibinfo {volume}
  {117}},\ \bibinfo {pages} {038103} (\bibinfo {year} {2016})}\BibitemShut
  {NoStop}%
\bibitem [{\citenamefont {Farrell}\ and\ \citenamefont
  {Ioannou}(2003)}]{FarrellIoannou2003}%
  \BibitemOpen
  \bibfield  {author} {\bibinfo {author} {\bibfnamefont {B.~F.}\ \bibnamefont
  {Farrell}}\ and\ \bibinfo {author} {\bibfnamefont {P.~J.}\ \bibnamefont
  {Ioannou}},\ }\href
  {https://doi.org/10.1175/1520-0469(2003)060<2101:SSOTJ>2.0.CO;2} {\bibfield
  {journal} {\bibinfo  {journal} {Journal of the Atmospheric Sciences}\
  }\textbf {\bibinfo {volume} {60}},\ \bibinfo {pages} {2101} (\bibinfo {year}
  {2003})}\BibitemShut {NoStop}%
\bibitem [{\citenamefont {Buzs\'aki}\ and\ \citenamefont
  {Draguhn}(2004)}]{buzsaki2004}%
  \BibitemOpen
  \bibfield  {author} {\bibinfo {author} {\bibfnamefont {G.}~\bibnamefont
  {Buzs\'aki}}\ and\ \bibinfo {author} {\bibfnamefont {A.}~\bibnamefont
  {Draguhn}},\ }\href {https://doi.org/10.1126/science.1099745} {\bibfield
  {journal} {\bibinfo  {journal} {Science}\ }\textbf {\bibinfo {volume}
  {304}},\ \bibinfo {pages} {1926} (\bibinfo {year} {2004})}\BibitemShut
  {NoStop}%
\bibitem [{\citenamefont {Buzs\'aki}\ and\ \citenamefont
  {Moser}(2013)}]{buzsaki2013}%
  \BibitemOpen
  \bibfield  {author} {\bibinfo {author} {\bibfnamefont {G.}~\bibnamefont
  {Buzs\'aki}}\ and\ \bibinfo {author} {\bibfnamefont {E.~I.}\ \bibnamefont
  {Moser}},\ }\href {https://doi.org/10.1038/nn.3304} {\bibfield  {journal}
  {\bibinfo  {journal} {Nat. Neurosci.}\ }\textbf {\bibinfo {volume} {16}},\
  \bibinfo {pages} {130} (\bibinfo {year} {2013})}\BibitemShut {NoStop}%
\bibitem [{\citenamefont {Mukherji}\ \emph {et~al.}(2013)\citenamefont
  {Mukherji}, \citenamefont {Kobiita}, \citenamefont {Ye},\ and\ \citenamefont
  {Chambon}}]{mukherji2013}%
  \BibitemOpen
  \bibfield  {author} {\bibinfo {author} {\bibfnamefont {A.}~\bibnamefont
  {Mukherji}}, \bibinfo {author} {\bibfnamefont {A.}~\bibnamefont {Kobiita}},
  \bibinfo {author} {\bibfnamefont {T.}~\bibnamefont {Ye}},\ and\ \bibinfo
  {author} {\bibfnamefont {P.}~\bibnamefont {Chambon}},\ }\href
  {https://doi.org/10.1016/j.cell.2013.04.020} {\bibfield  {journal} {\bibinfo
  {journal} {Cell}\ }\textbf {\bibinfo {volume} {153}},\ \bibinfo {pages} {812}
  (\bibinfo {year} {2013})}\BibitemShut {NoStop}%
\bibitem [{\citenamefont {Heddes}\ \emph {et~al.}(2022)\citenamefont {Heddes},
  \citenamefont {Altaha}, \citenamefont {Niu}, \citenamefont {Reitmeier},
  \citenamefont {Kleigrewe}, \citenamefont {Haller},\ and\ \citenamefont
  {Kiessling}}]{heddes2022}%
  \BibitemOpen
  \bibfield  {author} {\bibinfo {author} {\bibfnamefont {M.}~\bibnamefont
  {Heddes}}, \bibinfo {author} {\bibfnamefont {B.}~\bibnamefont {Altaha}},
  \bibinfo {author} {\bibfnamefont {Y.}~\bibnamefont {Niu}}, \bibinfo {author}
  {\bibfnamefont {S.}~\bibnamefont {Reitmeier}}, \bibinfo {author}
  {\bibfnamefont {K.}~\bibnamefont {Kleigrewe}}, \bibinfo {author}
  {\bibfnamefont {D.}~\bibnamefont {Haller}},\ and\ \bibinfo {author}
  {\bibfnamefont {S.}~\bibnamefont {Kiessling}},\ }\bibfield  {journal}
  {\bibinfo  {journal} {Nat. Commun.}\ }\textbf {\bibinfo {volume} {13}},\
  \href {https://doi.org/10.1038/s41467-022-33609-x}
  {10.1038/s41467-022-33609-x} (\bibinfo {year} {2022})\BibitemShut {NoStop}%
\bibitem [{\citenamefont {Kuang}\ \emph {et~al.}(2019)\citenamefont {Kuang},
  \citenamefont {Wang}, \citenamefont {Li}, \citenamefont {Ye}, \citenamefont
  {Ruhn}, \citenamefont {Behrendt}, \citenamefont {Olson},\ and\ \citenamefont
  {Hooper}}]{kuang2019}%
  \BibitemOpen
  \bibfield  {author} {\bibinfo {author} {\bibfnamefont {Z.}~\bibnamefont
  {Kuang}}, \bibinfo {author} {\bibfnamefont {Y.}~\bibnamefont {Wang}},
  \bibinfo {author} {\bibfnamefont {Y.}~\bibnamefont {Li}}, \bibinfo {author}
  {\bibfnamefont {C.}~\bibnamefont {Ye}}, \bibinfo {author} {\bibfnamefont
  {K.~A.}\ \bibnamefont {Ruhn}}, \bibinfo {author} {\bibfnamefont {C.~L.}\
  \bibnamefont {Behrendt}}, \bibinfo {author} {\bibfnamefont {E.~N.}\
  \bibnamefont {Olson}},\ and\ \bibinfo {author} {\bibfnamefont {L.~V.}\
  \bibnamefont {Hooper}},\ }\href {https://doi.org/10.1126/science.aaw3134}
  {\bibfield  {journal} {\bibinfo  {journal} {Science}\ }\textbf {\bibinfo
  {volume} {365}},\ \bibinfo {pages} {1428} (\bibinfo {year}
  {2019})}\BibitemShut {NoStop}%
\bibitem [{\citenamefont {Reitmeier}\ \emph {et~al.}(2020)\citenamefont
  {Reitmeier}, \citenamefont {Kiessling}, \citenamefont {Clavel}, \citenamefont
  {List}, \citenamefont {Almeida}, \citenamefont {Ghosh}, \citenamefont
  {Neuhaus}, \citenamefont {Grallert}, \citenamefont {Linseisen}, \citenamefont
  {Skurk}, \citenamefont {Brandl}, \citenamefont {Breuninger}, \citenamefont
  {Troll}, \citenamefont {Rathmann}, \citenamefont {Linkohr}, \citenamefont
  {Hauner}, \citenamefont {Laudes}, \citenamefont {Franke}, \citenamefont
  {Roy}, \citenamefont {Bell}, \citenamefont {Spector}, \citenamefont
  {Baumbach}, \citenamefont {O’Toole}, \citenamefont {Peters},\ and\
  \citenamefont {Haller}}]{reitmeier2020}%
  \BibitemOpen
  \bibfield  {author} {\bibinfo {author} {\bibfnamefont {S.}~\bibnamefont
  {Reitmeier}}, \bibinfo {author} {\bibfnamefont {S.}~\bibnamefont
  {Kiessling}}, \bibinfo {author} {\bibfnamefont {T.}~\bibnamefont {Clavel}},
  \bibinfo {author} {\bibfnamefont {M.}~\bibnamefont {List}}, \bibinfo {author}
  {\bibfnamefont {E.~L.}\ \bibnamefont {Almeida}}, \bibinfo {author}
  {\bibfnamefont {T.~S.}\ \bibnamefont {Ghosh}}, \bibinfo {author}
  {\bibfnamefont {K.}~\bibnamefont {Neuhaus}}, \bibinfo {author} {\bibfnamefont
  {H.}~\bibnamefont {Grallert}}, \bibinfo {author} {\bibfnamefont
  {J.}~\bibnamefont {Linseisen}}, \bibinfo {author} {\bibfnamefont
  {T.}~\bibnamefont {Skurk}}, \bibinfo {author} {\bibfnamefont
  {B.}~\bibnamefont {Brandl}}, \bibinfo {author} {\bibfnamefont {T.~A.}\
  \bibnamefont {Breuninger}}, \bibinfo {author} {\bibfnamefont
  {M.}~\bibnamefont {Troll}}, \bibinfo {author} {\bibfnamefont
  {W.}~\bibnamefont {Rathmann}}, \bibinfo {author} {\bibfnamefont
  {B.}~\bibnamefont {Linkohr}}, \bibinfo {author} {\bibfnamefont
  {H.}~\bibnamefont {Hauner}}, \bibinfo {author} {\bibfnamefont
  {M.}~\bibnamefont {Laudes}}, \bibinfo {author} {\bibfnamefont
  {A.}~\bibnamefont {Franke}}, \bibinfo {author} {\bibfnamefont {C.~I.~L.}\
  \bibnamefont {Roy}}, \bibinfo {author} {\bibfnamefont {J.~T.}\ \bibnamefont
  {Bell}}, \bibinfo {author} {\bibfnamefont {T.}~\bibnamefont {Spector}},
  \bibinfo {author} {\bibfnamefont {J.}~\bibnamefont {Baumbach}}, \bibinfo
  {author} {\bibfnamefont {P.~W.}\ \bibnamefont {O’Toole}}, \bibinfo {author}
  {\bibfnamefont {A.}~\bibnamefont {Peters}},\ and\ \bibinfo {author}
  {\bibfnamefont {D.}~\bibnamefont {Haller}},\ }\href
  {https://doi.org/10.1016/j.chom.2020.06.004} {\bibfield  {journal} {\bibinfo
  {journal} {Cell Host Microbe}\ }\textbf {\bibinfo {volume} {28}},\ \bibinfo
  {pages} {258} (\bibinfo {year} {2020})}\BibitemShut {NoStop}%
\bibitem [{\citenamefont {Tuganbaev}\ \emph {et~al.}(2020)\citenamefont
  {Tuganbaev}, \citenamefont {Mor}, \citenamefont {Bashiardes}, \citenamefont
  {Liwinski}, \citenamefont {Nobs}, \citenamefont {Leshem}, \citenamefont
  {Dori-Bachash}, \citenamefont {Thaiss}, \citenamefont {Pinker}, \citenamefont
  {Ratiner}, \citenamefont {Adlung}, \citenamefont {Federici}, \citenamefont
  {Kleimeyer}, \citenamefont {Moresi}, \citenamefont {Yamada}, \citenamefont
  {Cohen}, \citenamefont {Zhang}, \citenamefont {Massalha}, \citenamefont
  {Massasa}, \citenamefont {Kuperman}, \citenamefont {Koni}, \citenamefont
  {Harmelin}, \citenamefont {Gao}, \citenamefont {Itzkovitz}, \citenamefont
  {Honda}, \citenamefont {Shapiro},\ and\ \citenamefont
  {Elinav}}]{tuganbaev2020}%
  \BibitemOpen
  \bibfield  {author} {\bibinfo {author} {\bibfnamefont {T.}~\bibnamefont
  {Tuganbaev}}, \bibinfo {author} {\bibfnamefont {U.}~\bibnamefont {Mor}},
  \bibinfo {author} {\bibfnamefont {S.}~\bibnamefont {Bashiardes}}, \bibinfo
  {author} {\bibfnamefont {T.}~\bibnamefont {Liwinski}}, \bibinfo {author}
  {\bibfnamefont {S.~P.}\ \bibnamefont {Nobs}}, \bibinfo {author}
  {\bibfnamefont {A.}~\bibnamefont {Leshem}}, \bibinfo {author} {\bibfnamefont
  {M.}~\bibnamefont {Dori-Bachash}}, \bibinfo {author} {\bibfnamefont {C.~A.}\
  \bibnamefont {Thaiss}}, \bibinfo {author} {\bibfnamefont {E.~Y.}\
  \bibnamefont {Pinker}}, \bibinfo {author} {\bibfnamefont {K.}~\bibnamefont
  {Ratiner}}, \bibinfo {author} {\bibfnamefont {L.}~\bibnamefont {Adlung}},
  \bibinfo {author} {\bibfnamefont {S.}~\bibnamefont {Federici}}, \bibinfo
  {author} {\bibfnamefont {C.}~\bibnamefont {Kleimeyer}}, \bibinfo {author}
  {\bibfnamefont {C.}~\bibnamefont {Moresi}}, \bibinfo {author} {\bibfnamefont
  {T.}~\bibnamefont {Yamada}}, \bibinfo {author} {\bibfnamefont
  {Y.}~\bibnamefont {Cohen}}, \bibinfo {author} {\bibfnamefont
  {X.}~\bibnamefont {Zhang}}, \bibinfo {author} {\bibfnamefont
  {H.}~\bibnamefont {Massalha}}, \bibinfo {author} {\bibfnamefont
  {E.}~\bibnamefont {Massasa}}, \bibinfo {author} {\bibfnamefont
  {Y.}~\bibnamefont {Kuperman}}, \bibinfo {author} {\bibfnamefont {P.~A.}\
  \bibnamefont {Koni}}, \bibinfo {author} {\bibfnamefont {A.}~\bibnamefont
  {Harmelin}}, \bibinfo {author} {\bibfnamefont {N.}~\bibnamefont {Gao}},
  \bibinfo {author} {\bibfnamefont {S.}~\bibnamefont {Itzkovitz}}, \bibinfo
  {author} {\bibfnamefont {K.}~\bibnamefont {Honda}}, \bibinfo {author}
  {\bibfnamefont {H.}~\bibnamefont {Shapiro}},\ and\ \bibinfo {author}
  {\bibfnamefont {E.}~\bibnamefont {Elinav}},\ }\href
  {https://doi.org/10.1016/j.cell.2020.08.027} {\bibfield  {journal} {\bibinfo
  {journal} {Cell}\ }\textbf {\bibinfo {volume} {182}},\ \bibinfo {pages}
  {1441} (\bibinfo {year} {2020})}\BibitemShut {NoStop}%
\bibitem [{\citenamefont {Voigt}\ \emph {et~al.}(2014)\citenamefont {Voigt},
  \citenamefont {Forsyth}, \citenamefont {Green}, \citenamefont {Mutlu},
  \citenamefont {Engen}, \citenamefont {Vitaterna}, \citenamefont {Turek},\
  and\ \citenamefont {Keshavarzian}}]{voigt2014}%
  \BibitemOpen
  \bibfield  {author} {\bibinfo {author} {\bibfnamefont {R.~M.}\ \bibnamefont
  {Voigt}}, \bibinfo {author} {\bibfnamefont {C.~B.}\ \bibnamefont {Forsyth}},
  \bibinfo {author} {\bibfnamefont {S.~J.}\ \bibnamefont {Green}}, \bibinfo
  {author} {\bibfnamefont {E.}~\bibnamefont {Mutlu}}, \bibinfo {author}
  {\bibfnamefont {P.}~\bibnamefont {Engen}}, \bibinfo {author} {\bibfnamefont
  {M.~H.}\ \bibnamefont {Vitaterna}}, \bibinfo {author} {\bibfnamefont {F.~W.}\
  \bibnamefont {Turek}},\ and\ \bibinfo {author} {\bibfnamefont
  {A.}~\bibnamefont {Keshavarzian}},\ }\href
  {https://doi.org/10.1371/journal.pone.0097500} {\bibfield  {journal}
  {\bibinfo  {journal} {PLoS ONE}\ }\textbf {\bibinfo {volume} {9}},\ \bibinfo
  {pages} {e97500} (\bibinfo {year} {2014})}\BibitemShut {NoStop}%
\bibitem [{\citenamefont {Deaver}\ \emph {et~al.}(2018)\citenamefont {Deaver},
  \citenamefont {Eum},\ and\ \citenamefont {Toborek}}]{deaver2018}%
  \BibitemOpen
  \bibfield  {author} {\bibinfo {author} {\bibfnamefont {J.~A.}\ \bibnamefont
  {Deaver}}, \bibinfo {author} {\bibfnamefont {S.~Y.}\ \bibnamefont {Eum}},\
  and\ \bibinfo {author} {\bibfnamefont {M.}~\bibnamefont {Toborek}},\
  }\bibfield  {journal} {\bibinfo  {journal} {Front. Microbiol.}\ }\textbf
  {\bibinfo {volume} {9}},\ \href {https://doi.org/10.3389/fmicb.2018.00737}
  {10.3389/fmicb.2018.00737} (\bibinfo {year} {2018})\BibitemShut {NoStop}%
\bibitem [{\citenamefont {Altaha}\ \emph {et~al.}(2022)\citenamefont {Altaha},
  \citenamefont {Heddes}, \citenamefont {Pilorz}, \citenamefont {Niu},
  \citenamefont {Gorbunova}, \citenamefont {Gigl}, \citenamefont {Kleigrewe},
  \citenamefont {Oster}, \citenamefont {Haller},\ and\ \citenamefont
  {Kiessling}}]{altaha2022}%
  \BibitemOpen
  \bibfield  {author} {\bibinfo {author} {\bibfnamefont {B.}~\bibnamefont
  {Altaha}}, \bibinfo {author} {\bibfnamefont {M.}~\bibnamefont {Heddes}},
  \bibinfo {author} {\bibfnamefont {V.}~\bibnamefont {Pilorz}}, \bibinfo
  {author} {\bibfnamefont {Y.}~\bibnamefont {Niu}}, \bibinfo {author}
  {\bibfnamefont {E.}~\bibnamefont {Gorbunova}}, \bibinfo {author}
  {\bibfnamefont {M.}~\bibnamefont {Gigl}}, \bibinfo {author} {\bibfnamefont
  {K.}~\bibnamefont {Kleigrewe}}, \bibinfo {author} {\bibfnamefont
  {H.}~\bibnamefont {Oster}}, \bibinfo {author} {\bibfnamefont
  {D.}~\bibnamefont {Haller}},\ and\ \bibinfo {author} {\bibfnamefont
  {S.}~\bibnamefont {Kiessling}},\ }\href
  {https://doi.org/10.1016/j.molmet.2022.101628} {\bibfield  {journal}
  {\bibinfo  {journal} {Mol. Metab.}\ }\textbf {\bibinfo {volume} {66}},\
  \bibinfo {pages} {101628} (\bibinfo {year} {2022})}\BibitemShut {NoStop}%
\bibitem [{\citenamefont {Thaiss}\ \emph {et~al.}(2016)\citenamefont {Thaiss},
  \citenamefont {Levy}, \citenamefont {Korem}, \citenamefont {Dohnalová},
  \citenamefont {Shapiro}, \citenamefont {Jaitin}, \citenamefont {David},
  \citenamefont {Winter}, \citenamefont {Gury-BenAri}, \citenamefont
  {Tatirovsky}, \citenamefont {Tuganbaev}, \citenamefont {Federici},
  \citenamefont {Zmora}, \citenamefont {Zeevi}, \citenamefont {Dori-Bachash},
  \citenamefont {Pevsner-Fischer}, \citenamefont {Kartvelishvily},
  \citenamefont {Brandis}, \citenamefont {Harmelin}, \citenamefont {Shibolet},
  \citenamefont {Halpern}, \citenamefont {Honda}, \citenamefont {Amit},
  \citenamefont {Segal},\ and\ \citenamefont {Elinav}}]{thaiss2016transcr}%
  \BibitemOpen
  \bibfield  {author} {\bibinfo {author} {\bibfnamefont {C.~A.}\ \bibnamefont
  {Thaiss}}, \bibinfo {author} {\bibfnamefont {M.}~\bibnamefont {Levy}},
  \bibinfo {author} {\bibfnamefont {T.}~\bibnamefont {Korem}}, \bibinfo
  {author} {\bibfnamefont {L.}~\bibnamefont {Dohnalová}}, \bibinfo {author}
  {\bibfnamefont {H.}~\bibnamefont {Shapiro}}, \bibinfo {author} {\bibfnamefont
  {D.~A.}\ \bibnamefont {Jaitin}}, \bibinfo {author} {\bibfnamefont
  {E.}~\bibnamefont {David}}, \bibinfo {author} {\bibfnamefont {D.~R.}\
  \bibnamefont {Winter}}, \bibinfo {author} {\bibfnamefont {M.}~\bibnamefont
  {Gury-BenAri}}, \bibinfo {author} {\bibfnamefont {E.}~\bibnamefont
  {Tatirovsky}}, \bibinfo {author} {\bibfnamefont {T.}~\bibnamefont
  {Tuganbaev}}, \bibinfo {author} {\bibfnamefont {S.}~\bibnamefont {Federici}},
  \bibinfo {author} {\bibfnamefont {N.}~\bibnamefont {Zmora}}, \bibinfo
  {author} {\bibfnamefont {D.}~\bibnamefont {Zeevi}}, \bibinfo {author}
  {\bibfnamefont {M.}~\bibnamefont {Dori-Bachash}}, \bibinfo {author}
  {\bibfnamefont {M.}~\bibnamefont {Pevsner-Fischer}}, \bibinfo {author}
  {\bibfnamefont {E.}~\bibnamefont {Kartvelishvily}}, \bibinfo {author}
  {\bibfnamefont {A.}~\bibnamefont {Brandis}}, \bibinfo {author} {\bibfnamefont
  {A.}~\bibnamefont {Harmelin}}, \bibinfo {author} {\bibfnamefont
  {O.}~\bibnamefont {Shibolet}}, \bibinfo {author} {\bibfnamefont
  {Z.}~\bibnamefont {Halpern}}, \bibinfo {author} {\bibfnamefont
  {K.}~\bibnamefont {Honda}}, \bibinfo {author} {\bibfnamefont
  {I.}~\bibnamefont {Amit}}, \bibinfo {author} {\bibfnamefont {E.}~\bibnamefont
  {Segal}},\ and\ \bibinfo {author} {\bibfnamefont {E.}~\bibnamefont
  {Elinav}},\ }\href {https://doi.org/10.1016/j.cell.2016.11.003} {\bibfield
  {journal} {\bibinfo  {journal} {Cell}\ }\textbf {\bibinfo {volume} {167}},\
  \bibinfo {pages} {1495} (\bibinfo {year} {2016})}\BibitemShut {NoStop}%
\bibitem [{\citenamefont {Paulose}\ \emph {et~al.}(2016)\citenamefont
  {Paulose}, \citenamefont {Wright}, \citenamefont {Patel},\ and\ \citenamefont
  {Cassone}}]{paulose2016}%
  \BibitemOpen
  \bibfield  {author} {\bibinfo {author} {\bibfnamefont {J.~K.}\ \bibnamefont
  {Paulose}}, \bibinfo {author} {\bibfnamefont {J.~M.}\ \bibnamefont {Wright}},
  \bibinfo {author} {\bibfnamefont {A.~G.}\ \bibnamefont {Patel}},\ and\
  \bibinfo {author} {\bibfnamefont {V.~M.}\ \bibnamefont {Cassone}},\ }\href
  {https://doi.org/10.1371/journal.pone.0146643} {\bibfield  {journal}
  {\bibinfo  {journal} {PLoS ONE}\ }\textbf {\bibinfo {volume} {11}},\ \bibinfo
  {pages} {e0146643} (\bibinfo {year} {2016})}\BibitemShut {NoStop}%
\bibitem [{\citenamefont {Bishehsari}\ \emph {et~al.}(2020)\citenamefont
  {Bishehsari}, \citenamefont {Voigt},\ and\ \citenamefont
  {Keshavarzian}}]{zheng2020review}%
  \BibitemOpen
  \bibfield  {author} {\bibinfo {author} {\bibfnamefont {F.}~\bibnamefont
  {Bishehsari}}, \bibinfo {author} {\bibfnamefont {R.~M.}\ \bibnamefont
  {Voigt}},\ and\ \bibinfo {author} {\bibfnamefont {A.}~\bibnamefont
  {Keshavarzian}},\ }\href {https://doi.org/10.1038/s41574-020-00427-4}
  {\bibfield  {journal} {\bibinfo  {journal} {Nat. Rev. Endocrinol.}\ }\textbf
  {\bibinfo {volume} {16}},\ \bibinfo {pages} {731} (\bibinfo {year}
  {2020})}\BibitemShut {NoStop}%
\bibitem [{\citenamefont {Teichman}\ \emph {et~al.}(2020)\citenamefont
  {Teichman}, \citenamefont {O’Riordan}, \citenamefont {Gahan}, \citenamefont
  {Dinan},\ and\ \citenamefont {Cryan}}]{teichman2020}%
  \BibitemOpen
  \bibfield  {author} {\bibinfo {author} {\bibfnamefont {E.~M.}\ \bibnamefont
  {Teichman}}, \bibinfo {author} {\bibfnamefont {K.~J.}\ \bibnamefont
  {O’Riordan}}, \bibinfo {author} {\bibfnamefont {C.~G.}\ \bibnamefont
  {Gahan}}, \bibinfo {author} {\bibfnamefont {T.~G.}\ \bibnamefont {Dinan}},\
  and\ \bibinfo {author} {\bibfnamefont {J.~F.}\ \bibnamefont {Cryan}},\ }\href
  {https://doi.org/10.1016/j.cmet.2020.02.008} {\bibfield  {journal} {\bibinfo
  {journal} {Cell Metab.}\ }\textbf {\bibinfo {volume} {31}},\ \bibinfo {pages}
  {448} (\bibinfo {year} {2020})}\BibitemShut {NoStop}%
\bibitem [{\citenamefont {Voigt}\ \emph {et~al.}(2016)\citenamefont {Voigt},
  \citenamefont {Forsyth}, \citenamefont {Green}, \citenamefont {Engen},\ and\
  \citenamefont {Keshavarzian}}]{voigt2016review}%
  \BibitemOpen
  \bibfield  {author} {\bibinfo {author} {\bibfnamefont {R.}~\bibnamefont
  {Voigt}}, \bibinfo {author} {\bibfnamefont {C.}~\bibnamefont {Forsyth}},
  \bibinfo {author} {\bibfnamefont {S.}~\bibnamefont {Green}}, \bibinfo
  {author} {\bibfnamefont {P.}~\bibnamefont {Engen}},\ and\ \bibinfo {author}
  {\bibfnamefont {A.}~\bibnamefont {Keshavarzian}},\ }\href
  {https://doi.org/10.1016/bs.irn.2016.07.002} {\bibfield  {journal} {\bibinfo
  {journal} {Int. Rev. Neurobiol.}\ ,\ \bibinfo {pages} {193}} (\bibinfo {year}
  {2016})}\BibitemShut {NoStop}%
\bibitem [{\citenamefont {Chaumeil}\ \emph {et~al.}(2019)\citenamefont
  {Chaumeil}, \citenamefont {Mussig}, \citenamefont {Hugenholtz},\ and\
  \citenamefont {Parks}}]{chaumeil2019gtdbtk}%
  \BibitemOpen
  \bibfield  {author} {\bibinfo {author} {\bibfnamefont {P.-A.}\ \bibnamefont
  {Chaumeil}}, \bibinfo {author} {\bibfnamefont {A.~J.}\ \bibnamefont
  {Mussig}}, \bibinfo {author} {\bibfnamefont {P.}~\bibnamefont {Hugenholtz}},\
  and\ \bibinfo {author} {\bibfnamefont {D.~H.}\ \bibnamefont {Parks}},\ }\href
  {https://doi.org/10.1093/bioinformatics/btz848} {\bibfield  {journal}
  {\bibinfo  {journal} {Bioinformatics}\ }\textbf {\bibinfo {volume} {36}},\
  \bibinfo {pages} {1925} (\bibinfo {year} {2019})}\BibitemShut {NoStop}%
\end{thebibliography}%

\clearpage

\onecolumngrid

\setcounter{table}{0}
\renewcommand{\thetable}{S\arabic{table}}

\appendix
\section{Supplementary Material\\\ No persistent circadian oscillator at genome resolution:\\ pseudo-coherence in gut microbiome dynamics}

\noindent This Supplementary Material provides the full per-MAG mode-loading tables, the genus-level aggregation, the multiple-comparison-corrected surrogate results, and the mode-concentration statistics referenced in the main text. All loadings are mean absolute components of the unit-norm reaction ($\hat r$) and non-normal ($\hat n$) modes over the calibration windows of Mouse~A (118 cyc7plus-labelled MAGs, 241 windows of length $L=3$). Signs of the modes are calibration gauges and are not reported.

\FloatBarrier
\subsection{Top reaction-mode loadings (Mouse A)}
\begin{table}[h]\centering\small
\caption{Top-25 MAGs by mean absolute reaction-mode loading $\langle|r_i|\rangle_t$ (which genomes absorb the amplified excursion). Cluster labels are the cyc7plus assignment of Maskawa et al.}
\begin{tabular}{rllll}
\toprule
Rank & $\langle|r_i|\rangle_t$ & Cluster & Genus & Family \\
\midrule
1 & 0.108 & 2 & \emph{MGBC157735} & Pumilibacteraceae \\
2 & 0.104 & 2 & \emph{Sporofaciens} & Lachnospiraceae \\
3 & 0.102 & 2 & \emph{Pseudobutyricicoccus} & Butyricicoccaceae \\
4 & 0.099 & 2 & \emph{Sporofaciens} & Lachnospiraceae \\
5 & 0.099 & 2 & \emph{UBA7109} & Lachnospiraceae \\
6 & 0.096 & 1 & \emph{Fimisoma} & Anaerovoracaceae \\
7 & 0.095 & 2 & \emph{Avidehalobacter} & UBA5755 \\
8 & 0.093 & 2 & \emph{Sporofaciens} & Lachnospiraceae \\
9 & 0.092 & 1 & \emph{Bacteroides} & Bacteroidaceae \\
10 & 0.090 & 1 & \emph{CAG-485} & Muribaculaceae \\
11 & 0.089 & 2 & \emph{Caccovicinus} & Lachnospiraceae \\
12 & 0.089 & 1 & \emph{Duncaniella} & Muribaculaceae \\
13 & 0.088 & 2 & \emph{MGBC131033} & Lachnospiraceae \\
14 & 0.087 & 1 & \emph{Prevotella} & Bacteroidaceae \\
15 & 0.087 & 1 & \emph{CAG-485} & Muribaculaceae \\
16 & 0.086 & 2 & \emph{UBA1405} & Ruminococcaceae \\
17 & 0.085 & 2 & \emph{Eubacterium\_F} & Lachnospiraceae \\
18 & 0.085 & 2 & \emph{UBA3402} & Lachnospiraceae \\
19 & 0.082 & 1 & \emph{Bacteroides} & Bacteroidaceae \\
20 & 0.082 & 2 & \emph{UBA3402} & Lachnospiraceae \\
21 & 0.081 & 2 & \emph{Dysosmobacter} & Oscillospiraceae \\
22 & 0.079 & 2 & \emph{CAG-317} & Lachnospiraceae \\
23 & 0.079 & 2 & \emph{Coproplasma} & Borkfalkiaceae \\
24 & 0.078 & 1 & \emph{Emergencia} & Anaerovoracaceae \\
25 & 0.078 & 2 & \emph{CAG-95} & Lachnospiraceae \\
\bottomrule
\end{tabular}
\end{table}

\FloatBarrier
\subsection{Top non-normal-mode loadings (Mouse A)}
\begin{table}[h]\centering\small
\caption{Top-25 MAGs by mean absolute non-normal-mode loading $\langle|n_i|\rangle_t$ (which genomes drive the amplified excursion).}
\begin{tabular}{rllll}
\toprule
Rank & $\langle|n_i|\rangle_t$ & Cluster & Genus & Family \\
\midrule
1 & 0.122 & 1 & \emph{Prevotella} & Bacteroidaceae \\
2 & 0.120 & 2 & \emph{Pseudobutyricicoccus} & Butyricicoccaceae \\
3 & 0.118 & 1 & \emph{CAG-485} & Muribaculaceae \\
4 & 0.112 & 1 & \emph{Fimisoma} & Anaerovoracaceae \\
5 & 0.111 & 2 & \emph{MGBC157735} & Pumilibacteraceae \\
6 & 0.108 & 2 & \emph{UBA1405} & Ruminococcaceae \\
7 & 0.107 & 1 & \emph{CAG-485} & Muribaculaceae \\
8 & 0.103 & 2 & \emph{UBA7109} & Lachnospiraceae \\
9 & 0.100 & 1 & \emph{Duncaniella} & Muribaculaceae \\
10 & 0.094 & 2 & \emph{Sporofaciens} & Lachnospiraceae \\
11 & 0.094 & 1 & \emph{CAG-485} & Muribaculaceae \\
12 & 0.094 & 2 & \emph{Sporofaciens} & Lachnospiraceae \\
13 & 0.092 & 2 & \emph{Coproplasma} & Borkfalkiaceae \\
14 & 0.092 & 2 & \emph{Avidehalobacter} & UBA5755 \\
15 & 0.091 & 1 & \emph{Alistipes} & Rikenellaceae \\
16 & 0.091 & 1 & \emph{Bacteroides} & Bacteroidaceae \\
17 & 0.091 & 2 & \emph{Coproplasma} & Borkfalkiaceae \\
18 & 0.087 & 2 & \emph{MGBC131033} & Lachnospiraceae \\
19 & 0.086 & 1 & \emph{Bacteroides} & Bacteroidaceae \\
20 & 0.086 & 2 & \emph{UBA3402} & Lachnospiraceae \\
21 & 0.083 & 1 & \emph{CAG-873} & Muribaculaceae \\
22 & 0.083 & 2 & \emph{Sporofaciens} & Lachnospiraceae \\
23 & 0.082 & 1 & \emph{Emergencia} & Anaerovoracaceae \\
24 & 0.082 & 1 & \emph{CAG-485} & Muribaculaceae \\
25 & 0.081 & 2 & \emph{Pseudobutyricicoccus} & Butyricicoccaceae \\
\bottomrule
\end{tabular}
\end{table}

\FloatBarrier
\subsection{Genus-level aggregation}
\begin{table}[h]\centering\small
\caption{Top-15 genera by aggregated squared reaction-mode loading $\sum_i |r_i|^2$ over their MAGs.}
\begin{tabular}{rllll}
\toprule
Rank & Genus & \# MAGs & $\sum_i |r_i|^2$ & Family \\
\midrule
1 & \emph{Sporofaciens} & 5 & 0.0326 & Lachnospiraceae \\
2 & \emph{CAG-485} & 5 & 0.0283 & Muribaculaceae \\
3 & \emph{COE1} & 6 & 0.0243 & Lachnospiraceae \\
4 & \emph{UBA3402} & 4 & 0.0238 & Lachnospiraceae \\
5 & \emph{Bacteroides} & 4 & 0.0219 & Bacteroidaceae \\
6 & \emph{Dysosmobacter} & 5 & 0.0210 & Oscillospiraceae \\
7 & \emph{Choladocola} & 7 & 0.0208 & Lachnospiraceae \\
8 & \emph{Pseudobutyricicoccus} & 3 & 0.0198 & Butyricicoccaceae \\
9 & \emph{CAG-873} & 6 & 0.0189 & Muribaculaceae \\
10 & \emph{Muribaculum} & 6 & 0.0158 & Muribaculaceae \\
11 & \emph{Coproplasma} & 3 & 0.0150 & Borkfalkiaceae \\
12 & \emph{UBA3282} & 6 & 0.0148 & Lachnospiraceae \\
13 & \emph{Ligilactobacillus} & 6 & 0.0137 & Lactobacillaceae \\
14 & \emph{Duncaniella} & 3 & 0.0136 & Muribaculaceae \\
15 & \emph{CAG-95} & 3 & 0.0128 & Lachnospiraceae \\
\bottomrule
\end{tabular}
\end{table}

\FloatBarrier
\subsection{Mode concentration vs a random-unit-vector null}
\begin{table}[h]\centering\small
\caption{Concentration of the unit-norm modes, time-averaged: cumulative energy ($p_i=r_i^2$) in the top-5/10/20 MAGs, and inverse participation ratio $N_{\rm eff}=1/\sum_i p_i^2$. A random unit vector in $N$ dimensions has $N_{\rm eff}\approx N/3$. The empirical modes are about twice as concentrated as this null, on a few tens of MAGs across the two guilds.}
\begin{tabular}{lllll}
\toprule
Mouse / mode & top-5 & top-10 & top-20 & $N_{\rm eff}$ \\
\midrule
A ($N=118$) random null & 24.3\% & 39.0\% & 59.1\% & 40.7 \\
A reaction mode & 39.9\% & 57.9\% & 77.2\% & 21.9 \\
A non-normal mode & 41.1\% & 59.0\% & 78.0\% & 20.3 \\
\midrule
B ($N=181$) random null & 17.9\% & 29.5\% & 46.4\% & 61.7 \\
B reaction mode & 33.8\% & 48.9\% & 66.7\% & 29.0 \\
B non-normal mode & 33.3\% & 48.2\% & 66.0\% & 28.9 \\
\bottomrule
\end{tabular}
\end{table}

\FloatBarrier
\subsection{Surrogate spectral test (Benjamini--Hochberg corrected)}
\begin{table}[h]\centering\small
\caption{Frequency bins exceeding the amplitude-adjusted Fourier surrogate null ($N_{\rm surr}=250$) after Benjamini--Hochberg correction at FDR $0.05$, on the full record. Mouse~A: no significant bin in either window. Mouse~B (full record): significant in two bands; none survives after removing the first $72$\,h.}
\begin{tabular}{lllll}
\toprule
Mouse / window & significant band & period & empirical $p$ & BH $p$ \\
\midrule
A, full \& post-72h & none & --- & --- & --- \\
B, full record & $0.035$--$0.046\,\mathrm{h}^{-1}$ & $\sim$22--28\,h & 0.004 & 0.033 \\
B, full record & $0.010$--$0.013\,\mathrm{h}^{-1}$ & $\sim$79--95\,h & 0.004 & 0.033 \\
B, post-72h & none & --- & --- & --- \\
\bottomrule
\end{tabular}
\end{table}

\end{document}